\pdfoutput=1
\documentclass[12pt]{article}
\usepackage[
top = 2.52cm, 
bottom = 2.52cm, 
left = 2.52cm, 
right = 2.52cm]{geometry}

\usepackage{amsmath, amssymb, amscd, amsthm, amsfonts, amstext, mathrsfs}
\usepackage{physics}
 \usepackage{color}
 \usepackage{graphicx}
\usepackage{tensor}

\usepackage{latexsym} 
\usepackage{verbatim}
\usepackage{tikz}
\usepackage{dsfont}
\usepackage{xcolor}
\usepackage{mathtools}
\usepackage{bm,bbm}

\definecolor{DarkBlue}{rgb}{0.0, 0.0, 0.55}
\usepackage[colorlinks=true, allcolors=DarkBlue, linktoc=all]{hyperref}

\makeatletter
\g@addto@macro\normalsize{%
    \setlength\abovedisplayskip{12.5pt}
    \setlength\belowdisplayskip{12.5pt}
    \setlength\abovedisplayshortskip{12.5pt}
    \setlength\belowdisplayshortskip{12.5pt}
}
\makeatother

\usepackage[bbgreekl]{mathbbol}

\newcommand{\vecxi}{{\bbxi}}

\newcommand{\vecz}{{\mathbb z}}

\newcommand{\veceta}{\bbeta}

\makeatletter
\newcommand{\doublewidetilde}[1]{{%
\mathpalette\double@widetilde{#1}%
}}
\newcommand{\double@widetilde}[2]{%
\sbox\z@{$\m@th#1\widetilde{#2}$}%
 \ht\z@=.9\ht\z@
 \widetilde{\box\z@}%
}
\makeatother

\newcommand{\be}{\begin{equation}}
\newcommand{\ee}{\end{equation}}
\def\nn{\nonumber}
\newcommand{\ord}{{\mathscr O}}

\usepackage{tikz}
\newcommand*{\boxcolor}{black}
\makeatletter
\renewcommand{\boxed}[1]{\textcolor{\boxcolor}{%
\tikz[baseline={([yshift=-1ex]current bounding box.center)}] \node [rectangle, minimum width=1ex,rounded corners,draw] {\normalcolor\m@th$\;\,\displaystyle#1\;\,$};}}
 \makeatother

\usetikzlibrary{shapes.geometric}

\newcommand{\dia}{\tikz[baseline={([yshift=0.3mm]cross diamond.south)}]{
        \node[draw, line width=0.25mm, scale=0.55, diamond,
        path picture={\fill[gray] 
                (path picture bounding box.south) 
                -- (path picture bounding box.east) 
                -- (path picture bounding box.north) 
                -- (path picture bounding box.west) 
                -- cycle;}
        ](cross diamond){};
    }}
     
\newcommand{\diaPast}{\tikz[baseline={([yshift=0.3mm]cross diamond.south)}]{
        \node[draw, line width=0.25mm, scale=0.55, diamond,
        path picture={\fill[gray] 
                (path picture bounding box.center) 
                -- (path picture bounding box.-45) 
                -- (path picture bounding box.south) 
                -- (path picture bounding box.220) 
                -- cycle;
        \draw  (path picture bounding box.220) -- (path picture bounding box.center) -- (path picture bounding box.-45);}
        ](cross diamond){};
    }}

\newcommand{\diaUpper}{\tikz[baseline={([yshift=0.3mm]cross diamond.south)}]{
        \node[draw, line width=0.25mm, scale=0.55, diamond,
        path picture={\fill[gray] 
                (path picture bounding box.west) 
                -- (path picture bounding box.east) 
                -- (path picture bounding box.north) 
                -- cycle;
        \draw  (path picture bounding box.west) -- (path picture bounding box.east) -- (path picture bounding box.north);}
        ](cross diamond){};
    }}
    
\newcommand{\diaLower}{\tikz[baseline={([yshift=0.3mm]cross diamond.south)}]{
        \node[draw, line width=0.25mm, scale=0.55, diamond,
        path picture={\fill[gray] 
                (path picture bounding box.west) 
                -- (path picture bounding box.east) 
                -- (path picture bounding box.south) 
                -- cycle;
        \draw  (path picture bounding box.west) -- (path picture bounding box.east) -- (path picture bounding box.south);}
        ](cross diamond){};
    }}
    
\newcommand{\diasmall}{\tikz[baseline={([yshift=0.2mm]cross diamond.south)}]{
        \node[draw, line width=0.2mm, scale=0.4, diamond,
        path picture={\fill[gray] 
                (path picture bounding box.south) 
                -- (path picture bounding box.east) 
                -- (path picture bounding box.north) 
                -- (path picture bounding box.west) 
                -- cycle;}
        ](cross diamond){};
    }}
    
    \newcommand{\diasmallPast}{\tikz[baseline={([yshift=0.2mm]cross diamond.south)}]{
        \node[draw, line width=0.2mm, scale=0.4, diamond,
        path picture={\fill[gray] 
                (path picture bounding box.center) 
                -- (path picture bounding box.-45) 
                -- (path picture bounding box.south) 
                -- (path picture bounding box.220) 
                -- cycle;
        \draw  (path picture bounding box.220) -- (path picture bounding box.center) -- (path picture bounding box.-45);}
        ](cross diamond){};
    }}

\begin{document}
\thispagestyle{empty}

\begin{titlepage}

\begin{flushright} 
\end{flushright}

\begin{center} 
\vspace{0.7cm}  

{{ \fontsize{16pt}{0pt}
\bf
Virasoro OPE Blocks, Causal Diamonds,\\ \vspace{0.4cm}  
and Higher-Dimensional CFT
}}
\vspace{1.2cm}  
\\
 {\fontsize{12pt}{12pt}{Felix M.\ Haehl and Kuo-Wei Huang 
}}   
\\ 
\vspace{0.7cm} 
{\fontsize{11pt}{0pt}{\it 
School of Mathematical Sciences $\&$ STAG Research Centre, \\
University of Southampton, SO17 1BJ, U.K.
}}\\
\end{center}
\vspace{2cm} 

\begin{center} 
{\bf Abstract}
\end{center} 
{\noindent  
In two-dimensional Conformal Field Theory (CFT), multi-stress tensor exchanges between probe operators give rise to the Virasoro identity conformal block, which is fixed by symmetry. The analogous object, and the corresponding organizing principles, in higher dimensions are less well understood. In this paper, we study the Virasoro identity OPE block, which is a bilocal operator that projects two primaries onto the conformal family of multi-stress tensor states. Generalizing a known construction of global OPE blocks, our formalism uses integrals over nested causal diamonds associated with two timelike-separated insertions. We argue that our construction is adaptable to higher dimensions, and use it to provide a new derivation of the single-stress tensor exchange contribution to a four-point correlator in both three and four dimensions, to leading order in the lightcone limit. We also comment on a potential description using effective reparametrization modes in four dimensions.  
}

\end{titlepage}


\setcounter{page}{1}
\setcounter{tocdepth}{2}
\setcounter{footnote}{0}

\setlength{\parskip}{3 pt}
\jot=2 ex 

\hypersetup{linktocpage=true}
\tableofcontents

\hypersetup{linkcolor=DarkBlue}

\newpage

\section{Introduction}
\label{sec:intro}

A basic important 
 fact in physics is that the behavior of physical phenomena and observables, 
as well as the mathematical tools used to describe them, can depend sensitively 
on spacetime dimensions.
Intuitions and computational tools developed from the study of lower-dimensional 
physics do not necessarily extend directly to higher dimensions. To better 
manage the computational complexities inherent in higher-dimensional systems, 
especially when interactions are strong, it is advantageous to develop approaches 
that allow one to reorganize the computations of lower-dimensional observables 
without relying on special properties known to exist only in lower dimensions. 

Contemporary research in QFT is 
closely tied to the exploration of the Renormalization Group and critical phenomena 
through Conformal Field Theories (CFTs), which are vital across various 
subfields of physics, including condensed matter, particle physics, and 
gravitational physics. Conformal symmetry provides greater theoretical 
control over physical observables compared to general field theories. 
In this article, we focus on the primary tool in CFTs -- 
the operator product expansion (OPE). 

As an alternative to the traditional Lagrangian approach, the concept of the OPE was first 
introduced in QFT \cite{Wilson:1969zs, Wilson:1972ee} as an asymptotic short-distance 
expansion. The conformal OPE  is complete in a Hilbert space with a radius of 
convergence determined by the distance to the next-nearest operator insertion; for further 
discussion, see \cite{Pappadopulo:2012jk}. The convergent series expansion at finite distance 
separation has a powerful consequence: higher-point conformal correlators can, 
in principle, be determined by recursively applying the OPE, provided that the CFT data, 
$i.e.$, the spin and dimension of primary operators, as well as 
OPE coefficients, are known. Considering a four-point function, the OPE convergence implies 
crossing symmetry which serves as the organizing principle of the conformal bootstrap 
for constraining the CFT data; see \cite{Poland:2018epd} for a general review.

As noted back in 1971 \cite{Ferrara:1971vh}, the OPE convergence enables the global 
construction of manifestly conformally covariant OPEs valid at finite distances 
between two operators; for early references, see 
\cite{PhysRevD.5.3102, Ferrara:1972xe, Ferrara:1972ay, Ferrara:1972uq, Ferrara:1972kab}. 
Building on more recent work \cite{Czech:2016xec,deBoer:2016pqk}, we refer to these non-local CFT objects 
as bilocal OPE blocks. Motivated by the AdS/CFT correspondence
 \cite{Maldacena:1997re, Witten:1998qj, Gubser:1998bc} and the Ryu-Takayanagi formula \cite{Ryu:2006bv, Ryu:2006ef} 
of entanglement entropy, the OPE blocks can be interpreted as fields 
on the kinematic space of pairs of CFT points, $i.e.$, the space of causal diamonds \cite{Czech:2016xec, deBoer:2016pqk}.
The bilocal OPE block naturally arises in the study of entanglement entropy as the 
modular Hamiltonian \cite{Casini:2011kv}; 
see, $e.g.$, \cite{Lashkari:2013koa, Faulkner:2013ica, deBoer:2015kda, deBoer:2016pqk} 
for related discussions. In contrast to the local, short-distance expansion of the OPE 
where the insertion of two nearby local operators ${\cal O}_i(x_1)$ and ${\cal O}_j(x_2)$ 
in a conformal correlator is replaced by a series of local operators 
${\cal O}_k(y)$, the bilocal OPE blocks, 
$\mathfrak{B}_{ijk}(x_1, x_2)$, provide an integral representation of the OPE structure. 
In other words, they represent different ways to organize the OPE into (global) primary blocks:
 \begin{align}
\label{localex}
&{\rm{Local}}:~~~~~{\cal O}_i(x_1) {\cal O}_j(x_2)  
= \sum_{k} C_{ijk} \, \hat P_{ijk} (x_{12}, \partial_y)  {\cal O}_k(y) \,,\\
\label{globalex}
&{\rm{Bilocal}}:~~~{\cal O}_i(x_1) {\cal O}_j(x_2)  
=  \sum_{k} {1 \over |x_{12}|^{\Delta_i +\Delta_j}}\, {\mathfrak B}_{ijk}(x_1, x_2)  
\end{align} 
where $x_{12}= x_1-x_2$ and the spin index contractions are implicit. 
In the local form \eqref{localex}, the differential operator $\hat P_{ijk} (x_{12}, \partial_y)$ is 
determined by conformal symmetry and takes the form of an infinite power series in $\partial_y$, 
generating all the descendants in the conformal multiplet of ${\cal O}_k$. Different choices 
of the location $y$ are related through the Taylor expansion of ${\cal O}_k$, which can be 
incorporated into $\hat P_{ijk} (x_{12}, \partial_y)$. Typically, it is most convenient to set $y = 0$. 
On the other hand, in the form \eqref{globalex} the two points $x_1$ and $x_2$ do not need to be 
close to one another, despite the terminology of OPE. 

In this work, we consider CFTs with a large central charge $C_T$ and focus on the stress-tensor sector of the theory which includes the stress tensor itself 
and an infinite number of composite primary operators known as multi-stress tensors.  In two-dimensional CFTs, where states can 
be decomposed into irreducible representations of the Virasoro algebra, the Virasoro vacuum block 
encapsulates the resummation of all multi-stress tensors. 
A closed form at large central charge was obtained in \cite{Fitzpatrick:2014vua, Fitzpatrick:2015zha}; see also \cite{Perlmutter:2015iya}. 
The Virasoro identity block plays an important role in studies of, $e.g.$,  
two-dimensional entanglement entropy \cite{Hartman:2013mia, Asplund:2014coa}, 
chaos and information scrambling \cite{Roberts:2014ifa, Anous:2019yku}, 
and black hole information loss in three-dimensional gravity \cite{Fitzpatrick:2016ive, Chen:2017yze}. 
The dynamics and physical implications of multi-stress tensors in higher dimensions have been less explored, 
but recent years have seen progress driven by computations via 
holography 
\cite{Fitzpatrick:2019zqz, Karlsson:2019qfi, Li:2019tpf, Kulaxizi:2019tkd, Fitzpatrick:2019efk, Parnachev:2020fna,  
Fitzpatrick:2020yjb, Belin:2020lsr, Rodriguez-Gomez:2021pfh, Dodelson:2022yvn, Karlsson:2022osn, Dodelson:2022yvn, Huang:2022vet, Esper:2023jeq, Abajian:2023jye, Dodelson:2023nnr, Ceplak:2024bja}, 
the conformal bootstrap approach \cite{Karlsson:2019dbd, Li:2019zba, Karlsson:2020ghx, Li:2020dqm}, 
the ambient space formalism \cite{Parisini:2022wkb, Parisini:2023nbd}, 
and the differential equation approach \cite{Huang:2023ikg, Huang:2024wbq}.   

The motivation for the present paper is to establish a CFT approach for organizing the structure 
of multi-stress tensor operators and their contributions to a four-point scalar correlator by constructing 
{\it bilocal multi-stress tensor OPE blocks}. The approach does not require knowing any 
algebraic structure a priori, and may allow for a streamlined generalization to higher dimensions.

As a concrete example, we will apply our construction of the bilocal OPE blocks to compute a 
four-point correlator with two pairs of identical scalar primaries, taking
two of them to be heavy scalars ${\cal O}_{\Delta_H}$ with $\Delta_{H}\sim C_T$, 
and two light scalars ${\cal O}_{\Delta_L}$ with $\Delta_{L} \sim C_T^0$.
Such a heavy-light correlator is a useful observable for exploring the dynamics of multi-stress tensors.   
The presence of heavy scalars enhances contributions that would otherwise be suppressed at large 
$C_T$. The ratio of the dimension of the heavy scalar to the central charge is fixed, which can be 
viewed as an expansion parameter that counts the number of exchanged stress tensors.  
The CFT multi-stress tensors correspond to the exchanges of multi-gravitons in dual AdS gravity, 
where the black-hole thermal background can be interpreted as being sourced by the heavy operators.  
The heavy-light four-point function can be viewed as a thermal two-point function at large-N, which 
is an important observable in probing the interior structure of black holes, $e.g.$, \cite{Kraus:2002iv, 
Festuccia:2005pi, Fidkowski:2003nf, Hubeny:2006yu, Hashimoto:2018okj, Dodelson:2020lal, Grinberg:2020fdj,
Rodriguez-Gomez:2021pfh, Dodelson:2023nnr, Ceplak:2024bja, Belin:2025nqd, Chen:2025cee}.

In the conformal block decomposition, we will focus on the direct-channel (t-channel)
 \begin{align}
 {\cal O}_i  {\cal O}_i \to \Big(1 + T + [TT]_{n, l}+ [TTT]_{n, l}+ \dots  \Big) \to  {\cal O}_j {\cal O}_j 
\end{align} 
where $1$ represents the vacuum or identity contribution as the zeroth-order term 
in the stress-tensor sector.  Schematically, the multi-stress tensor operators are given by 
 \begin{align}
\label{multiT}
[T^k]_{n, l} =  T \cdots T \partial^{2n} \partial_{\alpha_1}  \partial_{\alpha_2} \dots \partial_{\alpha_l} T
\end{align}  
where $k$ is the number of the stress tensors, $n=0, 1, 2, 3...$, and $l=0, 2, 4, \ldots$. 
While there is only one term ($k=1$) for the single stress tensor exchange corresponding to 
the global conformal block, there are infinitely many multi-stress tensors for each fixed $k>1$.
Note that in the schematic form \eqref{multiT}, we do not specify the normal ordering, 
the various contractions of the Lorentz indices, the suitable antisymmetrization, 
or any necessary subtractions of lower powers of stress tensors required to 
construct primary operators. In the special case of two-dimensional CFTs, 
besides defining the Virasoro primaries, one can construct 
quasi-primary (holomorphic) multi-stress 
tensors by employing the point-splitting regularization procedure, $e.g.$, 
\cite{DiFrancesco:1997nk}.\footnote{See Appendix C.2 of \cite{Kulaxizi:2019tkd} for some 
examples of quasi-primary double-stress tensors labeled by spin.} In dimensions 
greater than two, the explicit forms of  composite primary multi-stress tensor operators, 
to our knowledge, have not  been constructed. The computational method based on 
the bilocal OPE blocks developed in this paper will not require any explicit forms of 
the multi-stress tensors as the OPE blocks will allow us to directly obtain 
the resummed contributions to the correlator for fixed $k$.

The simplest case is the single-stress tensor exchange in two dimensions. 
The corresponding global OPE block is the same as the modular Hamiltonian 
\cite{Casini:2011kv} in the entanglement entropy context. 
Generalizing the shadow operator formalism \cite{Ferrara:1972uq, Dolan:2011dv, Simmons-Duffin:2012juh},
 the {\it single-stress tensor OPE block} ${\cal B}^{(\cal O)}_T$ for timelike separated operators was constructed in 
\cite{Czech:2016xec, deBoer:2015kda, deBoer:2016pqk} and written as an integral over a causal diamond:
 \begin{align}
\label{globalBT2d}
 {\mathfrak B}_{{\cal O}{\cal O} T}(\vecz_1;\vecz_2) \equiv {\cal B}^{(\cal O)}_T(\vecz_1;\vecz_2) = n_{T} \,\int_{\diasmall_2^1} d^2\vecxi \,  
\frac{\langle \widetilde{T}(\vecxi) {\cal O}(\vecz_1) {\cal O}(\vecz_2) 
\rangle}{\langle {\cal O}(\vecz_1) {\cal O}(\vecz_2)\rangle}
  \, T(\xi)   
\end{align}  
where $\vecz= (z, \bar z)$ and $\vecxi=(\xi,\bar \xi)$.
The integration range is a causal diamond $\dia_2^1$ whose past and future tips are given by the
positions of the scalars, $\vecz_1$ and $\vecz_2$. 
The overall normalization $n_{T}$ will be determined by a short-distance expansion, 
which is discussed in the main text. The shadow of the stress tensor, denoted $\widetilde{T}$, 
is a formal auxiliary operator with zero conformal dimension, which is defined in terms of the holomorphic stress tensor $T$ via a nonlocal relation.
The shadow operator does not actually exist in the theory; only the identity 
operator has zero dimension. Nonetheless, it will serve as a useful computational tool
thanks to the fact that its two-point function with $T$ is a Dirac delta-function. This allows for the
construction of a Hilbert space projection onto the conformal family of the stress tensor; see, 
$e.g.$, \cite{Simmons-Duffin:2012juh}.  By construction, the bilocal stress-tensor 
OPE block includes contributions from the stress tensor and all its descendants.

The integration over a causal diamond is crucial for excluding unphysical shadow operator contributions: 
if one computes the correlator using bilocal OPE blocks by integrating over the entire 
spacetime, the result would be contaminated by contributions from shadow blocks, which 
are the conformal blocks associated with the shadow operators.
Restricting to a causal region associated with the operator insertion points 
removes the shadow contributions by means of a boundary condition in the limit where the size 
of the diamond shrinks to zero. The boundary condition fixes the normalization $n_T$ and 
thereby ensures the correct ``monodromy" 
($i.e.$, short-distance scaling behavior) \cite{Simmons-Duffin:2012juh,Czech:2016xec}.
Since a causal diamond is defined only in Lorentzian signature, caution is required 
when beginning with a Euclidean OPE and performing 
analytic continuation before evaluating the diamond integrals. We will observe that 
the integration kernels for multi-stress tensor exchanges 
are meromorphic functions. This makes the analytic continuation straightforward to perform.

The four-point function can be expressed as a series of the correlators of bilocal {\it multi-stress tensor OPE blocks}:
\begin{align}
{\langle {\cal O}_i(x_1){\cal O}_i(x_2) {\cal O}_j(x_3){\cal O}_j(x_4)\rangle \over \langle {\cal O}_i(x_1){\cal O}_i(x_2) \rangle 
 \langle {\cal O}_j(x_3){\cal O}_j(x_4) \rangle}\Big|_{T\text{-sector}}  
& = 1+ \sum_{n=1}^{\infty} \langle{\cal B}^{({\cal O}_i)}_{T^n}(x_1,x_2) {\cal B}^{({\cal O}_j)}_{T^n}(x_3,x_4)\rangle 
\end{align} 
where we focus on the stress-tensor sector. 
In two-dimensional CFTs at large central charge, the resummation yields the 
Virasoro vacuum block.
Our motivation is to develop a CFT approach for computing the corresponding structures in 
higher-dimensional CFTs at large central charge. To this end, we will first revisit the two-dimensional 
computation, approaching it as if we were unaware of the Virasoro symmetry. 
In this paper, we will extend the construction of the single-$T$ OPE blocks of \cite{Czech:2016xec, deBoer:2016pqk} 
to define the bilocal multi-stress tensor 
(or Virasoro) OPE blocks. Interestingly, we find that after integrating over the 
anti-holomorphic coordinates in our construction of the bilocal multi-stress tensor OPE blocks, 
the result exactly reproduce the semiclassical Virasoro OPE blocks derived using Chern-Simons 
Wilson line networks, which were motivated by AdS$_3$ gravity \cite{Fitzpatrick:2016mtp}. 
Since our approach does not depend on Chern-Simons theories, we hope that our construction 
may offer a formulation that is generalizable to higher dimensions. As the simplest starting point 
towards higher dimensions, we will compute the four-dimensional single-stress 
tensor OPE block in the lightcone limit using the causal diamond prescription. 
This computation provides an automatic monodromy projection, thus removing the shadow 
block contributions; to our knowledge, this has not been worked out in higher dimensions. We also show that the formalism generalizes to odd spacetime dimensions.

Another motivation for studying the dynamics of multi-stress tensor operators is the development of an 
effective field theory for quantum chaos based on reparametrization modes, whose Lorentzian cousins are also referred to as `scramblons'. 
In one dimension, the Schwarzian effective theory describes the low-energy limit of the SYK model 
\cite{1993SY, AKtalk, Maldacena:2016hyu} as well as the boundary degrees of freedom 
in Jackiw-Teitelboim gravity \cite{Almheiri:2014cka, Jensen:2016pah, Maldacena:2016upp}. Building 
on the Alekseev-Shatashvili theory \cite{Alekseev:1988ce}, reparametrization modes have 
been proposed as an effective description of quantum chaos and Virasoro vacuum blocks 
in two-dimensional CFTs \cite{Turiaci:2016cvo,Cotler:2018zff, Haehl:2019eae}. 
Extending a similar effective field theory 
for reparametrization modes to higher dimensions is outside the scope of this work, but we will 
present some preliminary results by focusing on the near-lightcone bilocal stress-tensor OPE 
block and the corresponding heavy-light correlator in four dimensions.

\subsubsection*{Summary of Results}

\begin{figure}
\begin{center}
\includegraphics[width=.4\textwidth]{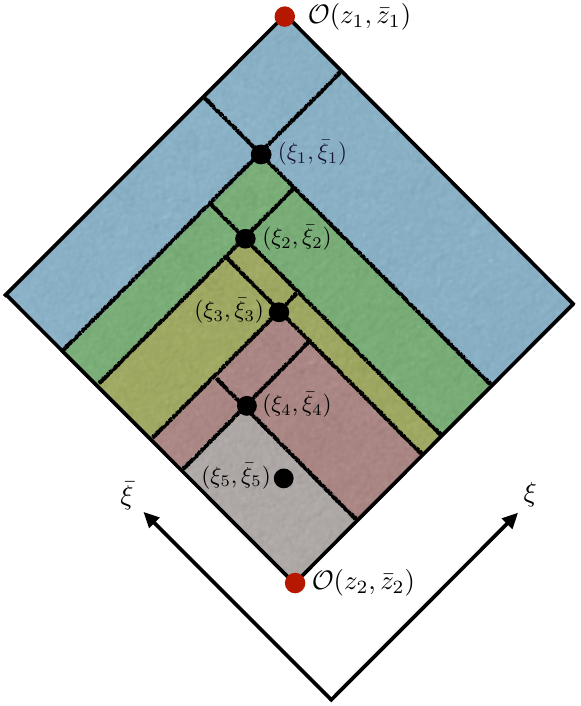}
\end{center}
\caption{The nested causal diamonds for quintuple stress-tensor exchanges, 
which are captured by the bilocal OPE block ${\cal B}^{({\cal O})}_{T^5}$. 
The red points at the top and bottom indicate the locations of the scalar operators. 
Five (causally ordered) shadow stress tensor operators 
are integrated over the nested diamonds as a tool to construct suitable integration kernels.}
\label{fig:diaintro}
\end{figure}

We propose the following expression for the bilocal Virasoro identity OPE blocks:
\be
\begin{split}
\mathcal{V}_{{\cal OO}\mathbf{1}}(\vecz_1,\vecz_2)  &= 1 + \sum_{n\geq 1}  {\cal B}^{({\cal O})}_{T^n}(\vecz_1,\vecz_2) 
 \end{split}
 \label{eq:multiTintro}
\ee 
where
{\small
\be
\begin{split}
 {\cal B}^{({\cal O})}_{T^n}(\vecz_1,\vecz_2) &\propto \int_{\diasmall_2^1} d^2 \vecxi_1\int_{\diasmallPast_2^{\vecxi_1}} d^2 \vecxi_2 \cdots \int_{\diasmallPast_2^{\vecxi_{n-1}}} d^2 \vecxi_n \, \frac{\big\langle \{\widetilde{T}(\vecxi_1) \cdots \widetilde{T}(\vecxi_n) \} {\cal O}(\vecz_1) {\cal O}(\vecz_2) \big\rangle}{\langle {\cal O}(\vecz_1) {\cal O}(\vecz_2)  \rangle} \,[ T(\xi_1) \cdots T(\xi_n) ]
 \nn
 \end{split}
\ee
}where $\vecz= (z, \bar z)$, $\vecxi= (\xi, \bar \xi) $. The overall normalization factor 
will be fixed based on a short-distance expansion.
The two-dimensional stress tensor is holomorphic, and its shadow depends on both 
coordinates $\xi$ and $\bar \xi$. This construction a priori does not require using the Virasoro algebra. 
In order to extend the bilocal single-stress tensor (global) OPE block \eqref{globalBT2d} to 
the bilocal multi-stress tensor OPE blocks \eqref{eq:multiTintro}, we introduce the following prescriptions:

\begin{itemize}
\item {\it Integration Region:} the integrals cover {\it nested} causal diamonds as the regions in which
causally ordered stress tensors are inserted; see Figure \ref{fig:diaintro} for 
illustration. The diamond integrals serve to remove 
the unphysical contributions due to shadow conformal blocks. 
We find that after integrating over the anti-holomorphic coordinates, 
the proposed formula reproduces the Virasoro OPE 
blocks in the Chern-Simons Wilson line formalism \cite{Fitzpatrick:2016mtp}. 
The computation developed in the present work does not explicitly rely on the Virasoro algebra
or the Chern-Simons theory.  
\item {\it Integration Kernel:} the integrals in \eqref{eq:multiTintro} have lightcone singularities 
whenever any of the
operator insertions become null separated. 
We find that some of these singularities are unphysical in the sense that they lead to ill-defined projectors; therefore, they should be excluded from the OPE block.
To construct the integration kernels, we subtract all the singular terms due to the coincidental limit of the anti-holomorphic 
coordinates of the shadow stress tensor
operators -- this is what the notation $\{\widetilde{T} \cdots \widetilde{T}\}$ means. 
From the viewpoint of the $T(z_1)T(z_2)$ OPE, this means that we discard all singular 
terms which include the central term $\sim c/z_{12}^4$, 
along with the terms $T(z_2)/z_{12}^2$ and $\partial T(z_2)/z_{12}$.  Consequently, in the integration kernels we only include the singular terms of the TO OPE (or, after applying the shadow transform, of the $\widetilde{T}{\cal O}$ OPE). 
Such a prescription will turn out to be necessary for properly defining multi-stress tensor projectors.

 A useful computational tool we develop along the way is the following recursion relation 
involving a shadow stress tensor:
\be
\begin{split}
& \langle \widetilde T(\vecz_0) {\cal O}_1(\vecz_2) \cdots {\cal O}_n(\vecz_n) \rangle 
 = -\frac{1}{2\pi}\sum_{i=1}^n \left( 4h\, \frac{z_{0i}}{\bar z_{0i}} -2\, \frac{z_{0i}^2}{\bar z_{0i}} \, \partial_i \right) \langle  {\cal O}_1(\vecz_2) \cdots {\cal O}_n(\vecz_n) \rangle + \ord(c) \,.
 \end{split}
\ee 
This recursion relation enables us to efficiently define the kernel 
$\big\langle \{\widetilde{T}\cdots \widetilde{T}\} {\cal O} {\cal O} \big\rangle$ 
for any number of $\widetilde{T}$ insertions. 
This recursion relation is simply a rewriting of the well-known TO OPE. 
We will first compute the kernels in Euclidean signature and then analytically continue 
to Lorentzian space before performing the integrals. From the recursion relation, 
we see that no branch cuts appear, making the analytic continuation straightforward. 
\item {\it Short-Distance Regularization:} the operator $[T(\xi_1) \cdots T(\xi_n)]$ is a stress tensor product with suitably regularized short-distance singularities. We adopt the regularization of \cite{Fitzpatrick:2016mtp}, which we will elaborate on. In particular, in the main text we will discuss how the fundamental properties -- idempotence and orthogonality -- of the projectors for multi-stress tensors in our approach require different regularizations.

\end{itemize}

While we primarily focus on the large central charge limit, 
we will also demonstrate that our proposal enables us to reproduce some $1/c$ corrections 
to the Virasoro vacuum block previously obtained using the Virasoro algebra 
\cite{Fitzpatrick:2015dlt, Chen:2016cms} and Chern-Simons Wilson 
lines \cite{Fitzpatrick:2016mtp}.

To present the simplest non-trivial extension to higher dimensions, in this work we will 
concentrate on the bilocal single-stress tensor OPE block in four dimensions.
Define
 \be
\begin{split}
\label{4dBTintro}
 {\cal B}_T(x_f,x_p) &= n^{(4d)}_{T} \int_{\diasmall_p^f} d^4 {\vecxi} \,   
  \frac{ \langle \widetilde T^{\mu\nu}(\xi^+,\xi^-,r,\theta) {\cal O}(x_f) {\cal O}(x_p)\rangle}{\langle  {\cal O}(x_f) {\cal O}(x_p)\rangle} \, T_{\mu\nu}(\xi^+,\xi^-,r,\theta)  
  \end{split}
 \ee 
where for the four-dimensional stress tensor we adopt coordinates $\vecxi= (\xi^+,\xi^-,r,\theta)$.
The two time-like separated scalars have causal cones intersecting on 
a codimension-two sphere, and the four-dimensional coordinates $x_f$ and $x_p$ 
represent the future and past tips of the causal diamond.  As we are interested in 
a four-point correlator, it is natural to set the transverse coordinates of the scalars 
to zero by taking $x_p^\mu=(x^+,x^-,0,0)$, $x_f^\mu=(0,0,0,0)$.  

The kernel in \eqref{4dBTintro} simplifies in the lightcone limit $x^- \rightarrow 0$.  
Interestingly, we observe that in the lightcone limit, the scalar  enforces the stress tensor to depend on only one coordinate 
-- we effectively 
have a ``holomorphic" stress tensor within the bilocal OPE block. 
As a check, we will reproduce the 
single-stress tensor exchange contribution to the near-lightcone four-point 
correlator in four dimensions. This global block computation does not 
require a large central charge.  

The bilocal stress-tensor OPE block near the lightcone encapsulates the contributions to 
the ${\cal O} {\cal O}$ OPE from the $d=4$ stress tensor lightcone component $T_{++}$ 
and its conformal descendants. Denoting this component as ${\cal B}_{T_{++}}^\times (x^\pm)$ and further considering a short-distance expansion in $x^+$ 
(on top of the lightcone limit), we find 
\begin{align}
&\lim_{x_+\rightarrow 0}  \lim_{x^- \to 0}{\cal B}_{T_{++}}^\times (x^\pm) \\
&= \frac{C_{T{\cal O}{\cal O}}}{2\pi C_T} \, x^-  (x^+)^3 \left[ T^\circ_{++}(0) + \frac{1}{2} \, x^+  \partial_+ T^\circ_{++}(0) + \frac{1}{7} \, (x^+)^2  \partial_+^2 T^\circ_{++}(0)
 + \ord\big( (x^+)^4 \big) \right] 
+ \ord\big( (x^-)^2 \big)   \nn
\end{align}  
where $T_{++}^\circ(\xi^+) \equiv \int_0^{2\pi} d\theta \, T_{++}(\xi^+,\xi^-\!=\!0,r\!=\!0,\theta)$.
We observe that this near-lightcone structure in $d=4$ is identical 
(up to an overall normalization) to the $d=2$ OPE block for the exchange of a spin-3 conserved current 
$W_3$. 

Drawing inspiration from this analogy, we will present a preliminary 
analysis on how one might rewrite the non-local $d=4$ stress-tensor OPE 
block as an un-integrated bilocal expression by introducing an auxiliary spin-3 
mode; this is similar to how the non-local $d=2$ stress-tensor OPE block can be expressed  
in an un-integrated bilocal form via a spin-2 reparametrization mode. 
This formalism leads to a very simple calculation of the near-lightcone single-stress tensor exchange. 
It is suggestive of an effective field theory framework for near-lightcone multi-stress tensor exchanges 
which is parametrically controlled by a large central charge as well as the near-lightcone expansion.

The rest of this paper is organized as follows.
 In Sec.\ \ref{sec:review} we review the shadow operator formalism and the kinematic space. 
Sec.\ \ref{sec:2d} contains our proposal for obtaining the bilocal Virasoro identity OPE block in two dimensions in terms of the nested integrals over causal diamonds. 
We provide evidence that the construction generalizes to four dimensions in Sec.\ \ref{sec:4d} by verifying that it furnishes a simple computation of the single-$T$ exchange in the near-lightcone limit.  
We confirm in Sec.\ \ref{sec:threedim} the validity of our near-lightcone OPE block prescription in three dimensions. In Sec.\ \ref{sec:reparametrizations}, we leverage the structure of the four-dimensional OPE block to suggest a reformulation in terms of the reparametrization modes. We end with some open questions in Sec.\ \ref{sec:discussion}.

\section{Bilocal OPE Blocks and Shadow Operators}
\label{sec:review}

In this section, we briefly review how bilocal OPE blocks are related to 
shadow operators and conformal four-point correlators. For more detailed discussions, 
see, $e.g.$, \cite{Dolan:2011dv, Simmons-Duffin:2012juh, Czech:2016xec, 
deBoer:2015kda, deBoer:2016pqk}. 

Consider a product of two primary operators with the short-distance expansion
\begin{align}
\label{local}
{\cal O}_i (x) {\cal O}_j (y) 
&= \sum_{k} {C_{ijk}\over |x-y|^{\Delta_i +\Delta_j - \Delta_k}} 
\Big(1+ a_1 x^\mu \partial_\mu +  a_2 x^\mu  x^\nu \partial_\mu \partial_\nu  + \cdots \Big){\cal O}_k(y)
\end{align} 
where constants $a_i$ are fixed by conformal symmetry. The sum over $k$ could either refer to a sum over global conformal primaries, or over Virasoro primaries (when $d=2$). In this section, we take $k$ to label global quasi-primary operators inserted at Euclidean points. But we remind the reader that one goal of this paper will be to give a new representation of Virasoro conformal blocks (and in Lorentzian signature).

For the sum over global primaries, one can equivalently write an integral representation of the OPE:
\begin{align}
\label{global}
{\cal O}_i (x) {\cal O}_j (y) 
= \sum_{k} {1 \over |x-y|^{\Delta_i +\Delta_j}} \,{\mathfrak B}_{ijk}(x, y) = \sum_{k}  \int d^dz ~S_{ijk} (x,y| z) {\cal O}_k(z) \ .
\end{align} 
The objects ${\mathfrak B}_{ijk}$ are the bilocal OPE blocks 
which depend on two insertion points.
In this non-local representation, the smearing function (kernel), $S_{ijk}(x, y| z)$, is fixed by conformal symmetry. To determine the kernel, it is convenient to introduce shadow operators.
Consider a conformal transformation as an invertible map $x^\mu \mapsto x'^\mu(x)$ that 
leaves the metric invariant up to local rescaling: 
$\eta_{\mu\nu} \mapsto \eta_{\mu\nu}'= \Omega(x)^{2}\, \eta_{\mu\nu}$.  
The proper distance 
$s^2_{12} = (x_1-x_2)^2$ transforms as $s^2_{12} \mapsto s'^2_{12} = \Omega(x_1)\Omega(x_2) s^2_{12}$.  
 A scalar primary of dimension $\Delta$ transforms as ${\cal O}(x) \mapsto {\cal O}'(x') = \Omega(x)^{-\Delta} {\cal O}(x) $. By the definition \eqref{global}, the OPE blocks thus transforms as follows:
\begin{align}
{\mathfrak B}_{ijk}(x_1, x_2)   \mapsto  {\mathfrak B}'_{ijk}(x'_1, x'_2)  &= \left(\frac{\Omega(x_2)  }{\Omega(x_1)}\right)^{{\Delta_i -\Delta_j\over 2}}{\mathfrak B}_{ijk}(x_1, x_2) \ .
\end{align}  
For identical external operators the OPE blocks transform invariantly.
Turning to the smeared representation in \eqref{global}, we see that the smearing function must transform as
\begin{align}
&S_{ijk}(x,y|z) \to S'_{ijk}(x', y'|z') = { 1 \over \Omega (x)^{\Delta_i} \Omega (y)^{\Delta_j} \Omega (z)^{d-\Delta_k }} \, S_{ijk}(x,y|z) 
\end{align}  
which can formally be considered as a three-point function involving 
a shadow operator $\widetilde {\cal O}_k$ of dimension $d-\Delta_k$. 
The shadow operator is not an actual local operator in the theory, 
so its dimension being below the unitarity bound does 
not cause an issue. 
We may then write
\begin{align}
{\mathfrak B}_{ijk}(x, y) = n_{ijk}   |x-y|^{\Delta_i +\Delta_j} \int d^dz ~ \big\langle {\cal O}_i(x) {\cal O}_j(y) \widetilde  {\cal O}_k(x) \big\rangle  {\cal O}_k(z) \ . 
\end{align} 
The normalization constants $n_{ijk}$ will be specified below. 

To incorporate tensor contributions, one promotes ${\cal O}_k$ to ${\cal O}^{\mu\nu \dots}_k$. 
Focusing on a symmetric-traceless operator ${\cal O}_{\Delta}$ 
its shadow operator is defined as
\begin{align}
\label{eq:QOhigherd}
   \widetilde{{\cal O}}^{\mu_1 \cdots \mu_\ell}(x) &=\frac{k_{\Delta,\ell} }{\pi^{d/2}}
    \int d^d y  \frac{ 1 }{(x-y)^{2(d-\Delta+\ell)} } \Big[ \prod_{i=1}^\ell \left( \delta^{\mu_i\nu_i} (x-y)^2 - 2 (x-y)^{\mu_i}(x-y)^{\nu_i}\right) \Big]  {\cal O}_{\nu_1\cdots \nu_\ell} (y) 
\end{align}
where the normalization constant 
\begin{align}
k_{\Delta,\ell} = {\Gamma(\Delta-1)\over \Gamma(\Delta+\ell-1)} {\Gamma(d-\Delta+\ell)\over \Gamma(\Delta- {d\over 2})}
\end{align}
is defined such that $\doublewidetilde{{\cal O}} = {\cal O}$; see $e.g.$, \cite{Dolan:2011dv}. This relation also applies to conserved currents with $\Delta_J = d+\ell-2$.
The shadow operator, $\widetilde{{\cal O}}$, has a corresponding conformal block denoted as 
$G_{d-\Delta}^{(\ell)}$. The conformal blocks of an operator and its shadow satisfy the same 
Casimir equation; their Casimir eigenvalues are equal, given by 
$C_{\widetilde{\Delta},\ell}=C_{\Delta,\ell}$.\footnote{The $SO(2,d)$ quadratic 
Casimir operator ${\cal C}$ commutes with all the global conformal generators, and 
it has eigenvalue $C_{\Delta,\ell} = \Delta(\Delta - d) + \ell(\ell+d-2)$, 
which classifies irreducible representations of the global conformal group. 
The stress tensor block has $C_{d,2} = 2d$. 
The shadow stress tensor has spin $2$ with vanishing dimension, 
$i.e.$, $\widetilde \Delta_T = d - \Delta_T = 0$.} 

Next, consider a four-point function with two pairs of identical scalar primaries.  
In terms of the conformal invariant cross-ratios 
\begin{align}
(u, v) =  \big(z \bar z, (1-z)(1-\bar z)\big)={1\over x_{13}^{2} x_{24}^{2}}\big( x_{12}^2 x_{34}^2 , x_{14}^2 x_{23}^2 \big) 
\end{align} where $x_{ij}= x_i - x_j$, 
the correlator can be written as a sum over irreducible multiplets, $i.e.$, conformal blocks
\begin{align}
{\langle {\cal O}_i(x_1){\cal O}_i(x_2) {\cal O}_j(x_3){\cal O}_j(x_4)\rangle \over \langle {\cal O}_i(x_1){\cal O}_i(x_2) \rangle  \langle {\cal O}_j(x_3){\cal O}_j(x_4) \rangle}  
&= \sum_{{\cal O}_{k}} C_{iik}\,  C_{jjk} \, G_{\Delta_k}^{(\ell_k)}(u,v) 
\end{align}
where ${\cal O}_{k}$ are the exchanged primaries. $G_{\Delta_k}^{(\ell_k)}(u,v) $ 
are the conformal blocks, and $C_{iik}$ and $C_{jjk}$ are OPE coefficients.  
The correlator of the bilocal OPE blocks gives a combination of physical blocks and shadow blocks:
\begin{align}
\label{eq:blockplusshadow}
\langle\mathfrak{B}_{iik}(x_1,x_2) \mathfrak{B}_{jjk}(x_3,x_4) \rangle =  
\alpha_{1} G_{\Delta_k}^{(\ell_k)}(u,v) + \alpha_{2} G_{d-\Delta_k}^{(\ell_k)}(u,v) \ .
\end{align} 
Both pieces are conformally invariant functions with the Casimir eigenvalue $C_{\Delta_k,\ell_k}$. 
The coefficients $\alpha_{1}$ and $ \alpha_{2}$ are related to 
the OPE coefficients and their ratio is such that \eqref{eq:blockplusshadow} is single-valued in the Euclidean $z$-plane. The two blocks behave differently at short distances:
\begin{align}
 \lim_{u \rightarrow 0, v \rightarrow 1}  \big(G_\Delta^{(\ell)},  G^{(\ell)}_{d-\Delta} \big)  
=  \big(u^{\frac{\Delta-\ell}{2}}  ,  u^{\frac{d-\Delta-\ell}{2}}  \big) (1-v)^\ell  + \dots
\end{align} 
One must remove the shadow blocks which have unphysical 
short-distance behavior in order to obtain a physical correlator.

Typically, there are two main challenges in the shadow operator formalism: 
$(i)$ evaluating the conformal integrals can be complicated, and $(ii)$ shadow blocks 
contain unphysical contributions. In \cite{Simmons-Duffin:2012juh}, 
the embedding formalism was employed to tackle the conformal integral, and a monodromy transformation corresponding to a specific integration 
contour was applied to eliminate the contributions from shadow blocks. 
In addition, note that the shadow operator formalism is set up to only compute {\it global} 
blocks. In this paper, focusing on the stress-tensor sector, we develop a generalized formalism which is in the spirit of the shadow operator formalism, but suitable for computing the Virasoro identity conformal (and OPE) blocks. This formalism has the monodromy projection onto the physical blocks automatically built in, and we will also demonstrate its potential applicability in higher dimensions.

\section{Two Dimensions}
\label{sec:2d}

In this section, we extend the bilocal OPE block formalism of the global OPE block 
\cite{Czech:2016xec, deBoer:2016pqk} to the Virasoro identity OPE blocks in two-dimensional CFTs.  
Specifically, we analyze the contributions arising from the stress tensor and its descendants 
to the four-point function of pairwise identical scalar operators:
\be
  \frac{\langle {\cal O} (\vecz_1){\cal O}(\vecz_2) {\cal O}'(\vecz_3) {\cal O}'(\vecz_4) \rangle}{\langle {\cal O} (\vecz_1){\cal O}(\vecz_2) \rangle\langle {\cal O}'(\vecz_3) {\cal O}'(\vecz_4) \rangle} \,.
 \label{eq:V4ptTnew} 
\ee
We often adopt the canonical configuration 
$\vecz_1 = (\infty,\infty)$, $\vecz_2=(1,1)$, $\vecz_3=(z,\bar z)$, $\vecz_4=(0,0)$. 
The objective is to use the bilocal OPE blocks to compute the correlator \eqref{eq:V4ptTnew} 
perturbatively in powers of $h$, $h'$, and $\frac{1}{c}$, without relying on the Virasoro algebra 
or Chern-Simons Wilson lines, thereby making this CFT approach potentially generalizable 
to higher dimensions.

We will first recall the global (or single-stress tensor) OPE block and use it to compute the contribution 
from a single-stress tensor exchange to the correlator, following \cite{Czech:2016xec, deBoer:2016pqk}.
Before extending the computation to the double-stress tensor exchanges, we will 
first outline the essential ingredients necessary to define the bilocal $T^2$ OPE block; 
we will then explicitly compute the double-stress tensor contributions to the correlator. 
The main result of this section is an all-order expression for the bilocal Virasoro 
OPE blocks. We will demonstrate  how it 
reproduces the known holomorphic 
expression derived from the approach based on the Chern-Simons Wilson lines.

\subsection{Single-Stress Tensor OPE Block}

Consider the timelike single-stress tensor OPE block in $d=2$ CFT for two scalar insertions at $\vecz_1=(z_1,\bar z_1)$ and $\vecz_2 = (z_2,\bar z_2)$, where  $\vecz_1$ is in the future lightcone of $\vecz_2$. 
We define this bilocal object  as an integral over the causal diamond $\dia_2^1$ whose future and past tips are $\vecz_1$ and $\vecz_2$:
\be
\label{eq:BTdef}
 {\cal B}^{(\cal O)}_T(\vecz_1;\vecz_2) = n_{T} \,\int_{\diasmall_2^1} d^2\vecxi \,  \frac{\langle \widetilde{T}(\vecxi) {\cal O}(\vecz_1) {\cal O}(\vecz_2) \rangle}{\langle {\cal O}(\vecz_1) {\cal O}(\vecz_2)\rangle}
  \, T(\xi) \,.
 \ee
The integral over the anti-holomorphic coordinate exhibits logarithmic divergence. 
However, as a consistency check, we will demonstrate that the normalization factor $n_{T}$ 
is independent of the operator dimension and cancels out the divergence.

The rationale behind integrating over a causal diamond is the following: 
the causal diamond is naturally selected if we impose that the kernel 
$\langle \widetilde{T}(\vecxi) {\cal O}(\vecz_1) {\cal O}(\vecz_2) \rangle$ is 
an {\it advanced} three-point correlator which has support only for a causally ordered configuration: 
\be
\vecz_1 \prec \vecxi \prec \vecz_2 \ .
 \ee
As will become clear below, this configuration enables us to consistently 
impose the appropriate short-distance limit on the bilocal OPE blocks. This 
is essential for the OPE block to accurately represent the stress-tensor OPE 
within a Lorentzian correlator. Moreover, in this context, we emphasize that 
the integration kernel in \eqref{eq:BTdef} behaves like a three-point function 
in a free theory -- it lacks any branch cuts. This characteristic renders the 
analytic continuation in the stress tensor sector trivial. The main task for 
us is to keep track of the causal order, which we will implement by 
appropriately restricting the integration region.\footnote{See \cite{Chen:2019fvi, Kobayashi:2020kgb} 
for discussions on Lorentzian OPE blocks and analytic continuation 
in a more general context.}

Let us first discuss the normalization of the global OPE block ${\cal B}_T$.  
Recall how the stress tensor appears in the short-distance OPE limit:
 \be
 \label{eq:bdyCond2D}
 \lim_{\vecz_2\rightarrow \vecz_1} \,  \frac{{\cal O}(\vecz_1) {\cal O}(\vecz_2)}{\langle {\cal O}(\vecz_1) {\cal O}(\vecz_2) \rangle} \sim 1 + \left(\frac{C_{T{\cal O}{\cal O}}}{2c}\, z_{12}^{2} \,T(z_1) +\text{anti-holomorphic} \right) + \ldots \ . 
 \ee In what follows, we will drop the anti-holomorphic stress tensor 
contributions. To compute the single-stress tensor contribution to the 
four-point function using the propagator of the OPE block
 \be
 \label{eq:4ptBB}
  \frac{\langle {\cal O}(\vecz_1) {\cal O}(\vecz_2) {\cal O}'(\vecz_3) {\cal O}'(\vecz_4) \rangle \big|_{T}}{\langle {\cal O} {\cal O}\rangle \langle {\cal O}' {\cal O}' \rangle} = \left\langle {\cal B}_T^{(\cal O)}(\vecz_1,\vecz_2){\cal B}^{(\cal O')}_T(\vecz_3,\vecz_4)\right\rangle \, \ ,
 \ee
 we impose the following boundary condition on ${\cal B}_T$:
 \be
 \label{eq:BTnorm}
  \lim_{\vecz_2\rightarrow \vecz_1} \;   \frac{{\cal B}_T^{(\cal O)}(\vecz_1;\vecz_2)}{(z_1-z_2)^{2}} =  \frac{C_{T{\cal O}{\cal O}}}{2c} \,T(z_1) = -\frac{h}{\pi c} \, T(z_1) \ .
 \ee  
This ensure \eqref{eq:4ptBB} is consistent with the 
short-distance limit of ${\cal O} \times {\cal O} \rightarrow T$ 
encoded in \eqref{eq:bdyCond2D}.

To explicitly compute the normalization factor, we need to 
regulate the anti-holomorphic integral. We achieve this by 
shrinking the causal diamond by a small amount $2\delta$ in the anti-holomorphic directions. As the result, we will integrate over the following ranges: 
\be
\begin{split}
\dia_2^1: \qquad \xi \in [z_2,z_1]  \, ,  \quad  \bar \xi \in [\bar z_2+\delta, \, \bar z_1-\delta] \, .
\end{split}
\ee 
The boundary condition \eqref{eq:BTnorm} then fixes 
the normalization in \eqref{eq:BTdef} to be 
 \be
\label{2dB1n}
 n_{T} = \frac{3 \pi}{2h \ln\delta}  \, \frac{C_{T{\cal O}{\cal O}}}{2c} = -\frac{3}{2c\ln\delta}\,.
\ee In the following, unless otherwise stated, the two-dimensional 
diamonds will be regularized in the same way.\footnote{The cutoff $\delta$ is introduced to avoid the singularity along the lightcone. If the sign of $\delta$ is changed, the singularity falls within the integration range.
Another way to regularize the integrals is to use 
an $i\varepsilon$ prescription, shifting $\widetilde{T}(\xi,\bar\xi) \rightarrow \widetilde{T}(\xi+i\varepsilon,\bar\xi-i\varepsilon)$ in the integration kernel.
The normalization can be determined by the same boundary condition \eqref{eq:BTnorm}. 
We do not explore different regularization schemes further here. 
See, $e.g.$, \cite{Besken:2017fsj, Besken:2018zro, DHoker:2019clx} 
for related discussions.}

 Since the two-dimensional stress tensor $T(\xi)$ is holomorphic, 
we can directly perform the $\bar\xi$-integral in \eqref{eq:BTdef}. We find
\be
\label{eq:BTdef2}
 {\cal B}_T^{(\cal O)}(\vecz_1;\vecz_2) =  \frac{3C_{T{\cal O}{\cal O}}}{c} \,\int_{z_2}^{z_1} d\xi \, \frac{(z_1-\xi)(\xi-z_2)}{(z_1-z_2)} 
  \, T(\xi) 
 \ee
 where the OPE coefficient has a canonical convention, $e.g.$, \cite{Osborn:1993cr}, $\frac{3C_{T{\cal O}{\cal O}}}{c}  = - \frac{6h}{\pi c}$. 
Notice that, upon integrating over $\bar\xi$, the resulting 
expression \eqref{eq:BTdef2} becomes independent of $\bar{z}_i$, 
making it manifest that the stress-tensor OPE block is holomorphic, 
as anticipated. We still formally use $\vecz$ 
on the left-hand side of \eqref{eq:BTdef2}, even if the result is holomorphic. 

Expanding this expression to higher orders in $(z_1-z_2)$ yields
\be
\label{eq:BTexpanded}
\lim_{\vecz_2 \rightarrow \vecz_1} {\cal B}_T^{(\cal O)}(\vecz_1;\vecz_2) = -\frac{h}{\pi c} \, z_{12}^2  \left[ T(z_2) + \frac{1}{2} \, \partial T(z_2) z_{12} + \frac{3}{20} \, \partial^2T(z_2) z_{12}^2 + \ldots \right] \, .
\ee  The global OPE block includes the contributions from 
the stress tensor and its descendants. 

The global conformal block associated with single-stress tensor 
exchange follows as a two-point function of the (linear-in-$T$) 
bilocal operators. Using the two-point function of the stress tensor, 
$\langle T(z_1) T(z_2) \rangle = c/(2z_{12}^4)$ with $C_T= 2c$,  
it is straightforward to find
\be
\label{eq:singleT}
\left\langle {\cal B}^{(\cal O)}_{T}(\vecz_1;\vecz_2) {\cal B}^{(\cal O')}_{T}(\vecz_3;\vecz_4) \right\rangle = \frac{C_{T{\cal O}{\cal O}} C_{T{\cal O}'{\cal O}'}}{4C_T} \,f_2(z) = \frac{hh'}{2\pi^2 c}\, f_2(z)\,, \qquad z \equiv \frac{z_{12} z_{34}}{z_{13}z_{24}}
\ee  where $f_a(z) = z^a \; {}_2F_1(a,a,2a,z)$. 
This represents the correct contribution of the single-stress tensor exchange to the correlator.
Let us reiterate that the monodromy projection in this approach is 
automatic -- the shadow block associated with the $\widetilde{T}$-exchange is absent.
\vspace{.3cm}

\noindent{\it Remarks on the projection in the shadow operator formalism}
 
It will prove useful to formulate the bilocal OPE block associated with the conformal 
family of the stress tensor in terms of a projector. We define
\be
\label{eq:TprojDef}
 |T| 
  \equiv \frac{3}{\pi c} \int d^2\vecxi\; |\widetilde{T}(\vecxi)\rangle \langle T(\xi)| 
  = \frac{3}{\pi c} \int d^2\vecxi\; |T(\xi)\rangle \langle \widetilde{T}(\vecxi)| 
  \equiv |\widetilde{T}| \,.
\ee
This projector satisfies $|T|^2= |T|$ (this follows from \eqref{eq:TTcorrsCollect}).
Furthermore, its insertion into correlation functions yields a linear combination 
of the global stress-tensor conformal block and its shadow  
if one chooses to integrate over the entire spacetime region rather than just the causal diamond:
\be
\label{eq:cpw2d}
  \frac{\langle {\cal O}(\vecz_1) {\cal O}(\vecz_2) |T| {\cal O}'(\vecz_3) {\cal O}'(\vecz_4) \rangle}{\langle {\cal O}(\vecz_1) {\cal O}(\vecz_2) \rangle\langle {\cal O}'(\vecz_3) {\cal O}'(\vecz_4) \rangle} = \frac{hh'}{2\pi^2 c} \,\left[ f_2(z) + 12 \, f_{-1}(z) \, f_1(\bar z) \right] \,.
\ee   
 This is the unique linear combination of the stress tensor block and its shadow, which is single-valued and thus features in harmonic analysis on the conformal group; see, $e.g.$, \cite{Dobrev:1977qv, Murugan:2017eto} for useful reviews.\footnote{A note on terminology: this special linear combination is sometimes referred to as conformal partial waves (CPWs) in the literature, $e.g.$, \cite{Dolan:2011dv}. However, CPWs can also refer to just the physical conformal blocks multiplied by certain distance factors, $x^2_{ij}$, $e.g.$, \cite{Simmons-Duffin:2012juh}.}
Extracting the physical block $\propto f_2(z)$ requires 
a {\it monodromy projection}: one needs to recognize that the above expression 
is a linear combination of two independent solutions to the conformal 
Casimir equation, and projects onto the solution with the correct 
short-distance behavior, $f_2(z\rightarrow 0) \sim z^2$ which is associated 
with the physical stress-tensor exchange. An advantage of the causal 
diamond approach is that it renders the monodromy projection automatic.  
We will revisit the projector in the next section.

\subsection{Recursion, Regularization, and Projectors}

To extend the global single-stress tensor OPE block construction 
to multi-stress tensor OPE blocks, in this section we derive a useful 
recursion relation involving the shadow stress tensor and discuss the  
regularization procedure for constructing the multi-stress tensor projectors.

\subsubsection{Shadow Stress Tensor and Recursion Relation}

We begin by recalling some identities in two dimensions.
Correlators with a stress tensor insertion can be computed recursively 
via \cite{Belavin:1984vu}
\be
\label{eq:Trecursion}
 \langle T(z_0) {\cal O}_1(\vecz_1) \cdots {\cal O}_n(\vecz_n) \rangle 
 = -\frac{1}{2\pi} \sum_{i=1}^n \left( \frac{h_i}{z_{0i}^2} + \frac{1}{z_{0i}} \, \partial_i \right) \langle  {\cal O}_1(\vecz_1) \cdots {\cal O}_n(\vecz_n) \rangle \,.
\ee
If one of ${\cal O}_i$ is itself a stress tensor, this recursion relation 
requires an additional $\ord(c)$ contribution, due to 
the stress-tensor OPE. For example:
\be
\begin{split}
\frac{ \langle T(z_0) {\cal O}(\vecz_1) {\cal O}(\vecz_2) \rangle }{\langle{\cal O}(\vecz_1) {\cal O}(\vecz_2) \rangle}
 &= -\frac{h}{2\pi}\, \frac{z_{12}^2}{z_{01}^2 z_{02}^2} \,,\\
\frac{ \langle T(z_0)T(z_0') \, {\cal O}(\vecz_1) {\cal O}(\vecz_2) \rangle }{\langle{\cal O}(\vecz_1) {\cal O}(\vecz_2) \rangle}
 &= \frac{h^2}{4\pi^2} \, \frac{z_{12}^4}{z_{01}^2 z_{02}^2 z_{0'1}^2 z_{0'2}^2} + \frac{h}{2\pi^2} \, \frac{z_{12}^2}{z_{00'}^2z_{01}z_{02} z_{0'1} z_{0'2}} + \frac{c}{2} \frac{1}{z_{00'}^4} \,.
 \end{split}
\ee

The shadow of the stress tensor is a non-local operator defined by 
 \be
  \widetilde{T}(\vecxi) = \frac{2}{\pi} \int d^2 \vecxi' \, \frac{(\xi-\xi')^2}{(\bar\xi-\bar\xi')^2} \, T(\xi') \,, 
 \ee 
which has dimensions $(h,\bar h)_{\widetilde{T}} = (-1,1)$. 
We may write this relation formally as $\partial^3\widetilde{T} = -4\,\bar\partial T$. 
Using the shadow transform, the recursion relation \eqref{eq:Trecursion} can 
be turned into a similar recursion relation\footnote{The required integrals are standard conformal 
integrals \cite{Dolan:2011dv} and their conformally covariant generalizations; 
see, $e.g.$, appendix B of \cite{Murugan:2017eto}.} describing an insertion of 
$\widetilde{T}$:
\be
\label{eq:Ttrecursion}
\begin{split}
& \langle \widetilde T(\vecz_0) {\cal O}_1(\vecz_1) \cdots {\cal O}_n(\vecz_n) \rangle 
 = -\frac{1}{2\pi}\sum_{i=1}^n \left( 4h_i\, \frac{z_{0i}}{\bar z_{0i}} -2\, \frac{z_{0i}^2}{\bar z_{0i}} \, \partial_i \right) \langle  {\cal O}_1(\vecz_1) \cdots {\cal O}_n(\vecz_n) \rangle\,.
 \end{split}
\ee 
For example, we have 
\be
\begin{split}
\label{eq:TtTtOO}
 &\frac{\langle \widetilde T(\vecz_0) {\cal O}(\vecz_1)  {\cal O}(\vecz_2) \rangle}{\langle  {\cal O}(\vecz_1)  {\cal O}(\vecz_2) \rangle }
 = -\frac{2h}{\pi} \, \frac{z_{01} z_{02} \bar z_{12}}{\bar z_{01} \bar z_{02} z_{12}}\,,\\
& \frac{\langle \widetilde T(\vecz_0)\widetilde T(\vecz_0') {\cal O}(\vecz_1)  {\cal O}(\vecz_2) \rangle}{\langle  {\cal O}(\vecz_1)  {\cal O}(\vecz_2) \rangle } 
= \frac{4h^2}{\pi^2}\, \frac{z_{01}z_{02}z_{0'1}z_{0'2} \bar z_{12}^2}{\bar z_{01}\bar z_{02}\bar z_{0'1}\bar z_{0'2} z_{12}^2}\\
 &\qquad\qquad\qquad  - \frac{2h}{\pi^2} \frac{\bar z_{12}}{z_{12}^2 \bar{z}_{01} \bar{z}_{02}} \left(\frac{z_{01}^2 z_{0'2}^2}{\bar{z}_{0'2}} -\frac{z_{0'1}^2 z_{02}^2}{\bar{z}_{0'1}} +\frac{z_{00'}z_{12} (z_{01}z_{0'2} + z_{02} z_{0'1})}{\bar{z}_{00'}}  \right) + \frac{2c}{3} \frac{z_{00'}^2}{\bar{z}_{00'}^2} \,. 
\end{split}
\ee

We remark that shadow operators are typically adopted as formal tools to define projectors onto global conformal irreducible representations, $e.g.$, \cite{Dolan:2011dv,Simmons-Duffin:2012juh}. Correlators with multiple insertions of shadow operators are not typically studied; they are unphysical. However, we will find it advantageous to {\it define} correlators with multiple insertions of the shadow stress tensor $\widetilde{T}$ using the recursion relation \eqref{eq:Ttrecursion}. Although these objects are not intended to represent physical correlations in the spectrum, they will serve a useful purpose as integration kernels with appropriate transformation properties. 

In addition, note that the normalized correlators \eqref{eq:TtTtOO} are meromorphic functions of the insertion points. Consequently, their analytic continuation to Lorentzian signature is trivial and we can simply Wick rotate the arguments without ambiguity. We will make use of this property when we adopt objects such as \eqref{eq:TtTtOO} as integration kernel in Lorentzian integrals.\footnote{The absence of branch cuts is special to stress tensor insertions -- for generic operators with non-integer dimensions, this would change; see, $e.g.$, \cite{Chen:2019fvi}.}

\subsubsection{Regularizations}
\label{sec:regularization}

To define the bilocal OPE blocks beyond the single-stress tensor case, and 
to compute the multi-stress tensor contributions to the four-point scalar correlators, 
we will introduce two separate regularization procedures. 
One is implemented when constructing the integration kernels for integrals over causal diamonds, 
while the other is adopted when computing the higher-point correlators of the 
stress tensor itself. 
We will justify these regularization schemes by the consistency with $(i)$ the basic 
properties of projectors, and $(ii)$ explicit computations of the correlators at large 
central charge.
\vspace{.3cm}

\noindent {\it Regulated Integration Kernel}

Correlators involving the shadow stress tensor \eqref{eq:TtTtOO} 
will play an important role as integration kernels in the 
bilocal multi-stress tensor OPE blocks.  It will prove essential to construct these kernels in such a way that they only 
exhibit OPE singularities with scalar primaries ${\cal O}(\vecz_i)$, 
rather than with other shadow stress tensors. 
We introduce {\it OPE-subtracted kernels} that involve multiple 
$\widetilde{T}$ and scalar primaries by subtracting the short-distance 
singularities associated the regularized operators:
\begin{align}
 &\langle   \{ \widetilde{T}(\vecxi) \} \, {\cal O}_1(\vecz_1) \cdots {\cal O}_n(\vecz_n) \big \rangle \equiv  \langle  \widetilde{T}(\vecxi)  \, {\cal O}_1(\vecz_1) \cdots {\cal O}_n(\vecz_n) \big\rangle \ ,  \\
 &\big\langle \{ \widetilde{T}(\vecxi_1) \cdots \widetilde{T}(\vecxi_m) \} \, {\cal O}_1(\vecz_1) \cdots {\cal O}_n(\vecz_n) \big\rangle\\
 &\qquad \equiv
 -\frac{1}{2\pi}\sum_{i=1}^n \left( 4h_i\, \frac{(\xi_1-z_i)}{(\bar\xi_1-\bar z_i)} -2\, \frac{(\xi_1-z_i)^2}{(\bar\xi_1-\bar z_i)} \, \partial_{z_i} \right)\big \langle   \{ \widetilde{T}(\vecxi_2) \cdots \widetilde{T}(\vecxi_m) \} \, {\cal O}_1(\vecz_1) \cdots {\cal O}_n(\vecz_n) \big\rangle \ . \nn
 \label{eq:modifiedRecTt}
\end{align}
The sum only runs over terms associated with the operators outside 
the regularized product $\{\,\cdot\,\}$. In particular, 
$\langle \{\widetilde{T}\cdots \widetilde{T}\} \rangle = 0$. 
Physically, the gravitational self-interactions are 
absent in these OPE-subtracted kernels, which, according to the definition, do not receive any $1/c$ corrections.

Take $\langle \{\widetilde{T} \widetilde{T}\} {\cal O} {\cal O}\rangle$ 
as an example. We employ the recursion \eqref{eq:Ttrecursion} 
but omit the term corresponding to the $\widetilde{T}\widetilde{T}$ 
OPE. This construction yields
\be
\label{eq:TtTtOOsplit}
\begin{split}
 &\frac{ \langle \{\widetilde  T(\vecz_0)\widetilde  T(\vecz_0')\} \, {\cal O}(\vecz_1) {\cal O}(\vecz_2) \rangle }{\langle{\cal O}(\vecz_1) {\cal O}(\vecz_2) \rangle} \\
 &\qquad =\frac{ \langle \widetilde  T(\vecz_0)\widetilde  T(\vecz_0') \, {\cal O}(\vecz_1) {\cal O}(\vecz_2) \rangle }{\langle{\cal O}(\vecz_1) {\cal O}(\vecz_2) \rangle} - \left( -\frac{2h}{\pi^2} \frac{\bar{z}_{12}z_{00'} ( z_{01}z_{0'2} + z_{0'1} z_{02})}{z_{12}\bar{z}_{01} \bar{z}_{02}\bar{z}_{00'}} + \frac{2c}{3} \frac{z_{00'}^2}{\bar{z}_{00'}^2} \right) \ .
 \end{split}
\ee
The terms in the large parentheses can be understood as the singular pieces of  
$\langle \widetilde{T}\widetilde{T}{\cal O}{\cal O}\rangle / \langle {\cal O}{\cal O}\rangle$ as $\bar z_0 \rightarrow \bar z_0'$; see also \eqref{eq:TtTtOO}. 
Evidently, applying the recursion relation and omitting the term corresponding to the $\widetilde{T} \times \widetilde{T}$ 
OPE is equivalent to removing the terms associated with the $\widetilde{T} \times \widetilde{T}$ 
OPE singularity from $\langle \widetilde{T} \widetilde{T} {\cal O}{\cal O}\rangle$. 

The subtraction denoted by $\{\,\cdot\,\}$ serves as a form of regularization 
in the sense that it removes the singularity in the integration kernel 
when a shadow stress tensor approaches another shadow operator. 
We will find that this singularity is integrable -- it does not lead to any 
divergence upon performing the diamond integrals. Therefore, 
the subtraction regulates the kernel but it is {\it not} intended to 
remove UV divergences from physical correlators.  Instead, we will show that 
the subtraction is required for the orthogonality of the projectors onto the space of $n$-stress tensor exchanges. 
In the case of double-stress tensor exchanges, starting with 
$\langle \{\widetilde{T} \widetilde{T}\} {\cal O} {\cal O}\rangle$ 
we will explicitly demonstrate that the piece we subtract corresponds to the 
two-to-one gravitational mixing that must be removed to 
construct an orthogonal basis for multi-stress tensor states.   

In addition, the subtraction $\{\,\cdot\,\}$ is unaffected by large $c$ corrections, 
distinguishing it from the regularization $[\,\cdot\,]$ employed for stress tensor 
correlators, discussed below.  The regularization for stress tensor correlators 
is more complicated, but essential 
for regulating the diamond integrals and ensuring both the idempotence and orthogonality of the projector.

\vspace{.3cm}

\noindent {\it Regulated Stress Tensor Correlators}

When computing the contributions from multi-stress tensor exchanges 
based on bilocal OPE blocks, we need to address the short-distance 
divergences that arise when the stress tensors collide in the diamond integrals.   
This type of divergence has been previously observed in the 
Chern-Simons Wilson-line network formalism \cite{Fitzpatrick:2016mtp}; 
see also, $e.g.$, \cite{Besken:2017fsj,Besken:2018zro,DHoker:2019clx} 
and related work 
\cite{Maxfield:2017rkn,Hikida:2018dxe,Blommaert:2018oro,Cotler:2018zff,Nguyen:2021jja,Nguyen:2022xsw}. 
Here, we will follow the regularization prescription first articulated in 
\cite{Fitzpatrick:2016mtp}. The basic idea is as follows: 
since all divergences come from the short-distance limit of stress-tensor operators, 
one introduces the {\it regulated-T product}, $[T(z_1) \cdots T(z_n)]$, 
which excludes all the singular terms in the short-distance limit from the operators inside the brackets; we will discuss its definition further. 
Note that this regularization is different from $\{\,\cdot\,\}$. 
In particular, the regularization $[\, \cdot \, ]$ is not directly defined at the operator level but, as a prescription, defined when computing the correlators.  

Considering correlators with one group of $n$ regularized product of stress 
tensors and $m$ non-regularized stress tensors, 
the regularization scheme requires that one must produce disconnected 
factors such that each factor contains just one of the regularized stress 
tensors. First, the regularization eliminates correlators of the following form:
\be
\label{eq:FKregDef0}
\langle [T(z_1) \cdots T(z_n) ] T(y_1)\cdots T(y_m) \rangle = 0  \qquad (n>m \geq 0)\,.
\ee 
This is also consistent with the fact that the vacuum 
expectation values of normal-ordered $T$-products ($i.e.$, when $m=0$) are zero. 

Next, for $m\geq n$, the regularization amounts to singling out  
disconnected factors where each factor contains exactly one of the regulated operators, $e.g.$, 
\begin{equation}
\label{eq:FKregDef}
\begin{split}
\langle [T(z_1) ] T(y_1)\cdots T(y_m) \rangle &= \langle T(z_1) T(y_1) \cdots T(y_m) \rangle \,,\\
  \langle [T(z_1)T(z_2)] T(y_1)T(y_2)\rangle &= \langle T(z_1)T(y_1) \rangle \langle T(z_2) T(y_2) \rangle + [y_1 \leftrightarrow y_2] \,,\\
  \langle [T(z_1)T(z_2)] T(y_1)T(y_2)T(y_3)\rangle  &=\\
  &\hspace{-8em} = \Big\{ \langle T(z_1)T(y_1) \rangle \langle T(z_2) T(y_2) T(y_3) \rangle 
+ [y_1 \leftrightarrow y_2] + [y_1 \leftrightarrow y_3] \Big\} + [z_1 \leftrightarrow z_2] \ .
  \end{split}
\end{equation}
Similarly, for more stress tensor insertions the regularization is defined by 
forming a sum over contractions of exactly one $T(z_i)$ with all possible 
``groups'' of $T(y_{i_1}) \cdots T(y_{i_k})$.
This manifestly preserves singularities in the OPE of any of the non-regularized 
stress tensors, while removing singularities among the regularized ones.

Finally, we will also need to consider the correlators of 
two groups of regularized products of stress tensors. 
Some important, and for our purposes sufficient, cases are
\be
\label{eq:regularizationT}
\begin{split}
 \langle [ T(z_1) \cdots T(z_n)][ T(y_1) \cdots T(y_m)] \rangle 
&= 0 \quad (n\neq m)\,,\\
  \langle [ T(z_1) \cdots T(z_n)][ T(y_1) \cdots T(y_n)] \rangle 
&= \langle [ T(z_1) \cdots T(z_n)]\,T(y_1) \cdots T(y_n) \rangle + \ord(c^{n-1}) \,.
\end{split} 
\ee  At leading large $c$, this regularization is nothing but the normal ordering. Explicit higher-order corrections in a $1/c$ expansion are complicated, but, in principle, they should be determined by the general properties described above -- see also Appendix \ref{app:projectors}. For a related discussion, see Appendix C in \cite{Fitzpatrick:2016mtp}.

Note that we regulate the integration kernels and the stress tensor 
correlators in Euclidean space. As there are no branch cuts in these objects, we 
can perform a Wick rotation to Lorentzian space before carrying 
out the causal diamond integrals.

\subsubsection{Projectors}
\label{sec:projectors}

In order to build the bilocal Virasoro identity OPE blocks 
order by order in the number of stress tensors,  
in this subsection, our goal is to construct a multi-$T$ 
generalization of the projector $|T|$, defined in \eqref{eq:TprojDef}, 
which we reproduce here for convenience:  
\be
\label{eq:TprojDefRepeat}
 |T| 
  \equiv \frac{3}{\pi c} \int d^2\vecxi\; |\widetilde{T}(\vecxi)\rangle \langle T(\xi)| 
  = \frac{3}{\pi c} \int d^2\vecxi\; |T(\xi)\rangle \langle \widetilde{T}(\vecxi)| 
  \equiv |\widetilde{T}| \,.
\ee
The issue with this conformally-invariant projector,  
 as mentioned, is that it contains shadow contributions. 
It therefore does not compute the correct global conformal 
block unless one performs a monodromy projection.
\vspace{.3cm}

\noindent{\it Projector for Single-Stress Tensor without Shadow Contribution}

We will adopt a modified projector $|T|$, which projects directly 
onto the OPE block associated with 
the physical stress-tensor exchange. 
To this end, consider the causal diamond $\dia_2^1$ associated 
with $\vecz_1$ and $\vecz_2$ and define two associated projections: 
\be
\begin{split}
 &|T|_{(\vecz_1,\vecz_2)} \equiv \frac{3}{\pi c} \int_{\diasmall_2^1} d^2\vecxi\;|\widetilde{T}(\vecxi)\rangle \langle T(\xi)|
       \,,  \\ 
 &|\widetilde{T}|_{(\vecz_1,\vecz_2)} \equiv \frac{3}{\pi c} \int_{\diasmall_2^1} d^2\vecxi\;  |T(\xi)\rangle \langle \widetilde{T}(\vecxi)|   \,.
\end{split}
\ee
Unlike the previous projectors defined in \eqref{eq:TprojDefRepeat}, 
integrating over a causal diamond region reveals that the projectors 
associated with $T$ and $\widetilde{T}$
are no longer identical. Nevertheless, both projectors remain idempotent. 
Furthermore, these projectors satisfy the following composition rule: 
\be
\label{eq:compositionrule}
 |T|_{(\vecz_1,\vecz_2)} \, |T|_{(\vecz_3,\vecz_4)} = \left\{ \begin{aligned}
 &0 \qquad\qquad\quad \text{if }\;\dia_{2}^1 \cap \dia_4^3 = \varnothing  \\
 &|T|_{(\vecz_i,\vecz_j)} \qquad \text{if }\;\dia_2^1 \cap \dia_4^3 = \dia_j^i
 \end{aligned}\right.
\ee
and similarly for $|\widetilde{T}|_{(\vecz_i,\vecz_j)}$.
These projectors will be used to project onto 
monodromy-projected (physical) OPE blocks associated with 
the conformal family of a stress-tensor. 
The single-$T$ OPE block is defined through
\be
\begin{split}
  \frac{\langle{\cal O}(\vecz_1) {\cal O}(\vecz_2) |T|_{(\vecz_1,\vecz_2)}}{\langle {\cal O}(\vecz_1) {\cal O}(\vecz_2) \rangle} = \frac{3}{\pi c} \int_{\diasmall_2^1} d^2\vecxi \; \frac{\langle \widetilde{T}(\vecxi) {\cal O}(\vecz_1) {\cal O}(\vecz_2) \rangle}{\langle {\cal O}(\vecz_1) {\cal O}(\vecz_2) \rangle } \,  \langle T(\xi)| = - \frac{2\ln\delta}{\pi} \, \langle {\cal B}^{({\cal O})}_T(\vecz_1,\vecz_2) |\,
\end{split}
\ee 
where we recall the normalization $n_T$ \eqref{2dB1n}.
It follows that the (global) single-stress tensor exchange can be 
computed by inserting the projectors into a four-point function, yielding
\be
\lim_{\delta\rightarrow 0} \; \left( - \frac{\pi}{2\ln \delta}\right)^2 \,
 \frac{\langle {\cal O}(\vecz_1) {\cal O}(\vecz_2) |T|_{(\vecz_1,\vecz_2)}\,|\widetilde{T}|_{(\vecz_3,\vecz_4)}\, {\cal O}'(\vecz_3) {\cal O}'(\vecz_4) \rangle }{\langle {\cal O}(\vecz_1) {\cal O}(\vecz_2) \rangle\langle {\cal O}'(\vecz_3) {\cal O}'(\vecz_4) \rangle }
  = \frac{hh'}{\pi^2 C_T} \; f_2(z) \, .
\ee 
This expression based on the projectors reproduces \eqref{eq:singleT}, 
using $C_T= 2c$ in two dimensions. 
\vspace{.3cm}

\noindent{\it Projector for Double-Stress Tensors}

Let us now apply the above construction to the case of bilocal double-$T$ OPE 
blocks, and subsequently to Virasoro OPE blocks. 

\begin{figure}
\begin{center}
\includegraphics[width=0.9\textwidth]{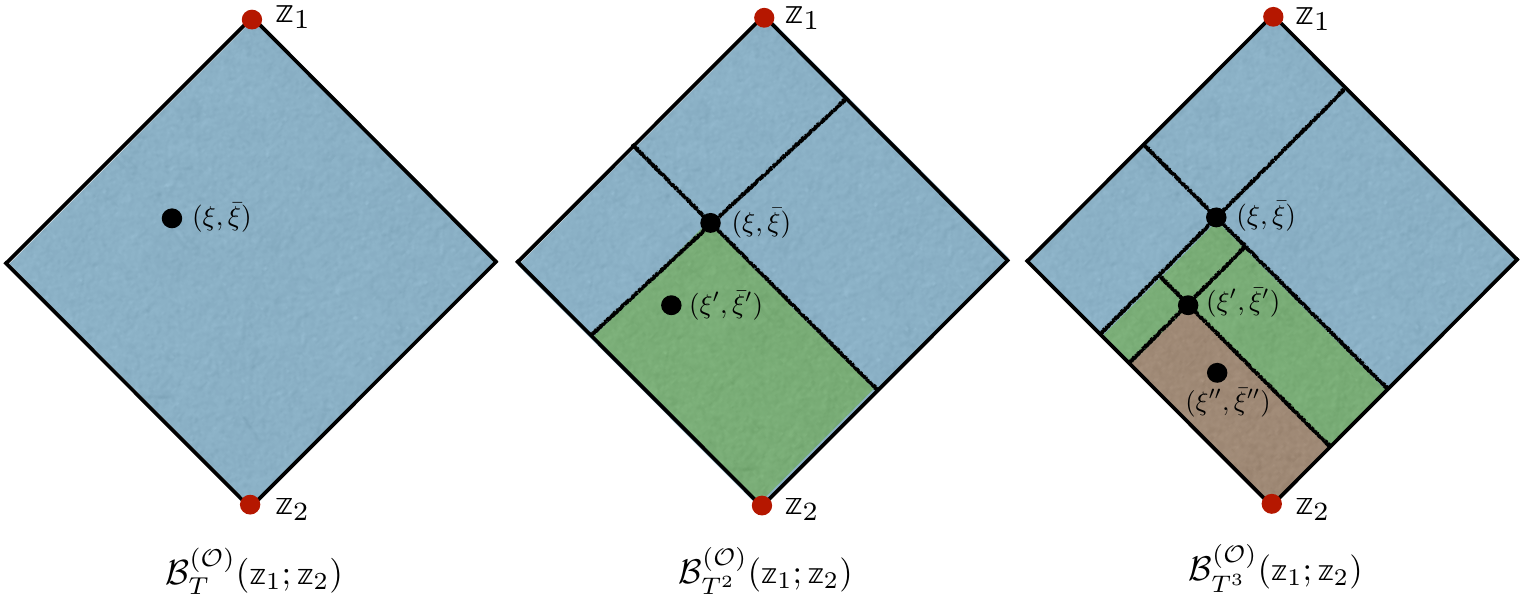}
\end{center}
\caption{Bilocal OPE blocks with the nesting of 
diamond-shaped integration regions, which are associated 
with causally ordered shadow stress tensor insertions 
with locations $(\xi_i, \bar \xi_i)$.}
\label{fig:diamonds2d}
\end{figure}

We define  
 \be
 \label{eq:proj2Def}
 |T^2|_{(\vecz_1,\vecz_2)} \equiv \frac{1}{2}\left(\frac{3}{\pi c}\right)^2 \int_{\diasmall_2^1} d^2\vecxi\int_{\diasmallPast_2^\vecxi} d^2\vecxi' \; \big|\{\widetilde{T}(\vecxi) \widetilde{T}(\vecxi')\}\big\rangle \big\langle [T(\xi)T(\xi')] \big| 
\ee as the projector onto the space of double-$T$ exchanges (similarly for $|\widetilde{T}^2|_{(\vecz_1,\vecz_2)}$). 
The inner diamond, denoted as $\diaPast_2^{\vecxi}$, has tips  $\vecz_2$ and $\vecxi$, 
corresponding to a restricted integration range. See the middle plot in Figure \ref{fig:diamonds2d}.  
We only integrate over configurations 
where the two insertions of $\widetilde{T}$ are timelike separated and causally ordered: 
\be
\begin{split}
&\dia_2^1 = \{ z_2 < \xi < z_1 \,,\;\; \bar z_2 < \bar\xi <\bar z_1 \} \,,\\
&\diaPast_2^\vecxi = \{z_2 < \xi' < \xi \,,\;\; \bar z_2 < \bar\xi' <\bar \xi \}  \,. 
\end{split}
\ee 
The projector \eqref{eq:proj2Def} has the following basic properties: 

\begin{enumerate}
\item Idempotence: 
the projector \eqref{eq:proj2Def}, defined with suitable regularizations, squares to itself. 
More generally, it satisfies the composition property analogous to \eqref{eq:compositionrule}: 
\be
\label{eq:compositionrule2}
 |T^2|_{(\vecz_1,\vecz_2)} \, |T^2|_{(\vecz_3,\vecz_4)} = \left\{ \begin{aligned}
 &0 \qquad\qquad\quad\; \text{if }\;\dia_{2}^1 \cap \dia_4^3 = \varnothing  \\
 &|T^2|_{(\vecz_i,\vecz_j)} \qquad \text{if }\;\dia_2^1 \cap \dia_4^3 = \dia_j^i
 \end{aligned}\right.
\ee 
and similarly for $|\widetilde{T}^2|_{(\vecz_1,\vecz_2)}$. 
This can be verified as a direct consequence of the regularization prescription 
for the stress tensor correlators, \eqref{eq:TTTtTtreg}. The idempotence and composition 
properties may be regarded as defining characteristics of the regularization scheme $[\,\cdot\,]$ 
introduced in the context of Chern-Simons Wilson lines \eqref{eq:FKregDef}, although in 
that formalism a projector with nested causal diamonds was not introduced.

\item Orthogonality: 
the double-$T$ projector is orthogonal to the single-$T$ projector 
and therefore computes independent contributions to the Virasoro 
OPE block (and hence to the conformal block). To verify this, 
we first find 
\be
|T^2|_{(\vecz_1,\vecz_2)}  |T|_{(\vecz_1,\vecz_2)} \propto  \int_{\diasmall_2^1} d^2\vecxi\int_{\diasmallPast_2^\xi} d^2\vecxi'  \int_{\diasmall_2^1} d^2\vecxi'' \; 
 \big|\{\widetilde{T}(\vecxi) \widetilde{T}(\vecxi')\}\big\rangle \big\langle [T(\xi)T(\xi')] \, \widetilde{T}(\vecxi'') \big\rangle \big\langle T(\xi'')| 
 \ee 
vanishes, due to the defining property $\langle [T^n] \, T^m \rangle = 0$ 
for $n>m$, which holds regardless of the insertion points. We next find
\be
|T|_{(\vecz_1,\vecz_2)}  |T^2|_{(\vecz_1,\vecz_2)}
\propto \int_{\diasmall_2^1} d^2\vecxi  \int_{\diasmall_2^1} d^2\vecxi' \int_{\diasmallPast_2^{\xi'}} d^2\vecxi''  \; \big|\widetilde{T}(\vecxi)\big\rangle \big\langle T(\xi) \, \{\widetilde{T}(\vecxi')\widetilde{T}(\vecxi'')\} \big\rangle \big\langle [T(\xi')T(\xi'')]| 
\ee
vanishes as well, due to the regularization $\{\widetilde{T}\widetilde{T}\}$ 
of the kernel. Explicitly, the unregulated correlator involving one stress 
tensor and two shadow stress tensors can be computed using the 
recursion relation \eqref{eq:Ttrecursion}, which yields
\be
\label{eq:TTtTt}
\begin{split}
\big\langle T(\xi) \widetilde{T}(\vecxi')\widetilde{T}(\vecxi'') \big\rangle
&= \frac{2c}{3} \, \frac{(\xi'-\xi'')^4}{(\xi-\xi')^2 (\bar\xi'- \bar\xi'')^2 (\xi''-\xi)^2} \,.
\end{split}
\ee
As $\{\widetilde{T}\widetilde{T}\}$ amounts to removing 
all singular terms associated with the $\widetilde{T}\times\widetilde{T}$ 
OPE, the regularization simply removes this correlator, $\langle T\{\widetilde{T}\widetilde{T}\}\rangle = 0$. See Appendix \ref{app:projectors} for more details.
\end{enumerate}

Since it meets the above properties, we expect that the projector $|T^2|_{(\vecz_1,\vecz_2)}$ can be used to define the bilocal OPE block associated with double-stress tensor exchanges. We therefore write this object as follows:
\be
\label{eq:OOproj2}
\begin{split}
  \Big\langle {\cal B}^{({\cal O})}_{T^2}(\vecz_1,\vecz_2) \Big| \propto \frac{\langle{\cal O}(\vecz_1) {\cal O}(\vecz_2) |T^2|_{(\vecz_1,\vecz_2)}}{\langle {\cal O}(\vecz_1) {\cal O}(\vecz_2) \rangle} \,.
\end{split}
\ee
In Sec. \ref{sec:TTblock}, we will determine the normalization factor and adopt this OPE block to compute the contribution from double-stress tensor exchanges to the four-point scalar correlator.
\vspace{.3cm}

\noindent{\it Projector for General Multi-Stress Tensors}

The construction above generalizes to more general $T^n$ exchanges 
in an analogous fashion: 
 \be
 \label{eq:projnDef}
 |T^n|_{(\vecz_1,\vecz_2)} \equiv \frac{1}{n!}\left(\frac{3}{\pi c}\right)^n \int_{\diasmall_2^1} d^2\vecxi_1 \cdots \int_{\diasmallPast_2^{\xi_{n-1}}} d^2\vecxi_n \; \big|\{\widetilde{T}(\vecxi_1) \cdots \widetilde{T}(\vecxi_n)\}\big\rangle \big\langle [T(\xi_1)\cdots T(\xi_n)] \big| \,.
\ee
As a justification of the regularization prescriptions introduced in 
Sec. \ref{sec:regularization}, we remark that if we had not yet introduced 
$\{\widetilde{T}\cdots \widetilde{T}\}$ and $[T\cdots T]$, we could still arrive at 
these regularizations by requiring that the projectors satisfy 
the following basic conditions:
\be
\label{eq:ProjProperties}
\begin{split}
\text{Idempotence:} \qquad |T^n|_{(\vecz_1,\vecz_2)} \,  |T^n|_{(\vecz_1,\vecz_2)} &=  |T^n|_{(\vecz_1,\vecz_2)} \,,\\
 \text{Orthogonality:} \qquad  |T^n|_{(\vecz_1,\vecz_2)} \,  |T^k|_{(\vecz_1,\vecz_2)} &= 0 \qquad (n\neq k) \,.
\end{split}
\ee 
These requirements are sufficient to motivate the regularizations we adopt.
For a detailed verification, see Appendix \ref{app:projectors}.

To summarize the above construction, we formally write the projection onto the Virasoro identity OPE block for operators inserted at $\vecz_1$ and $\vecz_2$  as follows:
\begin{equation}
 | \mathbf{1}_\text{Vir} |_{(\vecz_1,\vecz_2)}  \equiv 1 + \sum_{n\geq 1} \, |T^n|_{(\vecz_1,\vecz_2)} \,.
\end{equation}

\subsection{Double-Stress Tensor OPE Block}
\label{sec:TTblock}

We have gathered all the necessary elements to propose 
a generalization of the bilocal OPE block formalism that 
effectively captures Virasoro descendants without relying 
on the Virasoro algebra or techniques related to Chern-Simons 
theory. In this subsection, we will first analyze the double-stress 
tensor states and explicitly compute their contribution to a four-point 
scalar correlator. We will be able to construct OPE blocks that include 
more general multi-stress tensor exchanges, enabling us to build 
the Virasoro identity block order by order in terms of the number 
of exchanged stress tensors.
\vspace{0.3cm}

\noindent {\it Definition and Boundary Condition}

Using the projector \eqref{eq:OOproj2}, we define
\be
\label{eq:BTTdef}
 {\cal B}_{T^2}^{(\cal O)}(\vecz_1,\vecz_2) \equiv n_{T^2} \,\int_{\diasmall_{2}^1} d^2\vecxi \int_{\diasmallPast^\xi_{2}} d^2\vecxi' \, \frac{\langle \{\widetilde{T}(\vecxi) \widetilde{T}(\vecxi')\} {\cal O}(\vecz_1) {\cal O}(\vecz_2) \rangle}{\langle {\cal O}(\vecz_1) {\cal O}(\vecz_2) \rangle} \,  [{T}(\xi) {T}(\xi')] \, .
 \ee
Fixing the normalization factor $n_{T^2}$ requires a boundary condition. 
By employing a regulator $\delta$ that slightly shrinks the diamonds and 
recalling the expansion \eqref{eq:BTexpanded}, we find that
\be
\label{eq:Normalization2}
  n_{T^2}  =  \frac{9}{2\,c^2 (\ln \delta)^2} = 2 \left(  n_{T} \right)^2 
\ee
is the normalization needed to reproduce the well-known short-distance OPE  
expansion of 
${\cal O}{\cal O} \rightarrow T 
+ \text{conformal~descendants}$:\footnote{For 
the convention of the overall (constant) coefficient, 
see, for instance, (C.19) in Ref. \cite{Fitzpatrick:2016mtp}.}
\be
\lim_{\vecz_2 \rightarrow \vecz_1} \; \frac{{\cal B}_{T^2}^{(\cal O)}(\vecz_1,\vecz_2)}{(z_1-z_2)^4} =  \frac{2h(5h+1)}{5(2\pi)^2c^2} \; [T(z_1)^2] \,.
\ee 
We emphasize that the non-trivial, $h$-dependent factor $(5h+1)$ emerges 
from explicitly carrying out the diamond integrals with the regularized kernel.
\vspace{0.3cm}

\noindent {\it Connection to Wilson lines}

Before computing the conformal correlator, let us point out 
an interesting connection with the Chern-Simons Wilson 
line approach \cite{Fitzpatrick:2016mtp}. By performing the 
anti-holomorphic integrals, the double-stress tensor 
OPE block reduces to 
\be
\label{eq:BTTholomorphic}
\begin{split}
 {\cal B}^{(\cal O)}_{T^2}(\vecz_1,\vecz_2) 
 &=   {\cal B}_{T^2,h^2}^{(\cal O)}(\vecz_1,\vecz_2) + {\cal B}^{(\cal O)}_{T^2,h}(\vecz_1,\vecz_2) \,,
 \end{split}
 \ee
 where we split into an $\ord(h^2)$ and an $\ord(h)$ contribution: 
\begin{align}
\label{eq:BT2decompose}
  {\cal B}_{T^2,h^2}^{(\cal O)}(\vecz_1,\vecz_2) &= \left(\frac{6h}{\pi c}\right)^2 \int_{z_2}^{z_1} d\xi_1 d\xi_2 \, \frac{(z_1-\xi_1)(\xi_1-z_2)}{(z_1-z_2)} \frac{(z_1-\xi_2)(\xi_2-z_2)}{(z_1-z_2)} \, \Theta(\xi_1-\xi_2) [ T(\xi_1) T(\xi_2)] \,, \nn\\
  {\cal B}^{(\cal O)}_{T^2,h}(\vecz_1,\vecz_2) &= \frac{18h}{\pi^2 c^2} \int_{z_2}^{z_1} d\xi_1 d\xi_2 \, 
 \frac{(z_1-\xi_1)^2 (\xi_1-z_2)^2}{(z_1-z_2)^2}\,\Theta(\xi_1-\xi_2) \, [ T(\xi_1) T(\xi_2)]\,.
\end{align}  
The appearance of the step functions is the consequence of the 
causally ordered diamond integrals. 
Note that, prior to performing the diamond integrals, these expressions have been readily rotated to Euclidean signature.

We observe that these (holomorphic) expressions are identical to the path-ordered 
Wilson line operator discussed in \cite{Fitzpatrick:2016mtp}; to compare, 
see eq.\ (3.8) in that reference. Our approach here based on nested causal diamonds, essentially ``enhances" the Wilson line picture by incorporating specific 
anti-holomorphic dependencies through the use of regulated kernels. 

\vspace{0.3cm}

\noindent{\it Conformal Correlator and Quantum Correction}

The bilocal double-stress tensor OPE block enables us to compute 
various contributions to the four-point correlator \eqref{eq:V4ptTnew} 
at $\ord(\frac{1}{c^2})$. A useful limit is the so-called heavy-light limit 
where  two heavy scalars 
have large scaling dimension  $h_H \sim \ord(c)$, 
and two light scalars 
have $h_L \sim \ord(1)$. 
The semiclassical correlator can be organized 
through a double series expansion in terms of $\mu \sim \frac{h_H }{c}$ 
and $h_L$.
Terms with higher powers of $\frac{1}{c}$ not fitting into this double expansion 
are regarded as quantum corrections to the semiclassical 
result.

The heavy-light identity block ($i.e.$, the correlator) at leading order is normalized to 1. 
At $\ord(\frac{1}{c})$, it is given by the global block computed in \eqref{eq:singleT}.
At $\ord(\frac{1}{c^2})$ we expect the following pieces: 
$(i)$ the semiclassical 
contributions at $\ord\big(\mu^2 h^2_L\big)$ and $\ord\big( \mu^2 h_L\big)$, and 
$(ii)$ the quantum corrections to the $\ord\big(\frac{h_H}{c}\big)$ 
result \eqref{eq:singleT}.\footnote{In this work, we focus on the contributions from 
the stress tensor sector.  We do not consider the contributions from double traces made out of scalar operators.} 

The ``propagator'' of the double-stress tensor OPE block \eqref{eq:OOproj2} 
can be evaluated using nested causal diamonds and the regularized stress-tensor 
correlators, which at leading order at large $c$ is just a product of two-point functions:
\be
\begin{split}
 \langle [T(z_1) T(z_2)] [T(z_3) T(z_4)] \rangle =  \frac{c^2}{4} \left( \frac{1}{z_{13}^4 z_{24}^4}+ \frac{1}{z_{14}^4 z_{23}^4} \right)  + \ord(c) \, .
\end{split}
\ee
We obtain the complete $\ord({1\over c^2})$ contributions due to the double-stress tensor operators: 
{\normalsize 
\begin{align}
\label{eq:HHLLc2}
   &\left\langle {\cal B}_{T^2}^{(H)}(\infty;1) {\cal B}^{(L)}_{T^2}(\vecz;0)\right\rangle\Big|_{T^2} \\
   \quad &= \frac{1}{2} \left(\frac{h_H h_L}{c}  \, \frac{f_2(z)}{2\pi^2}\right)^2 + \frac{h^2_H h_L+ h_H h^2_L}{c^2} \frac{1}{8\pi^4}\Big( - f_2(z)^2 + \frac{6}{5} \, f_1(z) f_3(z) \Big)\nn\\
   \qquad &+ \frac{h_H h_L}{c^2} \frac{3}{4\pi^4} \Big(\frac{12(z-2)}{z} \, \text{Li}_2(z) + \frac{6(1-z)^2}{z^2} \, \ln(1-z)^2 + \frac{z-2}{z} \, \ln(1-z) + 16 \Big) +\ord\big({1\over c^3}\big)  \nn
\end{align}}where we recall the hypergeometric function  $f_a(z) = z^a \; {}_2F_1(a,a,2a,z)$.  
This expression represents the sum of an infinite number of 
double-stress tensors, constructed from two stress tensors and derivatives. 
The terms at order $\frac{h_H h_L^2}{c^2}$ and $\frac{h_H h_L}{c^2}$ are the quantum corrections.
This result \eqref{eq:HHLLc2}, including the quantum corrections, is consistent with the one 
previously computed using the Virasoro algebra and Chern-Simons Wilson 
lines \cite{Fitzpatrick:2015dlt, Chen:2016cms, Fitzpatrick:2016mtp}.

All the terms in \eqref{eq:HHLLc2} arise from the bilocal double-$T$ OPE block we defined in \eqref{eq:BTTdef}. 
Although we use the exact integration kernel $\langle \{\widetilde{T}\widetilde{T}\}{\cal O}{\cal O}\rangle/ \langle{\cal O}{\cal O}\rangle$, note that 
there are higher order $\frac{1}{c}$ quantum corrections to the correlator -- these are formally also
encoded in the definition of the OPE block, but computing these them explicitly 
requires a better understanding of the higher-order structures of the regularized 
stress-tensor correlators.

\vspace{0.3cm}

\noindent {\it Connection to Graviton Mixing}

We find it instructive to further decompose the expression \eqref{eq:HHLLc2} to develop a physical interpretation. 
To this end, and in order to keep the discussion simple, we shall ignore quantum corrections here. 

The first term in \eqref{eq:HHLLc2} at ${\cal O}(h^2_L)$ follows immediately from \eqref{eq:BT2decompose}.
This piece is unchanged by the regularization of the integration kernel; see \eqref{eq:TtTtOOsplit}.\footnote{This piece is related to the exponentiation of the correlator at large $h_L$ which can be computed as a geodesic in $AdS_3$ using holography.}

Next, we consider the more non-trivial piece at ${\cal O}(h_L)$, which  
has been regularized by subtracting the singular piece in the kernel. 
Let us separately evaluate the non-regularized version and the subtracted piece. Define 
\begin{align}
\label{eq:BTTdefSplit}
& {\cal B}_{T^2
}^{(L)}(\vecz_3,\vecz_4)\Big|_{\rm non-reg.} \equiv n_{T^2} \,\int_{\diasmall_4^3} d^2\vecxi \int_{\diasmallPast_4^\xi} d^2\vecxi' \, \frac{\langle \widetilde{T}(\vecxi) \widetilde{T}(\vecxi') {\cal O}(\vecz_3) {\cal O}(\vecz_4) \rangle}{\langle {\cal O}(\vecz_3) {\cal O}(\vecz_4) \rangle} \,  [{T}(\xi) {T}(\xi')]  \Big|_{h}  \,, \\
\label{eq:BTTdefSplit2}
&  {\cal B}_{T^2
}^{(L)}(\vecz_3,\vecz_4)\Big|_{\rm sub.} \equiv n_{T^2} \,\int_{\diasmall_4^3} d^2\vecxi \int_{\diasmallPast_4^\xi} d^2\vecxi' \, \left(-\frac{\langle \widetilde{T}(\vecxi) \widetilde{T}(\vecxi') {\cal O}(\vecz_3) {\cal O}(\vecz_4) \rangle}{\langle {\cal O}(\vecz_3) {\cal O}(\vecz_4) \rangle}\right)_{\text{sing.} \hspace{0.05cm}  \widetilde{T} \rightarrow  \widetilde{T}}
 [{T}(\xi) {T}(\xi')] \,
\end{align}
 where the singular piece 
was given in \eqref{eq:TtTtOOsplit}.
The contribution \eqref{eq:BTTdefSplit2} that we eliminate in our regularization scheme has a nice physical interpretation. To see this, we note how the two terms contribute to the four-point function: 
{\normalsize
\begin{align}
 &\left\langle {\cal B}_{T^2,h^2}^{(H)}(\infty;1) \Big({\cal B}^{(L)}_{T^2, h}(\vecz;0)\Big|_{\rm non-reg.}\Big) \right\rangle \nn\\
   &~~~~~~~~= \frac{h_H^2 h_L}{c^2} \frac{9}{2\pi^4} \Big(-\frac{16+(16-z)z}{(1-z)} - \frac{7(2-z)}{z}\,\ln(1-z)+ \frac{2(1-z)}{z^2}\,\ln(1-z)^2
 \Big)\ , \\
&\left\langle {\cal B}_{T^2,h^2}^{(H)}(\infty;1) \Big({\cal B}^{(L)}_{T^2, h}(\vecz;0)\Big|_{\rm sub.} \Big) \right\rangle \nn\\
   &~~~~~~~~ = \frac{h_H^2 h_L}{c^2} \frac{9}{2\pi^4}  \Big( \frac{12-z(12-z)}{z(1-z)}+ \frac{6(2-z)\ln(1-z)}{z} \Big) \ .
   \label{eq:subPiece}
\end{align}}We observe that these expressions precisely match the $2\rightarrow 2$ graviton scattering and $2\rightarrow 1$ graviton mixing, as computed using a basis defined by the Virasoro algebra -- for a detailed analysis in terms of Virasoro modes, see \cite{Fitzpatrick:2015qma} (specifically their (D.23) and (D.24)). In that context, the graviton mixing originates from the non-orthogonality of the two-graviton and one-graviton states defined in terms of Virasoro modes, making a subtraction necessary. It is interesting that our approach based on causul dimonds and shadow operators yields a one-to-one correspondence to the Virasoro algebra computation.\footnote{Note that one might interpret the ``cross term'' $\big\langle {\cal B}_{T^2}^{(H)}(\infty;1) {\cal B}_{T}^{(L)}(\vecz;0) \big\rangle$ as the $2\rightarrow 1$ graviton mixing. However, this term vanishes due to the regularization \eqref{eq:regularizationT} which imposes $\langle [TT] [T] \rangle=0$.} Geometrically, the $2\rightarrow 1$ piece \eqref{eq:subPiece} comes entirely from the singularity in the nested diamond integration where the two stress tensors in the ``ket'' become null separated.

\subsection{Virasoro OPE Blocks} 
\label{sec:virasoroOPE}

Once the generalization from the single-stress tensor OPE block to the double-stress tensor OPE block has been achieved, the extension to incorporate arbitrary multi-stress tensors is straightforward. Based on the nested causal diamonds and regularization prescriptions, we propose to write the {\it bilocal Virasoro identity OPE block} as 
\be
\label{eq:BvirasoroGeneral}
\mathfrak{B}^{({\cal O})}_{\mathbf{1}}(\vecz_1,\vecz_2)  = 1 + \sum_{n\geq 1}  {\cal B}^{({\cal O})}_{T^n}(\vecz_1,\vecz_2)  \ ,
\ee
{\small
\be
\begin{split}
 {\cal B}^{({\cal O})}_{T^n}(\vecz_1;\vecz_2) = n_{T^n} \int_{\diasmall_2^1} d^2 \vecxi_1\int_{\diasmallPast_2^{\xi_1}} d^2 \vecxi_2 \cdots \int_{\diasmallPast_2^{\xi_{n-1}}} d^2 \vecxi_n \, \frac{\big\langle \{\widetilde{T}(\vecxi_1) \cdots \widetilde{T}(\vecxi_n) \} {\cal O}(\vecz_1) {\cal O}(\vecz_2) \big\rangle}{\langle {\cal O}(\vecz_1) {\cal O}(\vecz_2)  \rangle} \, [ T(\xi_1) \cdots T(\xi_n) ]  \ . \nn
\end{split}
\label{eq:Bvirasoro}
\ee
}This can be obtained by acting with the $T^n$ projector 
\eqref{eq:projnDef} on the bra $\langle {\cal O}(\vecz_1) {\cal O}(\vecz_2)|$.
The integration regions correspond to successively nested 
causal diamonds, as indicated in Figure \ref{fig:diamonds2d}, 
which ensures the causal ordering $\vecz_2 \prec \vecxi_n \prec \cdots \prec \vecxi_1 \prec \vecz_1$.\footnote{The $n$ causal diamonds appearing in \eqref{eq:Bvirasoro} are, in the language 
of \cite{deBoer:2016pqk}, all null separated from each other in the kinematic space metric. The moduli space for the integrals in \eqref{eq:Bvirasoro} thus consists of ordered points along a null ray in kinematic space.}
This construction passes several immediate checks. 
We have already discussed ${\cal B}_{T}^{({\cal O})}$ 
and ${\cal B}_{T^2}^{({\cal O})}$ 
 in detail. To test the consistency to all orders, let us make 
a connection to the Virasoro OPE blocks obtained using 
Chern-Simons Wilson lines \cite{Fitzpatrick:2016mtp}; see also \cite{Besken:2018zro}.

We can use the shadow recursion relation \eqref{eq:Ttrecursion} and write 
{\small
\begin{align}
 {\cal B}^{({\cal O})}_{T^n}(\vecz_1;\vecz_2) &=n_{T^n} \left( \frac{-\ln \delta}{2\pi} \right)\int_{\diasmall_2^1} d^2 \vecxi_1\int_{\diasmallPast_2^{\xi_1}} d^2 \vecxi_2 \cdots \int_{\diasmallPast_2^{\xi_{n-2}}} d^2 \vecxi_{n-1}\int_{\xi_2}^{\xi_{n-1}} d\xi_n  \; \frac{1}{\langle {\cal O}(\vecz_1) {\cal O}(\vecz_2) \rangle}\\
 &\quad\;\; \times \left(2 (z_2-\xi_n)^2\partial_{z_2} +4h(z_2-\xi_{n}) \right) \big\langle \{\widetilde{T}(\vecxi_1) \cdots \widetilde{T}(\vecxi_{n-1}) \} {\cal O}(\vecz_1) {\cal O}(\vecz_2) \big\rangle \, [T(\xi_1) \cdots T(\xi_n)] \nn
\end{align}
}where we have integrated out an anti-holomorphic 
coordinate $\bar \xi_n$ from $\widetilde{T}(\vecxi_{n})$. 
All $z_1$ dependences vanish after performing the anti-holomorphic integral.
By repeating the process $n-1$ more times, we 
remove all the anti-holomorphic dependencies and obtain 
{\small
\begin{align}
\label{repW}
 {\cal B}^{({\cal O})}_{T^n}(\vecz_1;\vecz_2) 
 &= n_{T^n}  \times 2\left( \frac{-\ln \delta}{2\pi}  \right)^n \int_{z_2}^{z_1} d\xi_1 \int_{z_2}^{\xi_1} d\xi_2 \cdots \int_{z_2}^{\xi_{n-1}} d\xi_n \; \frac{1}{\langle {\cal O}(\vecz_1) {\cal O}(\vecz_2) \rangle} \nn\\
 &\quad\;\; \times \left( \prod_{i=1}^n \left( 2(z_2- \xi_i)^2 \partial_{z_2} + 4h (z_2 - \xi_i) \right) \langle {\cal O}(\vecz_1) {\cal O}(\vecz_2) \rangle\right) [T(\xi_1) \cdots T(\xi_n)] \ .
\end{align}
}Note the final integral over $\bar{\xi}_1$ produces an 
additional factor $2$ because the integration range now 
involves $\bar{z}_1$ as its upper bound, thus producing 
one more singularity than integrals over $\bar{\xi}_2,\ldots,\bar{\xi}_n$. 
It remains to fix the normalization of the OPE block:
\be
 n_{T^n} = 2^{n-1}\left(n_T \right)^n = \frac{1}{2}\left( - \frac{3}{c \, \ln \delta} \right)^n \,.
\ee  
When $n=2$, this reduces to \eqref{eq:Normalization2}. For general $n$, the correctness of this normalization can be confirmed by observing that it implies an exponentiated structure (eikonalization) of the OPE block. To see this, note that the highest power of $h$ is always canonical:\footnote{We can write \eqref{eq:BTnSingle} in a simpler form by summing over permutations of stress tensor insertions in Euclidean signature (to avoid possible ordering ambiguities in $[T^n]$):
\be
\label{2dexp}
{\cal B}_{T^n, \, Euc. }^{({\cal O})} (\vecz_1;\vecz_2) \Big|_{{\cal O}(h^n)} =   \frac{1}{n!} \left( {\cal B}_{T,  \, Euc. }^{({\cal O})} (\vecz_1;\vecz_2) \right)^n 
\ .
\ee }
{\small
\begin{align}
\label{eq:BTnSingle}
 &{\cal B}_{T^n}^{({\cal O})} (\vecz_1;\vecz_2) \Big|_{{\cal O}(h^n)} 
  = \left( -\frac{6h}{\pi c} \right)^n \int_{z_2}^{z_1} d\xi_1 \int_{z_2}^{\xi_1} d\xi_2 \cdots \int_{z_2}^{\xi_{n-1}} d\xi_n \left( \prod_{i=1}^n \frac{(z_1-\xi_i)(\xi_i-z_2)}{(z_2-z_1)} \right) [T(\xi_1) \cdots T(\xi_n)] \,.
\end{align}
}From this expression one can immediately compute the so-called ``light-light'' conformal identity block, $i.e.$, a four-point correlator where one takes $h \sim h' \sim \ord(\sqrt{c})$ and therefore only resums powers $\big(\frac{hh'}{c}\big)^n$ at leading order in large $c$:
\begin{equation}
\begin{split}
 \frac{\langle {\cal O}(\infty){\cal O}(1) {\cal O}'(\vecz) {\cal O}'(0)\rangle}{\langle {\cal O}(\infty){\cal O}(1)\rangle\langle {\cal O}'(\vecz) {\cal O}'(0)\rangle}
& = \sum_{n\geq 0} \left\langle \mathcal{B}_{T^n}^{({\cal O})}(\infty;1)\,{\cal B}_{T^n}^{({\cal O}')}(\vecz;0) \right\rangle \bigg|_{\ord\left( (\frac{hh'}{c})^n \right)} + \ldots\\
&= \text{exp} \left( \left\langle \mathcal{B}_{T}^{({\cal O})}(\infty;1)\,{\cal B}_{T}^{({\cal O}')}(\vecz;0) \right\rangle \right) \times (1 + \ldots)\\
&= \text{exp} \left( \frac{hh'}{2\pi^2 c} \, f_2(z) \right) \times (1+ \ldots)\,.
\end{split}
\end{equation}
This exponentiated structure is well-known; see, $e.g.$, \cite{Fitzpatrick:2014vua,Fitzpatrick:2015qma}. 
The structure \eqref{repW} and the normalization \eqref{eq:BTnSingle} are both consistent with the Wilson-line approach \cite{Fitzpatrick:2016mtp}.\footnote{The stress tensor here is defined with a factor of $-2\pi$ difference compared to the convention in \cite{Fitzpatrick:2016mtp}.}  
\vspace{.3cm}

\noindent{\it Symmetries of the OPE Blocks}

The OPE blocks manifest a global conformal symmetry $SL(2,\mathds{R})$. This is a consistency requirement such that correlators of OPE blocks indeed compute conformal correlation functions. A (right-moving) $SL(2,\mathds{R})$ transformation acts as follows:
\be
\label{eq:sl2Rtrf}
 (z,\bar z) \; \mapsto \; \left(\,\mathfrak{f}(z),\bar z\,\right) \equiv \left(\frac{az+b}{cz+d} \,,\; \bar z\, \right) \,.
\ee
For illustration, consider the single-stress tensor block ${\cal B}_T^{({\cal O})}(\vecz_1;\vecz_2)$. It transforms as follows:
\begin{align}
  {\cal B}_T^{({\cal O})}(\vecz_1;\vecz_2)
  \, & \mapsto\, n_T \, \int_{\mathfrak{f}(z_2)}^{\mathfrak{f}(z_1)} d\xi \int_{\bar z_2}^{\bar z_1} d\bar\xi \, \frac{\big\langle \widetilde{T}(\xi,\bar \xi) {\cal O}(\mathfrak{f}(z_1),\bar z_1){\cal O}(\mathfrak{f}(z_2),\bar z_2) \big\rangle}{\langle {\cal O}(\mathfrak{f}(z_1),\bar z_1){\cal O}(\mathfrak{f}(z_2),\bar z_2)\rangle} \, T(\xi)
  \nn\\
  &= n_T\, \int_{z_2}^{z_1} d\eta \, \mathfrak{f}'(\eta) \int_{\bar z_2}^{\bar z_1} d\bar\eta \, \frac{\big\langle \widetilde{T}(\mathfrak{f}(\eta),\bar\eta) {\cal O}(\mathfrak{f}(z_1),\bar z_1){\cal O}(\mathfrak{f}(z_2),\bar z_2) \big\rangle}{\langle {\cal O}(\mathfrak{f}(z_1),\bar z_1){\cal O}(\mathfrak{f}(z_2),\bar z_2)\rangle} \, \frac{T(\eta)- \frac{c}{12} \, \text{Sch}(\mathfrak{f},\eta)}{\mathfrak{f}'(\eta)^2}
  \nn\\
  &= {\cal B}_T^{({\cal O})}(\vecz_1;\vecz_2)\,
\end{align}
where we changed the integration variables to $(\eta,\,\bar\eta) = \big(\mathfrak{f}^{-1}(\xi),\,\bar\xi\,\big)$ and we used the usual stress tensor transformation rule involving a Schwarzian derivative. The Schwarzian derivative vanishes on $SL(2,\mathds{R})$ transformations and the integral then takes the original form.

A similar explicit calculation shows that the $SL(2,\mathds{R})$ invariance is also preserved by the multi-stress tensor OPE blocks \eqref{eq:BvirasoroGeneral}. This property is not immediately obvious due to the involvement of regularized kernels $\langle \{\widetilde{T}^k\} {\cal OO} \rangle/\langle{\cal OO}\rangle$. It is thus a consistency check on the formalism, which we have confirmed explicitly up to $k=4$. Not that, without using $\{\, \cdot \, \}$, this object remains $SL(2,\mathds{R})$ invariant, so the regularization is not fixed by the symmetry. 

\section{Four Dimensions}
\label{sec:4d}

We would like to leverage our construction of bilocal OPE blocks as integrals over causal diamonds and explore a generalization to higher dimensions. In this section, we focus specifically on heavy-light correlators in four-dimensional CFTs with near-lightcone kinematics although for the single-stress tensor exchange the heavy-light limit does not play a role.
We will demonstrate that the lightcone limit serves as a natural starting point for simplifying the structure of the correlator and providing a natural analog to the Virasoro OPE blocks --  in particular, we will be able to perform the monodromy projection automatically and compute the near-lightcone correlator in a manner similar to that of the two-dimensional case.

\subsection{Smeared Representation of Single-Stress Tensor OPE Block}

Analogous to the $d=2$ case \eqref{eq:BTdef}, in $d=4$ one may write the 
global OPE block as \cite{Dolan:2011dv,Czech:2016xec, deBoer:2016pqk}
 \be
\label{eq:BTdef4d}
\begin{split}
 {\cal B}_T(x_f,x_p) &= n_{T}^{(4d)} \int_{\diasmall_p^f} d^4\vecxi \,   
  \frac{ \langle \widetilde T^{\mu\nu}(\vecxi) {\cal O}(x_f) {\cal O}(x_p)\rangle}{\langle  {\cal O}(x_f) {\cal O}(x_p)\rangle} \, T_{\mu\nu}(\vecxi) 
  = \frac{n_{T}^{(4d)} \, C_{T{\cal O}{\cal O}}}{4} \int_{\diasmall_p^f} d^4\vecxi \,  \hat{X}^\mu \hat{X}^\nu \, T_{\mu\nu}(\vecxi) 
  \end{split}
 \ee
 where $x_f$ and $x_p$ are the future and past tips of the diamond and we are now dropping a superscript indicating the reference to ${\cal O}$. The integration kernel is expressed in terms of the conformal Killing vector associated with the causal diamond:
\be 
\label{eq:XhatDef}
  \hat{X}^\mu(\xi) = \frac{X^\mu(\xi)}{|X(\xi)|} \,,\quad X^\mu(\xi) = \frac{(\xi-x_f)^\mu}{(\xi-x_f)^2}-\frac{(\xi-x_p)^\mu}{(\xi-x_p)^2} \,.
\ee 
See Appendix \ref{app:conventions} for related conventions.

\begin{figure}
\begin{center}
\includegraphics[width=1\textwidth]{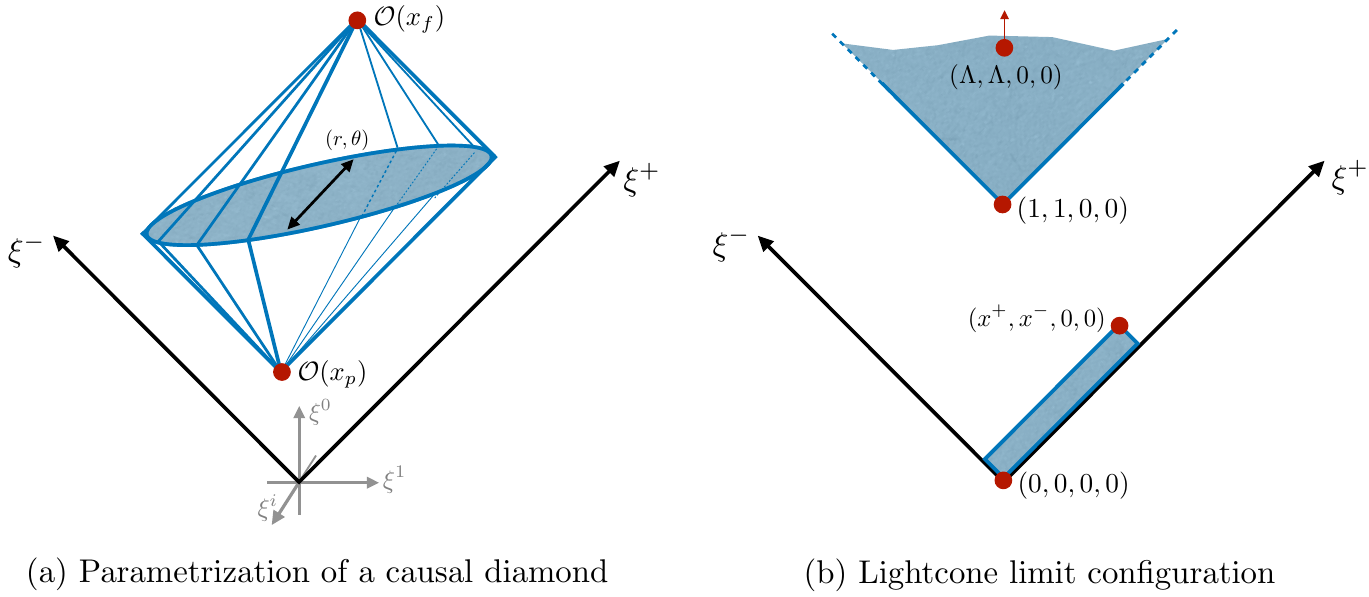}
\end{center}
\caption{Parametrization in four dimensions:  
(a) one causal diamond needed for defining the bilocal OPE block; 
(b) two causal diamonds used to compute the four-point correlator in the near-lightcone configuration ($x^- \rightarrow 0$).}
\label{fig:diamonds}
\end{figure}

For explicit computations, we will consider $x_i^\mu = (x_i^+,x_i^-,0,0)$ where the first two coordinates are null coordinates and we set transverse directions to zero.  We write $\xi^\mu = (\xi^+,\,\xi^-,\,r\cos \theta, \, r\sin\theta)$ to parametrize the diamond.
We can split the diamond in a lower half and an upper half, $i.e.$, $\dia = \diaLower \cup \diaUpper$,  with the following integration ranges: 
{\small
\be
\label{eq:diamondParametrization}
\begin{split}
\diaLower: \;\; \xi^+ &\in [x_p^+,x_f^+],\;\; \xi^- \in \bigg[x_p^-,\,x_f^- - \frac{x_{fp}^-}{x_{fp}^+}\,(\xi^+-x_p^+)\bigg] ,\;\; 
 \theta \in [0,2\pi) ,\;\; r^2 \in \left[0,(\xi^+-x_p^+) (\xi^--x_p^-)\right]\\
\diaUpper: \;\; \xi^+ &\in [x_p^+,x_f^+],\;\; \xi^- \in \bigg[x_f^- - \frac{x_{fp}^-}{x_{fp}^+}\,(\xi^+-x_p^+),\, x_f^-\bigg],\;\; \theta \in [0,2\pi) ,\;\; r^2 \in \left[0,(x_f^+-\xi^+) (x_f^--\xi^-)\right]
\end{split}
\ee
}Note that the radial integral needs to be regularized. We adopt a similar scheme as in the two-dimensional case:
we cut off the integral over $r$ at $r_\text{max}-\delta$, where $r_\text{max}$ are the respective upper bounds in \eqref{eq:diamondParametrization}. The parametrization is illustrated in Figure \ref{fig:diamonds}(a).

As in the two-dimensional case, we fix the normalization $n_T^{(4d)}$ in \eqref{eq:BTppExplicit} by demanding the correct short-distance OPE limit.  We find 
\be
\label{eq:TppBC}
\lim_{x_f\rightarrow x_p} \;{\cal B}_{T_{++}}(x_f,x_p)  \sim  - \frac{\pi}{120} \,  n_{T}^{(4d)}\,C_{T{\cal O}{\cal O}} \, \ln (\delta) \;   (x_{fp}^+)^2 \,x_{fp}^\mu x_{fp}^\nu\, T_{\mu\nu}(x_p)\,.
\ee
This has to be consistent with the ${\cal O} \times {\cal O} \rightarrow T$ OPE:
\be
\begin{split}
\lim_{x_f \rightarrow x_p} \; \frac{ {\cal O}(x_f) {\cal O}(x_p) }{\langle {\cal O}(x_f) {\cal O}(x_p) \rangle} 
&\sim 1 + \frac{C_{T{\cal O}{\cal O}}}{C_T} \,(x_{fp}^+)^2 \,x_{fp}^\mu x_{fp}^\nu\, T_{\mu\nu}(x_p) + \ldots
\end{split}
\ee
Consistency thus requires the following normalization:
\be
\label{eq:nTOO4d}
 n_{T}^{(4d)} = -\frac{120}{\pi \,C_T\,\ln \delta} \,.
\ee

\subsection{Application: Lightcone Limit and Heavy-Light Correlator}

Having described the basic ingredients, let us now consider the four-point function of pairwise identical operators in four dimensions. 
For concreteness, we focus on the heavy-light correlator in the lightcone limit (denoted by the symbol `$\times$'):
\be
 {\cal V}^{\bf \times}(x^+,x^-) \equiv \lim_{x^- \rightarrow 0}\frac{\langle {\cal O}_H (\infty){\cal O}_H(1)  {\cal O}_L(x^+,x^-)  {\cal O}_L(0) \rangle}{\langle  {\cal O}_H (\infty)  {\cal O}_H(1)\rangle\langle  {\cal O}_L(x^+,x^-)  {\cal O}_L(0) \rangle} \,
 \label{eq:VHHLL4d}
\ee 
where the explicit insertion points are given by
\be
\label{eq:4dinsertions}
 x_1^\mu = (\Lambda,\Lambda,0,0) ,\quad x_2^\mu = (1,1,0,0) , \quad x_3^\mu = (x^+,x^-,0,0) ,\quad x_4^\mu = (0,0,0,0) 
\ee
and the limit $\Lambda \rightarrow \infty$ is always implied; see, Figure \ref{fig:diamonds}(b).
We will construct a near-lightcone bilocal OPE block involving a stress tensor based on the causal diamond approach. Note that computing the single-stress tensor exchange does not require regularizations 
$[ \, \cdot \, ]$ and $\{\, \cdot \, \}$, a large central charge, or the heavy-light limit. Here we adopt the heavy-light limit primarily in order to align our discussion with the established computations in two dimensions.

We claim that ${\cal V}^\times(x^+,x^-)$ can be computed, to leading order in the $x^- \rightarrow 0$ limit, as
\be
{\cal V}^\times(x^+,x^-) = \lim_{x^- \rightarrow 0} \left\langle {\cal B}_T(x_1,x_2) \, {\cal B}^\times_{T_{++}} (x^\pm) \right\rangle \,
\ee
where the {\it lightcone OPE block} is defined as the following (non-covariant) component:
 \be
\label{eq:BTlcDef}
\begin{split}
 {\cal B}_{T_{++}}^\times(x^\pm) &\equiv n_{T}^{(4d)} \int_{\diasmall_0^\pm} d^4\veceta \,   
  \frac{ \langle \widetilde T^{++}(\veceta) {\cal O}(x_3) {\cal O}(x_4)\rangle}{\langle  {\cal O}(x_3) {\cal O}(x_4)\rangle} \, T_{++}(\veceta) \,.
  \end{split}
\ee
The diamond $\dia_0^\pm$ has future and past tips $x_3^\mu$ and $x_4^\mu$ and we use coordinates $\veceta$ instead of $\vecxi$ to label the lightcone OPE block.
The fact that the component $T_{++}$ dominates in the lightcone limit will become transparent below. Qualitatively, it can be understood as a consequence of the fact that every worldline within the causal diamond $\dia_0^\pm$ is almost null along the $\eta^+$ direction and energy transport therefore predominantly occurs in this direction.

\vspace{0.3cm}

\noindent {\it Lightcone OPE Block}

Let us consider the lightcone OPE block \eqref{eq:BTlcDef} in some more detail. 
The tips of the corresponding diamond are $x_3^\mu=(x^+,x^-,0,0)$ and $x_4^\mu=(0,0,0,0)$. Explicitly:
\be
\label{eq:BTppExplicit}
 {\cal B}_{T_{++}}(x^\pm) = \frac{n_T^{(4d)} \, C_{T{\cal O}{\cal O}}}{4\,x^+ x^-} \int_{\diasmall_0^\pm}d\eta^+ d\eta^- d\theta dr \, \frac{   r \left(r^2 x^+ - x^-  \eta^+(x^+ - \eta^+) \right)^2}{ \left(r^2 - (\eta^+-x^+)(\eta^--x^-)\right)  \left(r^2 - \eta^+\eta^-\right)}\,T_{++}(\veceta)\,.
\ee
It is instructive to define dimensionless variables
 \be 
\tilde\eta^\pm \equiv \eta^\pm/x^\pm \ , ~~~ \tilde r \equiv r/\sqrt{x^+ x^-} \,,
 \ee
 to scale out the $x^\pm$ dependence from \eqref{eq:BTppExplicit}:
{\small
\begin{align}
\label{eq:BTdef4d3}
& {\cal B}_{T_{++}}^\times(x^\pm) = \frac{n_T^{(4d)}\, C_{T{\cal O}'{\cal O}'}}{4}\,(x^+)^3x^-\int_0^{1}d\tilde\eta^+ \int_0^{1} d\tilde\eta^- \int_0^{2\pi} d\theta \int_0^{\tilde{r}_\text{max}-\delta}  \tilde r\,d\tilde r \\
& \qquad \times\frac{\left(\tilde r^2  - \tilde \eta^+(1-\tilde \eta^+)\right)^2}{ \left( \tilde  r^2 - (1-\tilde \eta^+)(1-\tilde \eta^-)\right)\left( \tilde r^2 - \tilde \eta^+ \tilde \eta^- \right)} \, T_{++}\big(\eta^+ = x^+\tilde \eta^+ \,, \, \eta^- = x^- \tilde \eta^- \,,\, r = \sqrt{x^+ x^-}\, \tilde r \,, \, \theta\big) \, \nn
  \end{align}
 }with $\tilde r_\text{max}^2 = \text{min}\big(\tilde \eta^+\tilde\eta^-,(1-\tilde \eta^+)(1-\tilde \eta^-)\big)$. All $x^\pm$ dependence is captured by the overall prefactor and the arguments of the stress tensor itself.
It is clear that, to leading order in $x^-$, the operator $T_{++}$ in \eqref{eq:BTdef4d3} will be evaluated at $\eta^- = r=0$ when contracted with any other operator at a finite separation. Using this observation, we can directly carry out the integrals over coordinates $\tilde \eta^-$ and $\tilde r$. 
We obtain
 \be
\label{eq:BTlc1}
\begin{split}
{\cal B}_{T_{++}}^\times (x^\pm)
&=  \frac{15 C_{T{\cal O}'{\cal O}'}}{\pi C_T}\frac{x^-}{(x^+)^2} \int_0^{x^+} d\eta^+ \, (\eta^+)^2 (x^+-\eta^+)^2\; T^\circ_{++}\big(\eta^+\big) + \ord\left((x^-)^2\right) \,
 \end{split}
 \ee
where 
 \be
T_{++}^\circ(\eta^+) \equiv \int_0^{2\pi} d\theta \, T_{++}(\eta^+,\,\eta^-=0,\,r=0,\,\theta)
 \ee
 is the stress tensor smeared around a transverse circle along the null ray $\eta^-=0$.\footnote{When considering correlators of operators at identical transverse locations, the smearing becomes trivial and simply yields a factor of $2\pi$.}

The expression \eqref{eq:BTlc1} is effectively one-dimensional, $i.e.$, written as a single integral along a null ray. This is reminiscent of the holomorphic expression in two dimensions, \eqref{eq:BTdef2}, up to the difference of powers in the integration kernel. To leverage this observation further, we will show in Sec.\ \ref{sec:reparametrizations} that \eqref{eq:BTlc1} in fact has the exact same structure as an OPE block for the exchange of a spin-3 conserved current in two-dimensional CFT (see \eqref{eq:BW2} below). This can also be observed by considering the OPE limit of the lightcone single-stress tensor OPE block at leading order in the $x^-$ expansion:
\begin{align}
\label{eq:OPE4dblock}
&\lim_{x_+\rightarrow 0}\lim_{x_-\rightarrow 0}  {\cal B}_{T_{++}}^\times(x^\pm) \nn \\
&\quad = \frac{C_{T{\cal O}'{\cal O}'}}{2\pi C_T} \, x^-  (x^+)^3 \left[ T^\circ_{++}(0) + \frac{1}{2} \, x^+  \partial_+ T^\circ_{++}(0) + \frac{1}{7} \, (x^+)^2  \partial_+^2 T^\circ_{++}(0)  + \ord \big( (x^+)^3 \big)\right]+\ord\big((x^-)^2\big)
\end{align}
In Sec. \ref{sec:reparametrizations} we will find that, up to an overall factor, this structure is identical (to all orders in $x^+$) to that encountered in two-dimensional CFTs with a spin-3 conserved current ($W_3(z)$) in the OPE ${\cal O}' \times {\cal O}' \rightarrow W_3 + \text{descendants}$.\footnote{In passing, we also note that this near-lightcone structure obtained in four dimensions exhibits a similarity with the two-dimensional $TT$ OPE, which forms a symmetry algebra.}
This observation will motivate us to consider an effective spin-3 reparametrization mode in four dimensions.

The $T_{++}$ component of the single-$T$ lightcone OPE block, ${\cal B}_{T_{++}}^\times(x^\pm)$, dominates the lightcone limit. To verify this, one can derive an expression similar to \eqref{eq:BTdef4d3} for all other components of the integrand $\langle \widetilde{T}^{\mu\nu}(\veceta){\cal O}(x_3){\cal O}(x_4)\rangle T_{\mu\nu}(\veceta)$:  every other component is at least $\ord\big( (x^-)^2 \big)$ and therefore subleading. 

\vspace{0.3cm}

\noindent{\it Lightcone Correlator from OPE Block: Single-Stress Tensor Exchange}

We now use \eqref{eq:BTlc1} to compute the four-point correlator at $\ord(x^-)$. Explicitly, we compute ${\cal V}^\times(x^+,x^-)$ as follows:
\begin{align}
\label{eq:BB4dCalc}
&\lim_{x^-\rightarrow 0}\, \left\langle  {\cal B}_{T}(\Lambda,1) \, {\cal B}_{T_{++}}^\times(x^\pm) \right\rangle = n_T^{(4d)}  \, \frac{15 C_{T{\cal O'O'}}}{\pi C_T} \frac{x^-}{(x^+)^2}\\
&\qquad\qquad\qquad \times \int_{\diasmall_1^\Lambda} d^4\vecxi \int_0^{x^+} d\eta^+ \, (\eta^+)^2 (x^+-\eta^+)^2 \,  \frac{\langle \widetilde{T}^{\mu\nu}(\vecxi){\cal O}(\Lambda){\cal O}(1)\rangle}{\langle{\cal O}(\Lambda){\cal O}(1)\rangle}  \, \left\langle T_{\mu\nu}(\vecxi\,) T^\circ_{++}(\eta^+) \right\rangle  \nn
\end{align}
with $\vecxi$ covering the following infinitely extended causal diamond ($c.f.$, Figure \ref{fig:diamonds}(b)):
\be
\xi^+ \in [1,\Lambda]\,,\quad \xi^- \in [1,\Lambda]\,,\quad \theta \in [0,2\pi] \,,\quad r^2 \in [0,r_\text{max}^2 - \delta]\,
\ee
with $r_\text{max}^2 \equiv \text{min}\left\{ (\xi^+-1) (\xi^--1), (\Lambda-\xi^+)(\Lambda-\xi^-)\right\}$. The quantities appearing in the integral are defined in Appendix \ref{app:conventions}. After summing over contractions, we obtain
{\small
\be
\begin{split}
&\frac{\langle \widetilde{T}^{\mu\nu}(\vecxi){\cal O}(\Lambda){\cal O}(1)\rangle}{\langle{\cal O}(\Lambda){\cal O}(1)\rangle}  \, \left\langle T_{\mu\nu}(\vecxi\,) T^\circ_{++}(\eta^+) \right\rangle \\
&\quad = \frac{ C_{T{\cal O O}}}{4} \, \frac{\left(r^4 + r^2 (\Lambda+\xi^-+\Lambda \xi^--2\xi^+\xi^-) - (\xi^-)^2(\Lambda-\xi^+)(\xi^+-1)\right)^2}{\left(r^2-(\Lambda-\xi^+)(\Lambda-\xi^-)\right)\left(r^2-(1-\xi^+)(1-\xi^-)\right)}\, \frac{C_T}{4\left(r^2-(\xi^+-\eta^+)\xi^-\right)^6} \ .
\end{split}
\ee} Carrying out all the integrals, we find\footnote{In this computation, it is essential to take the regulator $\delta \rightarrow 0$ {\it before} taking $\Lambda \rightarrow \infty$.}
\begin{align}
\label{eq:BTblockF}
{\cal V}^\times(x^+,x^-) = \lim_{x^-\rightarrow 0}\, \left\langle  {\cal B}_{T}(\Lambda,1) \, {\cal B}_{T_{++}}^\times(x^\pm) \right\rangle  = \frac{C_{T{\cal O}{\cal O}}C_{T{\cal O}'{\cal O}'}}{4C_T}\,f_3(x^+) \, x^- \, \ .
\end{align}
This is indeed the correct near-lightcone correlator in four dimensions \cite{Dolan:2000ut}, crucially without a spurious shadow contribution. It is possible to also compute higher orders in $x^-$ using \eqref{eq:BTlc1}, but we expect that other components of the stress tensor will start to contribute when moving away from the lightcone.

\vspace{0.3cm}

\noindent{\it Reduction from $d=4$ Lightcone Correlator to $T_{++}$ Exchange}

We conclude with the following observation: in the above calculation it was manifest that we only needed to keep the component $T_{++}$ in the lightcone OPE block, thanks to the small $x^-$ limit; but we considered all components $T_{\mu\nu}$ in the ``bra'' $\langle {\cal B}_T(\Lambda,1)|$. The latter is important because any individual component would lead to divergences as $\Lambda\rightarrow\infty$, thus signalling a breaking of conformal invariance. However, one can also obtain the correct result \eqref{eq:BTblockF} in a simpler way by keeping only the $T_{++}$ component in both the ket and the bra, and furthermore aligning the transverse location of the stress tensor operator in the bra with the lightcone limit. Let us denote this projected OPE block as follows:
\be
\label{eq:BbraSimplified}
 {\cal B}_{T_{++}}^{(\perp=0)}(\Lambda,1) \equiv 
n_T \int_{\diasmall_1^\Lambda} d^4\vecxi \;  \frac{\langle \widetilde{T}^{++}(\vecxi){\cal O}(\Lambda){\cal O}(1)\rangle}{\langle{\cal O}(\Lambda){\cal O}(1)\rangle}  \; T_{++}(\xi^+,\,\xi^-,\, r= 0 ,\, \theta) \,
\ee
where $T_{++}$ is projected onto $r=0$, thus aligning it with the lightcone limit of \eqref{eq:BTdef4d3}.\footnote{More explicitly, we have ($\Lambda\rightarrow \infty$ is implied, as always):
 \be
\label{eq:BbraSimplified2}
\begin{split}
& {\cal B}_{T_{++}}^{(\perp=0)}(\Lambda,1) = 
 \frac{15 C_{T{\cal OO}}}{\pi  C_T} \int_1^\Lambda d\xi^+  \int_1^\Lambda d\xi^-  \; \frac{(\Lambda+1-\xi^+-\xi^-)}{(\Lambda-1)} \\
 &\qquad\qquad \times \left[ (\xi^+-1)^2\, \Theta \left(\Lambda+1-\xi^+-\xi^-  \right) - (\Lambda-\xi^+)^2\, \Theta\left(\xi^++\xi^- - \Lambda-1\right)\right]
  \; \int_0^{2\pi} d\theta\, T_{++}(\xi^+,\xi^-, 0 , \theta) \,.
  \end{split}
\ee
}
One can confirm that this reduced object still computes the leading single-stress tensor exchange correctly:
\be
\label{eq:singleTsimplifiedResult}
{\cal V}^\times(x^+,x^-) = \lim_{x^-\rightarrow 0}\, \left\langle {\cal B}_{T_{++}}^{(\perp=0)}(\Lambda,1)  \, {\cal B}_{T_{++}}^\times(x^\pm) \right\rangle \,.
\ee
It would be interesting to understand this better. While it could be a coincidence, it seems suggestive of a ``collinear'' null-line symmetry, $e.g.,$ \cite{Braun:2003rp, Belin:2020lsr, Huang:2021hye}, protecting a sector of stress tensor components that dominate the lightcone limit. The lightcone stress-tensor exchange is then projected into a transverse plane and dominated by the $T_{++}$ exchange.

\section{Three Dimensions}
\label{sec:threedim}

We would like to test our OPE block formalism in odd-dimensional CFTs, where the conformal blocks are known to be more complicated than in even-dimensional cases; see, $e.g.$, 
\cite{Kos:2014bka, Kos:2013tga, Penedones:2015aga, Erramilli:2019njx, Erramilli:2020rlr, Fortin:2022fov}. 
From the perspective of the effective action, a particularly obstructive aspect in odd dimensions is the absence of a conformal anomaly, which seems to hinder the construction of an effective theory for stress tensor exchanges. In two dimensions, such an effective action (or the generating functional for stress tensor correlators) is given by the conformal anomaly action \cite{POLYAKOV1981207}. 
For concreteness, we focus on the three-dimensional case. In this section, we show that the near-lightcone OPE block approach indeed captures the three-dimensional near-lightcone correlator, suggesting that an effective theory may exist, at least in the near-lightcone regime.
\vspace{0.3cm}

\noindent {\it Bilocal Stress-Tensor OPE block in Three Dimensions} 

Consider a three-dimensional causal diamond with its past and future tips located at $x_p^\mu$ and $x_f^\mu$, respectively.
We denote such a diamond as $\dia_p^f = \diaLower \cup \diaUpper$ and parametrize it in the following piecewise way:
\be
\label{eq:diamondParametrization3d}
\begin{split}
\diaLower: \quad \xi^+ &\in [x_p^+,x_f^+]\,,\quad \xi^- \in \bigg[x_p^-,\,x_f^- - \frac{x_{fp}^-}{x_{fp}^+}\,(\xi^+-x_p^+)\bigg] \,,\quad
\xi^1 \in \left[-\xi_p^\text{max}+\delta, \, \xi_p^\text{max}-\delta\right] \,  \\
\diaUpper: \quad \xi^+ &\in [x_p^+,x_f^+]\,,\quad \xi^- \in \bigg[x_f^- - \frac{x_{fp}^-}{x_{fp}^+}\,(\xi^+-x_p^+),\, x_f^-\bigg]\,,\quad \xi^1 \in \left[-\xi_f^\text{max}+\delta, \, \xi_f^\text{max}-\delta\right]\,
\end{split}
\ee
where 
\be
\xi_p^\text{max} = \sqrt{(\xi^+-x_p^+) (\xi^--x_p^-)}\,,\qquad
\xi_f^\text{max} = \sqrt{(x_f^+-\xi^+) (x_f^--\xi^-)} \,.
\ee
The single-stress tensor OPE block can be defined covariantly as before:
 \be
\label{eq:BTdef3d}
\begin{split}
 {\cal B}_T(x_f,x_p)
  &= \frac{16 \, n_T^{(3d)} \, C_{T{\cal O}{\cal O}}}{9\pi} \int_{\diasmall_p^f} d^3\vecxi \;  \hat{X}^\mu \hat{X}^\nu \, T_{\mu\nu}(\vecxi) \,, \qquad n_T^{(3d)} =  \frac{18 i}{\sqrt{3\pi}  \, C_T \, \ln \delta} \,,
  \end{split}
 \ee
with $\hat{X}^\mu$ defined in \eqref{eq:XhatDef} which encodes the integration kernel $\langle \widetilde{T} {\cal O}{\cal O}\rangle / \langle {\cal O}{\cal O}\rangle$.  
The normalization factor $n_T^{(3d)}$ is determined using the same method as in even dimensions via the following short-distance OPE limit:
 \be
\label{eq:TppBC3d}
\lim_{x_f\rightarrow x_p} \, {\cal B}_T(x_f,x_p)   \sim \frac{2\,C_{T{\cal O}{\cal O}}}{\sqrt{3\pi} \, C_T} \, \sqrt{x_{fp}^2} \, x_{fp}^\mu \,x_{fp}^\nu\,T_{\mu\nu}(x_p)\,.
\ee Notice the imaginary value of the normalization in odd dimensions. This is to compensate for the sign in the Lorentzian stress-tensor correlator (listed below in \eqref{eq:TT3d}) so that  the OPE limit \eqref{eq:TppBC3d} is real.
\vspace{0.3cm}

\noindent {\it Single-Stress Tensor Exchange near the Lightcone in Three Dimensions}

We now use the three-dimensional OPE block \eqref{eq:BTdef3d} to compute the leading contribution to a four-point function in the lightcone limit $x_- \rightarrow 0$ (a configuration analogous to \eqref{eq:4dinsertions}). The stress-tensor exchange is then dominated by the $T_{++}$ component, as in four dimensions. One can check that the simplified prescription \eqref{eq:BbraSimplified} is valid in three dimensions. We can therefore focus on the $T_{++}$ exchange in the lightcone limit $x^- \rightarrow 0$: 
{\small
 \be
 \label{eq:BTdef3d2inf}
\begin{split}
 {\cal B}_{T_{++}}^{(\perp=0)}(\Lambda;1) &= \frac{16 \, n_T^{(3d)} \, C_{T{\cal O}{\cal O}}}{9\pi} \int_{\diasmall_1^\Lambda}d^3 \vecxi \, \frac{\left((\xi^+-1)(\Lambda-\xi^+) -(\xi^1)^2 \right)^2\; T_{++}(\xi^+,\xi^-,0)}{\left((\Lambda-\xi^-)(\Lambda-\xi^+)-(\xi^1)^2\right) \left((\xi^+-1)(\xi^--1)-(\xi^1)^2\right)} 
  \, \\
  {\cal B}^\times_{T_{++}}(x^\pm;0) &=  \frac{16 \, n_T^{(3d)} \, C_{T{\cal O}'{\cal O}'}}{9\pi} \int_{\diasmall_0^{\pm}}d^3 \veceta \, \frac{\left(\sqrt{\frac{x^-}{x^+}}\,(x^+-\eta^+)\eta^+ -\sqrt{\frac{x^+}{x^-}}\,(\eta^1)^2 \right)^2\; T_{++}(\eta^+,0,0 )}{\left((x^--\eta^-)(x^+-\eta^+)-(\eta^1)^2\right) \left(\eta^+ \eta^- -(\eta^1)^2\right)} \, 
  \end{split}
 \ee
 }where the second expression is the leading term in the lightcone limit. 
To compute the four-point scalar correlator, we use  
\be
\label{eq:TT3d}
 \langle T_{++}(\xi^+,\xi^-,0) \, T_{++}(\eta^+,0,0) \rangle = -\frac{C_T}{4} \, \frac{1}{(\xi^+-\eta^+)^5\, \xi^-} 
\ee 
which has been simplified in the same way as in four dimensions.

Putting everything together, and performing all integrals except the $\xi^+$ integral, we obtain
\be
\begin{split}
 &\lim_{\Lambda\rightarrow\infty}\,\lim_{x^- \rightarrow 0} \; \left\langle {\cal B}^{(\perp=0)}_{T_{++}}(\Lambda;1) {\cal B}^\times_{T_{++}}(x^\pm) \right\rangle 
 =\lim_{\Lambda\rightarrow\infty}\, \frac{16 \,C_{T{\cal O}{\cal O}} \, C_{T{\cal O}'{\cal O}'}}{3\pi^2\,C_T}  (x^-)^{\frac{1}{2}} (x^+)^{\frac{5}{2}} \\
 &\qquad\qquad\qquad \times \int_1^\Lambda d\xi^+ \, \frac{(\xi^+-1)^{\frac{3}{2}} \left[(\Lambda+1-\xi^+) \text{arctan}\left(\sqrt{\Lambda-\xi^+}\right) -\sqrt{\Lambda-\xi^+}\right]}{(\Lambda-1)(\xi^+(\xi^+-x^+))^{\frac{5}{2}}}\,.
\end{split}
\ee
The last integral is more involved, but it can be evaluated by implementing a systematic expansion in the large  $\Lambda$  limit, noting that only the leading term contributes. 
The resulting expression is given in terms of an elliptic integral, which can be written as the following Euler integral representation of the Gauss hypergeometric function:
\be
  \int_1^\infty d\xi^+ \; \frac{(\xi^+-1)^{\frac{3}{2}}}{(\xi^+ ( \xi^+ - x^+))^{\frac{5}{2}}} 
 = 
  \frac{\Gamma\left(\frac{5}{2}\right)^2}{\Gamma(5)} \; {}_2F_1\left( \frac{5}{2}, \frac{5}{2} , 5 ; x^+ \right) \,.
\ee
We obtain the leading result 
\be
\begin{split}
 &\lim_{\Lambda\rightarrow\infty}\,\lim_{x^- \rightarrow 0} \; \left\langle {\cal B}^{(\perp=0)}_{T_{++}}(\Lambda;1) {\cal B}^\times_{T_{++}}(x^\pm) \right\rangle 
 = \frac{C_{T{\cal O}{\cal O}} \, C_{T{\cal O}'{\cal O}'}}{4\,C_T} \; f_{\frac{5}{2}}(x^+) \, \sqrt{x^-}\,.
\end{split}
\ee 
As in the four-dimensional case, it is important to take the lightcone limit before taking the large $\Lambda$ limit.
Using the three-point function coefficient \eqref{eq:CTOOdef}, this final expression obtained via the OPE block matches the three-dimensional correlator previously obtained in \cite{Karlsson:2019dbd}.\footnote{We thank P. Tadi\'{c} for mentioning this existing expression in three dimensions, which allows us to cross-verify our result. To make a comparison, note $(C_T)_\text{here} = \frac{1}{4\pi^4} \, (C_T)_{\text{\cite{Karlsson:2019dbd}}}$ in three dimensions. }

\section{Comments on Reparametrization Modes}
\label{sec:reparametrizations}

In two dimensions, there is strong evidence that Virasoro identity conformal block is computable
using an effective field theory of reparametrization modes, $e.g.$, \cite{Turiaci:2016cvo, Caputa:2018kdj,Haehl:2018izb, Cotler:2018zff, Haehl:2019eae, Anous:2020vtw, Nguyen:2021jja, Blake:2021wqj, Nguyen:2021pdz, Vos:2021erh, Ebert:2022cle, Banerjee:2022wht, Nguyen:2022xsw, Choi:2023mab, Giombi:2023zte}. This approach is formulated in terms of a set of bi-local Feynman rules for the fields that capture the universal dressing of probe operators with stress tensors. The effective action governing the dynamics of these modes is the Alekseev-Shatashvili action arising in the quantization of Virasoro coadjoint orbits \cite{Alekseev:1988ce} (see also \cite{Witten:1987ty}) -- this is a model of lower-dimensional quantum gravity \cite{Polyakov:1987zb,Verlinde:1989ua}, or holography \cite{Cotler:2018zff,Altland:2025qqw}. In higher dimensions, such a formulation is complicated by the lack of holomorphic factorization and the difficulties encountered in formulating an effective action for the conformal anomaly, $e.g.$, \cite{Riegert:1984kt,Erdmenger:1996yc}.

It is natural to ask whether it is possible to re-express the bilocal OPE blocks
formulated in this work in an un-integrated form to connect to the formulation
in terms of reparametrizations. 
In this section, we comment on a way to achieve this in two dimensions for single-$T$ exchange 
that is generalizable to the less-understood four-dimensional scenario.

\subsection{Two Dimensions}

Here we first review the stress tensor case 
and then consider a spin-3 conserved current generalization, 
which we will find useful for the four-dimensional construction.

\vspace{0.5cm}

\noindent {\it Reparametrized Spin-2 OPE Block}

We briefly review the shadow operator formalism in two dimensions in terms of a regularizing operator 
$\hat\epsilon$ whose construction is motivated by the theory of reparametrizations \cite{Haehl:2019eae}.
The basic idea is to write the stress-tensor operator and its shadow (formally) as descendants of an auxiliary mode with negative dimension:
\be
\label{eq:2dTeps}
T \equiv -\frac{\pi c}{6}  \, \partial^3 \hat\epsilon\,,
\qquad \widetilde{T} = \frac{2\pi c}{3} \, \bar\partial \hat\epsilon\,.
\ee 
This turns out to furnish a convenient rewriting of the 
shadow operator formalism for stress tensor exchanges. The OPE block in the shadow
 formalism is a conformal integral \cite{Simmons-Duffin:2012juh}, and expressing it in terms of \eqref{eq:2dTeps} yields 
\be
\label{eq:BOOT2d}
\begin{split}
 \mathfrak{B}_{{\cal OO}T}(\hat\epsilon|\vecz_1,\vecz_2) 
 &= -\frac{1}{2}\int d^2\vecxi \; \frac{\langle {\cal O}(\vecz_1){\cal O}(\vecz_2) \widetilde{T}(\vecxi)\rangle}{\langle {\cal O}(\vecz_1){\cal O}(\vecz_2)\rangle} \, \partial^3 \hat\epsilon(\vecxi) \,.
\end{split}
\ee 
Let us first consider integrating the entire spacetime (without using a causal diamond). 
The writing in terms of $\hat\epsilon$ allows us to integrate by parts three times.
 From \eqref{eq:TtTtOO} we see that the first two partial integrations are trivial;
 the third yields contact terms (related to the anomalous Ward identity) due to the relation 
  $\partial_\xi \big(\frac{1}{\bar \xi - \bar z_i} \big) = \pi \delta^{(2)}(\vecxi-\vecz_i)$. 
 The delta functions localize the integral at the operator insertions, giving 
\be
\label{eq:BTblock}
 \mathfrak{B}_{{\cal OO}T}(\hat\epsilon|\vecz_1,\vecz_2) =-\frac{\pi C_{T{\cal O}{\cal O}}}{2} \,  \, \left( \partial \hat\epsilon(\vecz_1) + \partial \hat \epsilon(\vecz_2) - \frac{2}{z_{12}} (\hat\epsilon(\vecz_1) - \hat\epsilon(\vecz_2)) \right)\,.
\ee 
This expression has a familiar issue: it contains the short-distance monodromies
 associated with both the physical stress tensor and its shadow:
\be 
 \lim_{\vecz_2 \rightarrow \vecz_1} \,  \mathfrak{B}_{{\cal OO}T}(\hat\epsilon|\vecz_1,\vecz_2)
 = -\frac{\pi C_{T{\cal OO}}}{12}\left[ \left( \partial^3 \epsilon(\vecz_1) \, z_{12}^2 + \ldots \right)
 +\left(-12\,\bar\partial\epsilon(\vecz_1) \, \frac{\bar z_{12}}{z_{12}} + \ldots \right)\right]\,.
\ee
This may be remedied by suitably projecting out the shadow component by demanding $\bar{z}_1=\bar{z}_2$ (while $z_1 \neq z_2$).\footnote{We remark that this is also the condition required to make $\mathfrak{B}_{{\cal OO}T}(\hat\epsilon|\vecz_1,\vecz_2)$ invariant under {\it local} $SL(2,\mathds{R})$ transformations $\hat\epsilon(\vecz) \rightarrow \hat\epsilon(\vecz) + c_0(\bar z) + c_1(\bar z) \, z + c_2(\bar z) \, z^2$. In addition, for the multi-stress tensor OPE blocks the $SL(2,\mathds{R})$ invariance also becomes a local ($\bar z$-dependent) redundancy in the limit $\bar z_{12}\rightarrow 0$; 
see the related discussion in Sec. \ref{sec:virasoroOPE}.} 

Consider now a similar strategy by plugging \eqref{eq:2dTeps} into the 
OPE block \eqref{eq:BTdef} (or equivalently \eqref{eq:BTdef2}) with a causal diamond.  
This takes the same form as \eqref{eq:BOOT2d} but with an 
integration region restricted to a causal diamond $\dia_2^1$. Now, upon partial 
 integration, any potential contact terms have no support within the (regularized) 
 integration region. Nevertheless the diamond integral is not zero: integration by parts
 now yields {\it boundary terms} along the edges
 of the diamond. In the limit $\bar z_{12} \rightarrow 0$, we re-discover the corresponding limit of same expression as 
above, \eqref{eq:BTblock}:
\be
\label{eq:BTblockDiamond}
\begin{split}
\lim_{\bar z_2 \rightarrow \bar z_1 = \bar z} \, {\cal B}_T^{(\cal O)}(\hat\epsilon| \vecz_1,\vecz_2)
&= \frac{h}{2} \int_{z_2}^{z_1} d\xi \, \frac{(z_1-\xi)(\xi-z_2)}{z_1-z_2} \, \partial^3\hat\epsilon(\xi,\bar z) \\
&= -\frac{\pi C_{T{\cal OO}}}{2}\left( \partial\hat\epsilon(z_1,\bar z) + \partial\hat\epsilon(z_2,\bar z) - \frac{2}{z_{12}}\left( \hat\epsilon(z_1,\bar z) - \hat\epsilon(z_2,\bar z)\right)\right).
\end{split}
\ee
 
As a consistency check, we should reproduce the correct single-stress tensor exchange contribution to the correlator using  \eqref{eq:BTblockDiamond}.
To do the computation, we need the two-point function of $\hat\epsilon$, which can be computed using 
\be
  \langle T(z_1) T(z_2) \rangle = \frac{c}{2z_{12}^4} = \left(-  \frac{\pi c}{6}\right)^2 \partial_1^3 \partial_2^3 \langle \hat\epsilon(\vecz_1) \hat\epsilon(\vecz_2) \rangle  
\ee
which gives
\be
\label{eq:epseps2d}
\langle \hat\epsilon(\vecz_1) \hat\epsilon(\vecz_2) \rangle = \frac{3}{2\pi^2 c} \, z_{12}^2 \, \ln(z_{12} \bar{z}_{12}) + \ldots
\ee 
where any integration constants $\ldots$ do not contribute to physical 
observables as they are associated with $SL(2,\mathds{R})$ zero modes.
The resulting 
correlator of bilocal operators produces the global stress-tensor block:
\be
\label{eq:BTblockDiamondCalc}
  \left\langle\left( \lim_{\bar z_2 \rightarrow \bar z_1} {\cal B}_T^{({\cal O})}(\hat\epsilon| \vecz_1,\vecz_2) \right)\left(\lim_{\bar z_4 \rightarrow \bar z_3}{\cal B}_T^{({\cal O}')}(\hat\epsilon| \vecz_3,\vecz_4) \right)\right\rangle
= \frac{hh'}{2\pi^2 c} \; f_2(z) \,.
\ee

The above construction is connected to a representation of the Virasoro vacuum module in terms 
of reparametrization modes \cite{Cotler:2018zff,Haehl:2018izb,Nguyen:2021jja}, 
governed by the coadjoint orbit action \cite{Witten:1987ty,Alekseev:1988ce}. 
The starting point of such an approach is the generating 
functional for stress tensor correlators, sourced by a metric perturbation due to
a coordinate reparametrization
\be
z \rightarrow f(\vecz) = z + \epsilon(\vecz) + \ldots \ .
\ee
At the quadratic level, the induced generating functional is of the form
\be
\label{eq:W22d}
 W_2[\epsilon] = \frac{c}{24\pi} \int d^2z \; \bar\partial\epsilon \; \partial^3 \epsilon \,.
\ee
The coordinate transformation also acts covariantly on the conformal two-point function 
$\langle {\cal O}(\vecz_1) {\cal O}(\vecz_2)\rangle$. After normalizing the result, it transforms into
\be
 {\cal B}^{({\cal O})}(\epsilon|\vecz_1,\vecz_2) \equiv 
 z_{12}^{2h} \left( \frac{\partial f(\vecz_1) \, \partial f(\vecz_2)}{(f(\vecz_1)-f(\vecz_2))^2} \right)^h 
 = 1 + h \left( \partial \epsilon(\vecz_1) + \partial \epsilon(\vecz_2) - \frac{2}{z_{12}} (\epsilon(\vecz_1) - \epsilon(\vecz_2)) \right)  + \ldots 
\ee
which takes the same form as \eqref{eq:BTblock}.
A systematic study of this approach and its relation to OPE blocks will be given in \cite{toappear}.

\vspace{0.5cm}

\noindent {\it Reparametrized Spin-3 OPE Block}

It will be useful for Sec. \ref{sec:reparam4d} to extend the above analysis to  the exchange of a spin-3 conserved current. This corresponds to CFTs with an extended $W_3$ symmetry \cite{Zamolodchikov:1985wn}.\footnote{See, $e.g.$, \cite{Bouwknegt:1992wg, deBoer:2014sna, Perlmutter:2016pkf, Karlsson:2021mgg} for some reviews and discussions of higher-spin CFTs in two dimensions.} There is a (holomorphic) spin-3 current $W \equiv W_{zzz}$, which we normalize via 
\be
\label{eq:WW}
\langle W(z_1) W(z_2) \rangle = \frac{c}{3} \frac{1}{(z_1-z_2)^6} \,.
\ee
We define the following bilocal OPE block that describes the fusion into a spin-3 operator:
\be
\label{eq:BOW}
{\cal B}^{({\cal O})}_W(\vecz_1,\vecz_2) \equiv n_W  \int_{\diasmall_2^1} d^2\vecxi \; \frac{\langle \widetilde{W}(\vecxi){\cal O}(\vecz_1) {\cal O}(\vecz_2) \rangle}{\langle {\cal O}(\vecz_1) {\cal O}(\vecz_2) \rangle} \, W(\xi) \,
\ee
where   
\be
\frac{\langle \widetilde{W}(\vecxi){\cal O}(\vecz_1) {\cal O}(\vecz_2) \rangle}{\langle {\cal O}(\vecz_1) {\cal O}(\vecz_2) \rangle} \equiv C_{W{\cal O}{\cal O}} \, \frac{(z_1-\xi)^2(\xi-z_2)^2(\bar z_2-\bar z_1)}{(\bar z_1-\bar \xi)(\bar \xi-\bar z_2)( z_2-z_1)^2} \,  \ .
\ee 
As before, the short-distance limit implies the normalization:
\be
 \lim_{\vecz_2 \rightarrow \vecz_1} \; \frac{{\cal B}^{({\cal O})}_W(\vecz_1,\vecz_2)}{(z_1-z_2)^3} = \frac{3\,C_{W{\cal O}{\cal O}}}{c} \, W(z_1) 
 \qquad \Rightarrow \qquad  n_W = \frac{45}{c \, \ln \delta}\,.
 \ee 
Including the normalization, performing the anti-holomorphic integral of \eqref{eq:BOW} yields
\be \label{eq:BW2}
{\cal B}^{({\cal O})}_W(\vecz_1,\vecz_2) = \frac{90\, C_{W{\cal O}{\cal O}}}{c}  \int_{z_2}^{z_1} d\xi \; \frac{(z_1-\xi)^2(\xi-z_2)^2}{(z_1-z_2)^2} \, W(\xi) \,.
\ee Similar to the holomorphic spin-2 case, any anti-holomorphic dependence is eliminated.

Next we consider a rewriting in terms of a mode operator $\hat{w}$ defined via 
\be
\label{eq:Wwrelation}
  W = \frac{c}{12} \, \alpha_w \, \partial^5 \hat{w} \,
\ee
where the normalization $\alpha_w$ will be irrelevant in the following discussions. Formally, $\hat{w}$ has conformal weights $(h,\bar h)_{\hat{w}}=(-2,0)$, and it can be interpreted in analogy with the spin-2 case. In particular, it furnishes a useful formulation of the shadow operator formalism, where one would write $\widetilde{W} \propto \bar\partial \hat{w}$; this would then lead to an un-integrated expression for the projector $|W|$ onto the conformal family associated with $W$, etc.  
We shall not pursue these ideas further here. Instead, we are interested in obtaining an expression for the single-$W$ OPE block in terms of $\hat{w}$.  
Plugging the relation \eqref{eq:Wwrelation} into \eqref{eq:BW2}, the OPE block in terms of $\hat{w}$ again localizes to the tips of the diamond:
\be 
\label{eq:Bw2}
\lim_{\bar z_2\rightarrow\bar z_1}\,{\cal B}^{({\cal O})}(\hat{w}|\vecz_1,\vecz_2) = 15\,\alpha_w \, C_{W{\cal O}{\cal O}}\, \left[ \left(\partial^2\hat{w}(\vecz_1)  - \frac{6}{z_{12}} \,\partial \hat{w}(\vecz_1) + \frac{12}{z_{12}^2} \, \hat{w}(\vecz_1) \right) \; - \; (\vecz_1 \leftrightarrow \vecz_2) \right] \,.
\ee
This bilocal expression is the spin-3 generalization of \eqref{eq:BTblock}. Note that, in this case, the expression involves a second derivative acting on the reparametrization modes.\footnote{The spin-3 OPE block \eqref{eq:Bw2} is invariant under $\hat{w} \rightarrow \hat{w} + c_0 + c_1 z+ c_2 z^2 + c_3 z^3 + c_4 z^4$. In the limit $\bar z_{1} \rightarrow  \bar{z}_2 \equiv \bar z$ the coefficients $c_i$ can be made $\bar z$-dependent.}

As an application and consistency check, let us verify that these building blocks compute the 
$W_3$ global conformal block. To this end, we need the two-point function of $\hat{w}$, which follows directly from \eqref{eq:WW}:
\be
 \langle \hat{w}(\vecz_1) \hat{w}(\vecz_2) \rangle = \frac{1}{60\, \alpha_w^2\,c} \; z_{12}^4 \ln(z_{12}\,\bar z_{12}) + \ldots \,.
\ee
Using this, we obtain  
\be 
\left\langle \left(\lim_{\bar z_2\rightarrow \bar z_1}\,{\cal B}^{({\cal O})}(\hat w|\vecz_1,\vecz_2) \right)\left( \lim_{\bar z_4 \rightarrow \bar z_3}\, {\cal B}^{({\cal O}')}(\hat w|\vecz_3,\vecz_4) \right)\right\rangle = \frac{3 \, C_{W{\cal O}{\cal O}} \, C_{W{\cal O}'{\cal O}'}}{c} \; f_3(z) \,,
\ee
where $f_3(z) = z^3 \; {}_2F_1(3,3,6,z)$. This reproduces the known single-$W_3$ exchange contribution to the four-point correlator; see, $e.g.$, \cite{Karlsson:2021mgg}.

Let us conclude this subsection with a curious observation. 
By expanding the bilocal spin-3 OPE block \eqref{eq:BW2} to higher orders, we find
\be
\label{eq:OPE2dW3block}
 \lim_{\vecz_2 \rightarrow \vecz_1} \; {\cal B}^{({\cal O})}_W(\vecz_1,\vecz_2) =  \frac{3\,C_{W{\cal O}{\cal O}}}{c} \, z_{12}^3 \left[ W(z_2) +\frac{1}{2} \, \partial W(z_2)  z_{12} + \frac{1}{7} \, \partial^2 W(z_2) z_{12}^2 + \ldots \right] \,.
\ee
This expansion contains information about the operator $W$ and its conformal descendants in the ${\cal O}{\cal O}$ OPE. 
We observe that the short-distance expansion \eqref{eq:OPE2dW3block} in the case of the spin-3 current matches precisely -- to all orders -- the analogous expansion of the four-dimensional single-stress tensor OPE block at leading order in the lightcone limit given in \eqref{eq:OPE4dblock}. 
Motivated by this, we will show that introducing an effective spin-3 reparametrization mode
in four dimensions will enable us to rewrite the OPE block in an un-integrated form.

\subsection{Four Dimensions} 
\label{sec:reparam4d}

Now we consider four-dimensional CFTs. We first discuss some subtleties involved in generalizing the two-dimensional discussion, and then we propose a writing of the bilocal OPE block in terms of a reparametrization field, which determines the near-lightcone correlator.

\subsubsection{Obstructions Towards a Covariant Effective Description}

Can one express the four-dimensional single-stress tensor OPE block \eqref{eq:BTdef4d} in a simple, un-integrated form based on local reparametrization modes that automatically eliminate the shadow block contribution?  Developing an effective field theory for such reparametrizations in higher dimensions would be interesting, and we would like to draw lessons from the two-dimensional construction.
However, a covariant generalization faces obstacles, as we will review below. This motivates us, as a simpler starting point, to focus on the near-lightcone OPE block constructed in \eqref{eq:BTlc1}. It might then be possible to systematically generalize the computation order by order in a near-lightcone expansion; we leave such an exploration to future work.

Let us start with a brief review of the four-dimensional covariant construction based on \cite{Haehl:2019eae}, which re-expresses the shadow operator formalism for stress-tensor exchange.  
As in two dimensions, it turns out to be convenient to write the shadow of the stress tensor
 as a symmetric-traceless ``descendant'' of a formal negative-dimension operator $\hat\epsilon^\mu$:
\be
\label{eq:Tt4deps}
\widetilde{T}_{\mu\nu}(x) \equiv \frac{2\pi^2 C_T}{k_{0,2}}\, \mathbb{P}_{\mu\nu}^{\rho\sigma} \, \partial_\rho \hat{\epsilon}_\sigma(x)\,,
\ee
where $\mathbb{P}_{\mu\nu}^{\rho\sigma} = \frac{1}{2} \left( \delta_\mu^\rho \delta_\nu^\sigma + \delta_\mu^\sigma \delta_\nu^\rho \right) -\frac{1}{4} \, \eta_{\mu\nu} \eta^{\rho\sigma}$. From this definition, an expression for $T_{\mu\nu}(x)$ can be worked 
out via a shadow transform. The relevant components in the lightcone limit are
\be
\label{eq:TppEps}
T_{++}(x) \propto \left(  \partial_\mu \partial_+ - \frac{3}{4} \,\eta_{\mu+}\,\Box \right)\partial_+\Box  \hat\epsilon^\mu(x) \,,\qquad \widetilde{T}^{++} \propto \partial_- \hat\epsilon^+(x) \,.
\ee
Similar expressions can be motivated by writing a quadratic generating functional
sourced by a reparametrization 
$x^\mu \rightarrow x^\mu + \epsilon^\mu(x)$. One finds an expression analogous  
to \eqref{eq:W22d}:
\begin{equation}
\label{4daction}
   W_2[\epsilon]  \propto C_T \int d^4x \; \mathbb{P}_{\mu\nu}^{\rho\sigma} \partial_\rho \epsilon_\sigma(x) \; \,\left( \partial^\lambda\partial^{(\mu} - \frac{3}{4} \Box\,\eta^{\lambda(\mu}  \right)  \partial^{\nu)} \,\Box \epsilon_\lambda(x)  \, 
\end{equation} which features a six-derivative kinetic term for the reparametrization 
mode $\epsilon^\mu(x)$. 
The first issue with this higher-dimensional setup is the non-universality of higher order terms in the action: while the two-dimensional action \eqref{eq:W22d} can easily be extended to the non-linear level \cite{Witten:1987ty,Alekseev:1988ce,Cotler:2018zff}, no satisfactory prescription for a four-dimensional generalization has been put forward.\footnote{The situation is worse in odd dimensions, where even the guiding principle of anomaly actions is absent. This motivates our analysis in Sec.\ \ref{sec:threedim}.}

Let us turn to the second issue with this construction. The motivation for introducing the reparametrization mode is to enable a
convenient bilocal approach to the shadow operator formalism without the need to 
perform conformal (or diamond) integrals. 
Using covariant relations such as \eqref{eq:Tt4deps} and \eqref{eq:TppEps}, one can proceed as in two dimensions
and express the bilocal OPE block \eqref{eq:BTdef4d} as \cite{Haehl:2019eae}
\be
{\cal B}^{({\cal O})}_{\text{cov.}}(\hat \epsilon| x_1,x_2) \equiv \Delta_{\cal O} \left[ \frac{1}{4} \big(\partial_\mu \hat\epsilon^\mu(x_1)+\partial_\mu \hat\epsilon^\mu(x_2) \big) -2 \, \frac{(\hat\epsilon^\mu(x_1) - \hat\epsilon^\mu(x_2))^\mu(x_1-x_2)_\mu}{(x_1-x_2)^2} \right] \,.
\label{eq:B4dCov}
\ee
This object does not allow us to compute the physical conformal block, not even at the level of single-stress tensor exchange. Instead, it computes the single-valued linear combination of the stress-tensor conformal block and its shadow  \cite{Haehl:2019eae}. 
The monodromy projection onto the physical block -- which was easy to implement in two dimensions thanks to the holomorphic factorization -- presents a non-trivial constraint on the effective description in terms of the reparametrization modes in higher dimensions. We are not aware of any choice of correlator $\langle \hat\epsilon^\mu(x_1) \hat\epsilon^\nu(x_2) \rangle$ that would compute only the physical (monodromy-projected) global conformal block via $\langle {\cal B}^{({\cal O})}_{\text{cov.}}(\hat \epsilon| x_1,x_2){\cal B}^{({\cal O}')}_{\text{cov.}}(\hat \epsilon| x_3,x_4)\rangle$. 
It is unclear how to project out the shadow contribution using this description.

\subsubsection{Lightcone Effective Reparametrization Modes}

We will now develop an effective theory of stress-tensor
exchanges in four dimensions in lightcone kinematics. 
We find that analyzing the OPE block through an expansion in the lightcone limit 
circumvents the impasses discussed above.

First recall the correlator computation based on causal diamond integrals \eqref{eq:BB4dCalc}. As noted above, the dependence on the null coordinate $\xi^+$ plays a special role and controls the lightcone kinematics: we argued at the end of Sec.\ \ref{sec:4d} that the exchange of the $T_{++}$ component along a transverse plane faithfully captures the single-stress tensor exchange to leading order in $x^-$.
It is therefore natural to perform all integrals except for the $\xi^+$ and $\eta^+$ integrals in the lightcone OPE block correlator \eqref{eq:singleTsimplifiedResult}, which effectively controls the relevant exchange. After consistently taking the limit $\Lambda\rightarrow\infty$, we find
 \begin{align}
 \label{eq:4dIntCald}
 &\lim_{x^-\rightarrow 0}\, \left\langle  {\cal B}^{(\perp=0)}_{T_{++}}(\Lambda,1) \, {\cal B}_{T_{++}}^\times(x^\pm) \right\rangle  \\
&\qquad = \frac{15^2\,C_{T{\cal O}{\cal O}}C_{T{\cal O}'{\cal O}'}}{\pi^2C_T^2}\, \frac{x^-}{(x^+)^2}  \int_0^{x^+} d\eta^+ \int_1^{\infty}d\xi^+   \left[   \, (x^+-\eta^+)^2(\eta^+)^2(\xi^+-1)^2 \right] \, \frac{\pi^2C_T}{(\xi^+-\eta^+)^6}  \ . \nn
 \end{align}
To express these null-line integrals in terms of auxiliary modes analogous to reparametrizations, we draw inspiration from the observation that the short-distance OPE limit of the lightcone block, \eqref{eq:OPE4dblock}, precisely matches the structure of the two-dimensional OPE involving a spin-3 current, ${\cal O}\times{\cal O} \rightarrow W + \text{descendants}$.  This observation motivates us to formulate the four-dimensional calculation in analogy with the spin-3 current exchange in two dimensions.\footnote{A similar connection has been made previously, $e.g.$, \cite{Huang:2021hye, Korchemsky:2021htm, Karlsson:2021mgg}. In the present OPE block context, we see a new incarnation of this connection.}

We introduce an auxiliary operator $\hat t(\xi^+)$ to describe the dependence on lightcone integrals \eqref{eq:4dIntCald}:
\be
\label{eq:thathat}
 \langle \hat t(\xi^+) \hat t(\eta^+) \rangle  \equiv  \frac{\pi^2 C_T}{(\xi^+ - \eta^+)^6} \,.
\ee
This resembles the structure of a spin-3 current exchange in two dimensions, $i.e.$, \eqref{eq:WW}. 
Formally, $\hat{t} \equiv \hat{t}_{+++}$ has dimension $\Delta_t = 3$ and spin $\ell_t = 3$. 
We would like to interpret $\hat{t}$ as an effective ``holomorphic'' stress-tensor-like field in an effective description of the near-lightcone dynamics of the four-dimensional correlator.  We emphasize that we do not establish this rigorously -- the following discussion is meant to give some evidence supporting the usefulness of this perspective.
 
Using the analogy with two-dimensional spin-3 current exchange, in the four-dimensional context we introduce an effective reparametrization operator $\hat{w}(\xi^+)$ via the formal relation
\be
\label{eq:tTOw}
 \hat t(\xi^+) = \frac{C_T}{12} \, \frac{\alpha_t}{\sqrt{x^-}} \, \partial_+^5 \hat{w}(\xi^+) \,.
\ee 
The two-point function of $\hat{w}$ is derived from the defining property \eqref{eq:thathat}:
\be 
\label{eq:omegaomega}
 \left\langle \hat{w}(\xi^+) \hat{w}(\eta^+)\right\rangle = \frac{\pi^2\, x^-}{20\,\alpha_t^2 C_T} \; (\xi^+ - \eta^+)^4 \ln( \xi^+ - \eta^+) + \dots \,.
\ee
The coordinate-independent normalization constant $\alpha_t$ will not be important in the following discussion, but note that we chose the normalization in \eqref{eq:tTOw} to absorb the factor $x^-$ appearing in \eqref{eq:4dIntCald} into the two-point function of the effective reparametrization field \eqref{eq:omegaomega}.
This is motivated by the fact that near-lightcone multi-stress tensor exchanges are organized in powers of $\frac{x^-}{C_T}$;  
the propagator $\langle\hat{w}\hat{w}\rangle$ of a putative effective theory should reflect this organizing principle. The above rewriting suggests that $x^-$ plays the role of an energy scale that controls the validity of this near-lightcone effective field  theory. 

With the above setup, we can return to the individual OPE blocks and express them in terms of the auxiliary modes. By inspection of \eqref{eq:4dIntCald} and the identification \eqref{eq:thathat}, it is clear how the effective OPE blocks along the null-ray should be identified in a symmetric fashion:
\be
\begin{split}
  \left\langle {\cal B}_\text{eff}(\hat{w}|\infty,1) \right| &\equiv \frac{15 \, \alpha_t\, C_{T{\cal OO}}}{12\pi} \int_1^\infty d\xi^+ \; (\xi^+-1)^2 \, \partial_+^5 \hat{w}(\xi^+) \,,\\
    \left| {\cal B}_\text{eff}(\hat{w}|x^\pm,0) \right\rangle&\equiv \frac{15 \, \alpha_t\,C_{T{\cal O'O'}}}{12\pi} \, \frac{1}{(x^+)^2}\int_0^{x^+} d\eta^+ \; (x^+-\eta^+)^2(\eta^+)^2 \, \partial_+^5\hat{w}(\eta^+) \,.
\end{split}
\ee
This is analogous to \eqref{eq:BW2}.
We find that in the lightcone OPE block all integrals can be performed and we obtain an un-integrated bilocal expression:
\begin{align}
\label{eq:BTlc4D}
&  \left| {\cal B}_\text{eff}(\hat{w}|x^\pm,0) \right\rangle= \\
&=  \frac{5\alpha_t C_{T{\cal O}{\cal O}}}{2\pi}\left| \left(\partial_+^2\hat{w}(x^+) -  \partial_+^2 \hat{w}(0) \right)- \frac{6}{x^+} \left( \partial_+ \hat{w}(x^+)+ \partial_+ \hat{w}(0)\right) + \frac{12}{(x^+)^2} \left(  \hat{w}(x^+)- \hat{w}(0)\right) \right\rangle  . \nn
 \end{align} 
This takes the same form as the two-dimensional spin-3 current OPE block \eqref{eq:Bw2}. A similar procedure in the ``bra'' yields a simple expression (due to one point being inserted at $\infty$):
\be
\begin{split}
 \left\langle {\cal B}_\text{eff}(\hat{w}|\infty,1) \right|   &= \frac{5\alpha_t\,C_{T{\cal O}'{\cal O}'}}{2\pi}\left\langle -\partial_+^2 \hat{w}(1)  \right| \,.
  \end{split}
 \ee 
Finally, using the propagator \eqref{eq:omegaomega}, we reproduce the single-stress tensor exchange contribution to the four-dimensional near-lightcone correlator without having to compute any 
diamond integrals:
\be
\left\langle {\cal B}_\text{eff}(\hat{w}|\infty,1) \,\big|\, {\cal B}_\text{eff}(\hat{w}|x^\pm,0) \right\rangle=  \frac{C_{T{\cal O}{\cal O}} C_{T {\cal O}'{\cal O}'}}{4 C_T} \; f_3(x^+) \, x^- \,.
\ee  
This computation does not require a large central charge. The heavy-light limit can be trivially implemented by setting ${\cal O} = {\cal O}_H$ and ${\cal O}' = {\cal O}_L$.

\section{Discussion}
\label{sec:discussion}

We have presented a new perspective on bilocal OPE blocks, emphasizing the exchange of multi-stress tensor operators.
Central to our construction are nested causal diamonds and the regularization prescriptions introduced to define suitable projectors.
The bilocal Virasoro OPE block formulated here can be considered a generalization of the kinematic space construction of the global OPE block ($i.e.$, modular Hamiltonian) \cite{Casini:2011kv, Czech:2016xec, deBoer:2016pqk}.  Our main motivation lies in its potential to be generalized to higher dimensions. 
We have demonstrated that our construction automatically removes the unphysical shadow block contributions and enables the direct computation of the single-stress tensor exchange contribution to lightcone correlators in higher-dimensional CFTs.  We have also discussed an effective description in terms of reparametrization modes in four dimensions.

Let us conclude with several open questions and avenues for further exploration.
An immediate question is whether our construction of multi-stress tensor OPE blocks can be extended to higher dimensions. 
Focusing on CFTs at large $C_T$ (and with a large higher-spin gap), such a construction would enable a first-principle CFT computation of multi-stress tensor contributions to conformal correlators, applicable to both odd and even dimensions. One technical challenge in generalizing to higher dimensions is identifying the relevant recursion relations needed to compute suitable integration kernels. Considering a near-lightcone limit should allow for various simplifications. On the other hand, recall that in the two-dimensional case, our construction -- after integrating over anti-holomorphic coordinates -- reduces to the Chern-Simons Wilson line networks \cite{Fitzpatrick:2016mtp}. It would be interesting, for instance, to explore whether a Wilson line-like picture can be developed based on the four-dimensional global OPE block near the lightcone we constructed. 

We took initial steps towards formulating an effective theory of reparametrization modes in four dimensions in Sec. \ref{sec:reparametrizations}, focusing on the near-lightcone stress-tensor exchange. A motivating and ambitious goal in this direction is to identify a symmetry that might emerge near the lightcone in a class of large-N CFTs.  If the higher-dimensional near-lightcone correlators are governed by an enhanced symmetry, one could expect to develop an efficient description analogous to the two-dimensional coadjoint orbit action, which determines the dynamics of the reparametrization modes describing lowest-twist multi-stress tensors. Constructing such a description starting from a covariant anomaly effective action is notoriously difficult in higher dimensions. Indeed, as we emphasized, even for the single-stress tensor exchange, it is unclear how to project out the shadow block contribution from the covariant action described in \cite{Haehl:2019eae}.  An interesting observation from our construction of the global OPE block, in both four and three dimensions, is that there exists a smeared version of the stress tensor which is effectively chiral in the limit where an external scalar approaches the lightcone. 
Further investigation into this emergent chirality in the near-lightcone dynamics of higher-dimensional CFTs at large central charge would be interesting. On the other hand, restricting to the lightcone limit breaks general covariance but allows us to perform monodromy projections more easily. 
In the present work, we merely demonstrate how to overcome the technical challenges at the level of the single-stress tensor OPE block.  Along with the similarities to spin-3 current exchanges in two dimensions ($e.g.$, \eqref{eq:thathat} and \eqref{eq:BTlc4D}), hopefully these results could facilitate future work to scrutinize and further explore higher-dimensional generalizations via symmetry-based effective actions.

We emphasize that even in two dimensions, the story is far from complete. Indeed, a particularly thorny challenge is to extend the computation to include higher-order quantum corrections in a systematic way, as this would require explicitly calculating correlators involving multiple regulated stress tensors (such as $ \langle [TT] [TT] \rangle$ at higher orders in a $1/c$ expansion), for which we currently lack explicit expressions. To gain further insights into the regularization and renormalization prescriptions, it might be helpful to explore the connections between our approach and the computations of the Virasoro identity block beyond the semi-classical limit using Virasoro modes \cite{Fitzpatrick:2015dlt, Chen:2016cms}, as well as to establish links with the reparametrization mode approach \cite{toappear}.

It would also be interesting to extend our approach to compute the contributions from the exchanges of double-trace operators. In this case, the integration kernel generally develops branch cuts, and one has to be careful when performing the analytic continuation.

\subsection*{Acknowledgments}
 
We would like to thank V.\ Benedetti, P.\ Caputa, P.\ Kraus, J.\ Mag\'{a}n, A.\ Shekar, P.\ Tadi\'{c} and G.\ van der Velde for helpful discussions and comments. This work is supported in part by UK Research and Innovation (UKRI) under the UK government’s Horizon Europe Funding Guarantee EP/X030334/1. FH is grateful to the Institut Pascal at Universit\'{e} Paris-Saclay for support through the program {\it Investissements d’avenir} (ANR-11-IDEX0003-01), and to the Kavli Institute for Theoretical Physics (NSF PHY-2309135).

\newpage
\appendix 

 \section{Conventions}
 \label{app:conventions}

\paragraph{\it Coordinates.}
Instead of Minkowski coordinates $\xi^\mu = (t,x,\xi^1,\xi^2)$ we often use the lightcone coordinates:
\be
 x^\pm = t \pm x  \,, \qquad ds^2 = -dx^+ dx^- + (d\xi^i)^2 \,.
\ee
In two dimensions, we adopt $\vecz\equiv (z,\bar z)$ instead of $(x^+,x^-)$. These analytically continue to Euclidean complex conjugate coordinates.

\paragraph{\it Universal Correlators.} 
We use the following normalizations:
\be
\label{eq:Tcorrs}
\begin{split}
\langle T^{\mu\nu}(x_1) T^{\rho\sigma}(x_2) \rangle &= \frac{C_T}{x_{12}^{2d}} \, {\cal I}^{\mu\nu,\rho\sigma}(x_{12}) \,,\\
\langle T^{\mu\nu}(x_0) {\cal O}(x_1) {\cal O}(x_2) \rangle &= \frac{C_{T{\cal O}{\cal O}}}{x_{01}^d x_{02}^d x_{12}^{2\Delta-d}} \left( \frac{X^\mu X^\nu}{X^2} - \frac{\eta^{\mu\nu}}{d}\right) \,,\quad X^\mu = \frac{x_{01}^\mu}{x_{01}^2} - \frac{x_{02}^\mu}{x_{02}^2} \, ,
\end{split}
\ee
where 
\be
 {\cal I}^{\mu\nu,\rho\sigma} = \frac{1}{2} \left( I^{\mu\rho}I^{\nu\sigma} + I^{\mu\sigma}I^{\nu\rho}\right) - \frac{1}{d} \, \eta^{\mu\nu}\eta^{\rho\sigma} \,,\qquad I^{\mu\nu}(x) = \eta^{\mu\nu} - \frac{2}{x^2} \, x^\mu x^\nu \,.
\ee 
The shadow transform relevant for the OPE block construction is \cite{Dolan:2011dv}
\be
\label{eq:TtcorrsApp}
\begin{split}
\langle \widetilde{T}^{\mu\nu}(x_0) {\cal O}(x_1) {\cal O}(x_2) \rangle &= \frac{1}{\Gamma\left(\frac{d}{2}+1\right)^2}\,\frac{C_{T{\cal O}{\cal O}}}{x_{12}^{2\Delta}} \left( \frac{X^\mu X^\nu}{X^2} - \frac{\eta^{\mu\nu}}{d}\right) \,.
\end{split}
\ee

The two-point correlator of scalar primary operators is
\be
\label{eq:OOnormalization}
 \langle {\cal O}(x_1) {\cal O}(x_2) \rangle = \frac{C_{{\cal O}}}{x_{12}^{2\Delta}} \, ,
\ee
where we often set $C_{\cal O}=1$.
For the stress tensor, $C_T$ is the central charge appearing in \eqref{eq:Tcorrs}. The three-point function coefficient is
\be
\label{eq:CTOOdef}
C_{T{\cal O}{\cal O}} = -\frac{\Gamma(\frac{d}{2}+1)}{\pi^{d/2}(d-1)}\,\Delta_{\cal O} \,.
\ee

In two dimensions we use index-free notation: 
$T \equiv T_{zz}$ and $\overline{T} \equiv  T_{\bar z\bar z}$. Similarly, $\widetilde{T} \equiv \widetilde{T}^{zz}$.
The metric is $ds^2 = dz d\bar z$.
The central charge $c$ in two dimensions is given by $C_T = 2c$.\footnote{For comparison with Ref. \cite{Fitzpatrick:2016mtp}, note that $c_\text{here} = (2\pi)^{-2} c_\text{\cite{Fitzpatrick:2016mtp}}$ due to the different normalization of the stress tensor.}
The stress tensor correlators in two dimensions are normalized as
\be
 \langle T(z_1) T(z_2) \rangle = \frac{c}{2z_{12}^4}\,,\qquad
 \langle T(z_0) {\cal O}(\vecz_1) {\cal O}(\vecz_2) \rangle = \frac{ C_{T{\cal O}{\cal O}}}{4}\, \frac{z_{12}^2}{z_{01}^2 z_{02}^2} \frac{1}{z_{12}^h \bar{z}_{12}^{\bar h}} \,,
\ee
with $C_{T{\cal O}{\cal O}} = -\frac{1}{\pi}\,\Delta_{\cal O}$.

 \section{Regularization Schemes and Projectors}
 \label{app:projectors}
 
 In this appendix, we provide more details about the regularization schemes in two dimensions, and discuss how they relate to the multi-stress tensor OPE block projector defined in \eqref{eq:projnDef}.
 
 First we note that the idempotence property in \eqref{eq:ProjProperties} is equivalent to the following: 
\be
\label{eq:idem}
\begin{split}
 \big\langle [T(\xi_1) \cdots T(\xi_n)] \, \{ \widetilde{T}(\vecxi'_1) \cdots \widetilde{T}(\vecxi'_n) \} \big\rangle 
 & = \big\langle [T(\xi_1) \cdots T(\xi_n)] \, \widetilde{T}(\vecxi'_1) \cdots \widetilde{T}(\vecxi'_n)  \big\rangle \\
 & = \sum_{\pi \in S_n} \langle T(\xi_1) \widetilde{T}(\vecxi'_{\pi(1)}) \rangle \cdots \langle T(\xi_n) \widetilde{T}(\vecxi'_{\pi(n)}) \rangle \\
 & = \left(\frac{\pi c}{3} \right)^n \sum_{\pi\in S_n}  \delta^{(2)}(\xi_1 - \xi'_{\pi(1)}) \cdots  \delta^{(2)}(\xi_n - \xi'_{\pi(n)}) \,
\end{split}
\ee
where the sum is over permutations of $n$ labels, $i.e.$, overall Wick contractions. The reduction of $\langle [T^n] T^n \rangle$ (or $\langle [T^n] \widetilde{T}^n\rangle$) to Wick contractions is one of defining properties of the regularization $[T^n]$. 
It plays a crucial role because it is the only way to ensure a product of delta-functions --  
these collapse the $2n$ integrals over the causal diamonds to $n$ integrals, therefore implementing idempotence of the projector $|T^n|_{(\vecz_1,\vecz_2)}$. 

A straightforward generalization of this argument leads to an analog of the composition rule discussed in \eqref{eq:compositionrule} for $|T^n|_{(\vecz_1,\vecz_2)} \,  |T^n|_{(\vecz_3,\vecz_4)}$.

On the other hand, the orthogonality property in \eqref{eq:ProjProperties} is equivalent to the following requirements: 
\begin{align}
\label{eq:kGn}
&~n<k:~~~  \big\langle [T(\xi_1) \cdots T(\xi_n)] \, \{ \widetilde{T}(\vecxi'_1) \cdots \widetilde{T}(\vecxi'_k) \} \big\rangle   = \big\langle [T(\xi_1) \cdots T(\xi_n)] \,  \widetilde{T}(\vecxi'_1) \cdots \widetilde{T}(\vecxi'_k)  \big\rangle = 0 \nn\\
\end{align}
\begin{align}
\label{eq:nGk}
 &n>k:~~~ \big\langle [T(\xi_1) \cdots T(\xi_n)] \, \{ \widetilde{T}(\vecxi'_1) \cdots \widetilde{T}(\vecxi'_k) \} \big\rangle  \nn\\
 &~~~~~~~~ =\!\!\! \sum_{\substack{\text{groupings}\\ \{ j_1,\ldots,j_{n-1} \}}} \big\langle T(\xi_1) \, \{ \widetilde{T}(\vecxi_{i_1}) \cdots \widetilde{T}(\vecxi_{i_{j_1}}) \} \big\rangle \cdots
\big\langle T(\xi_n) \, \{ \widetilde{T}(\vecxi_{i_{j_{n-1}+1}}) \cdots \widetilde{T}(\vecxi_{i_k}) \} \big\rangle  = 0 \, \ .
\end{align}
The sum runs over all possible ways to contract every $T(\xi_i)$, with any number of (at least one) $\widetilde{T}(\vecxi_j)$. 
Such groupings then ensure that the correlator vanishes, precisely due to the regularization $\{ \widetilde{T}\cdots \widetilde{T}\}$: since for $n>k$ there will be at least one factor $\langle T \, \{ \widetilde{T} \cdots \widetilde{T} \} \rangle$ with more than one $\widetilde{T}$. This factor is zero because the only contributions to $\langle T  \widetilde{T} \cdots \widetilde{T} \rangle$ are OPE singularities of the $\widetilde{T}$ operators, which the regularization scheme removes. 

In sum, the three properties  -- \eqref{eq:idem}, \eqref{eq:kGn}, and \eqref{eq:nGk} -- define both regularizations, $[T^n]$ and $\{\widetilde{T}^k\}$. The first two properties impose precisely the contraction rules \eqref{eq:FKregDef0} and \eqref{eq:FKregDef} and thus define $[T^n]$. Once $[T^n]$ is defined, then the property \eqref{eq:nGk} imposes that $\{\widetilde{T}^k\}$ removes terms singular in the short-distance OPE limit.\\

\noindent{\it Useful formulas}

For convenience, here we collect some basic formulas. We first have the basic correlators:
\be 
\label{eq:TTcorrsCollect}
\langle T(z_1) T(z_2) \rangle = \frac{c}{2} \frac{1}{z_{12}^4} \,,\qquad \langle \widetilde{T}(\vecz_1)\widetilde{T}(\vecz_2) \rangle = \frac{2c}{3} \frac{z_{12}^2}{\bar z_{12}^2} \,,\qquad  \langle T(z_1) \widetilde{T}(\vecz_2) \rangle = \frac{c\pi}{3} \, \delta^{(2)}(\vecz_{12}) \,,
\ee  where we recall that the two-dimensional stress tensor is holomorphic, while its shadow depends on both $z$ and $\bar z$.
Without any regularization, some examples for the higher-point stress-tensor correlators are
{\small
\begin{align}
 \langle T(z_1) T(z_2) T(z_3) \rangle &= -\frac{c}{2\pi} \frac{1}{z_{12}^2 z_{23}^2 z_{31}^2}  \,,\qquad 
 \langle T(z_1) \widetilde{T}(\vecz_2) \widetilde{T}(\vecz_3) \rangle = \frac{2c}{3} \, \frac{z_{23}^4}{z_{12}^2\bar z_{23}^2 z_{31}^2}\,, \\
 \langle T(z_1) T(z_2) T(z_3) T(z_4) \rangle &=  \frac{c^2}{4} \left(\frac{1}{z_{12}^4 z_{34}^4}+ \frac{1}{z_{13}^4 z_{24}^4}+ \frac{1}{z_{14}^4 z_{23}^4} \right)+ \frac{c}{2\pi^2} \left( \frac{1}{z_{12}^2 z_{34}^2 z_{14}^2 z_{23}^2} - \frac{1}{z_{12} z_{34} z_{14} z_{23} z_{13}^2 z_{24}^2} \right) \ . \nn
\end{align}}With regularization, some correlators relevant for our analysis are 
\be
\begin{split}
 &
 \langle [T(z_1)] T(z_2) T(z_3)  \rangle  = -\frac{c}{2\pi} \frac{1}{z_{12}^2 z_{23}^2 z_{31}^2}  \,,\qquad 
 \langle [T(z_1)] \widetilde T(\vecz_2) \widetilde{T}(\vecz_3) \rangle = \frac{2c}{3} \, \frac{z_{23}^4}{z_{12}^2\bar z_{23}^2 z_{31}^2}\,, \\
& \langle [T(z_1) T(z_2)] T(z_3) \rangle = 0 \,,\qquad
 \langle [T(z_1) T(z_2)] T(z_3) T(z_4) \rangle =  \frac{c^2}{4} \left( \frac{1}{z_{13}^4 z_{24}^4}+ \frac{1}{z_{14}^4 z_{23}^4} \right) \,,\\
& \langle [T(z_1) T(z_2)] [T(z_3) T(z_4)] \rangle =  \frac{c^2}{4} \left( \frac{1}{z_{13}^4 z_{24}^4}+ \frac{1}{z_{14}^4 z_{23}^4} \right)  + \ord(c) \,.
\end{split}
\label{eq:TTTTcollect}
\ee
We emphasize that the subleading term $\ord(c)$ in the correlator of a pair of the regulated stress tensor product are crucial for the orthogonality of the projectors; see \eqref{eq:ProjProperties}. To see that the nontrivial subleading term must exist, notice that the first relation in the second line of \eqref{eq:TTTTcollect} can only be true (in a non-trivial way) if the regulated product receives certain corrections at higher orders. These corrections, unfortunately, are more difficult to compute explicitly; see \cite{Fitzpatrick:2016mtp} for a more detailed discussion on the regularization $[ \, \cdot  \, ]$.\footnote{For instance, in the coincident limit, it is shown in \cite{Fitzpatrick:2016mtp} that $[ T(0) T(0) ] =  \frac{c}{22+5c}(5\,(TT)(0)-\frac{3}{2}\,T''(0))$, where $(TT)(z)= \oint {dw \over 2 \pi i}{T(w)T(z)\over (w-z)}$.}

For pure contact term correlators, similar formulas apply. In particular,
\be
\label{eq:TTTtTtreg}
\begin{split}
&\big\langle [T(z_1)T(z_2)] \{\widetilde{T}(\vecz_3)\widetilde{T}(\vecz_4)\}\big\rangle =
\big\langle [T(z_1)T(z_2)] \,\widetilde{T}(\vecz_3)\widetilde{T}(\vecz_4)\big\rangle
\\
&\qquad = \left( \frac{\pi c}{3} \right)^2 \,\big(\delta^{(2)}(\vecz_{13})\delta^{(2)}(\vecz_{24})+\delta^{(2)}(\vecz_{14})\delta^{(2)}(\vecz_{23})\big) \,.
\end{split}
\ee 
The subtraction of singular terms via $\{ \, \cdot \,\}$ can also be applied to pure stress tensor correlators ($i.e.$ no scalar involved in the correlators). For example, we have 
\be
\begin{split}
 \langle \{ T(z_1) T(z_2) \} T(z_3) \rangle &= -\frac{c}{2\pi} \frac{(z_{23}+2z_{13})}{ z_{23}^5 z_{31}^2}\,,
 \qquad
 \langle \{ \widetilde{T} (\vecz_1) \widetilde{T}(\vecz_2) \} T(z_3) \rangle = 0\,.
\end{split}
\ee
The first expression serves to illustrate the sense in which $\{ \,\cdot\,\}$ removes the OPE singularity of the associated operators; the second identity is crucial for the orthogonality of the projectors discussed in Sec. \ref{sec:projectors}.

\section{Virasoro Ward Identity}
\label{sec:ward}

In this appendix we perform some additional consistency checks on the proposed Virasoro identity OPE block \eqref{eq:BvirasoroGeneral}. These checks are similar to calculations performed previously in \cite{Fitzpatrick:2016mtp,Besken:2018zro,DHoker:2019clx,Nguyen:2022xsw}, so we will be brief and only illustrate one particular complementary calculation.

Consider the following operator:
\be
\Psi^{(2n)}[\mu; \vecz_1,\ldots ,\vecz_{2n}] \equiv 
e^{i  \int  \mu T} \, \prod_{i=1}^n \langle {\cal O}(\vecz_{2i-1}){\cal O}(\vecz_{2i}) \rangle\,{\cal B}^{({\cal O})}(\vecz_{2i-1};\vecz_{2i})\,
\ee
where 
\be
\label{eq:Ballorders}
{\cal B}^{({\cal O})}(\vecz_1;\vecz_2) = 1 + {\cal B}_{T}^{({\cal O})}(\vecz_1;\vecz_2) + {\cal B}_{T^2}^{({\cal O})}(\vecz_1;\vecz_2) + \ldots
\ee
One can demonstrate that the Virasoro Ward Identity (VWI) is satisfied by $\Psi^{(2n)}$, which provides a formal proof that the Virasoro OPE blocks compute Virasoro identity conformal block to all orders in the $1/c$ expansion.  
\vspace{0.3cm}

\noindent{\it Stress Tensor Dressing of OPE blocks}

One way to verify the VWI is by observing that correlators between ${\cal O} {\cal O} $ and any number of stress tensors are computed by the bilocal \eqref{eq:Ballorders}. This then verifies the Ward identity for $n=1$. We claim 
\be
 \left\langle {\cal B}^{({\cal O})}(\vecz_1;\vecz_2)\,  T(y_1) \cdots T(y_m) \right\rangle = \frac{\langle {\cal O}(\vecz_1) {\cal O}(\vecz_2) T(y_1) \cdots T(y_m) \rangle}{\langle {\cal O}(\vecz_1) {\cal O}(\vecz_2)\rangle}  \,.
\ee
We will now check this statement explicitly for small $m$.
\\

\noindent\underline{$m=0$}: Without any stress tensor insertions, $\langle  {\cal B}^{({\cal O})}(\vecz_1,\vecz_2)\rangle = 1$ follows from the regularization used to define the OPE blocks.
\\

\noindent\underline{$m=1$}: For a single stress tensor we find
\be
\begin{split}
  \left\langle {\cal B}^{({\cal O})}(\vecz_1;\vecz_2)\,  T(y_1) \right\rangle
  &=  \left\langle {\cal B}^{({\cal O})}_T(\vecz_1;\vecz_2)\,  T(y_1) \right\rangle = \frac{\langle {\cal O}(\vecz_1) {\cal O}(\vecz_2) T(y_1)\rangle}{\langle {\cal O}(\vecz_1) {\cal O}(\vecz_2)\rangle} \,
\end{split}
\ee 
where the first equality follows from the fact that $\langle T \rangle = 0$ (for $n=0$ stress tensor insertions), $\langle [T(\xi)] T(y_1)\rangle = \langle T(\xi)T(y_1)\rangle$ (for $n=1$ stress tensor), and all ${\cal B}^{(\cal O)}_{T^k}$ for $k > 1$ lead to integrals over regulated correlators of the form $\langle [T(\xi_1) \cdots T(\xi_k)]\, T(y_1) \rangle = 0$. The second equality follows from explicit computation using the diamond integral \eqref{eq:BTdef}.
\\

\noindent\underline{$m=2$}: Similarly, for two stress tensors, we find
\be
\begin{split}
  \left\langle {\cal B}^{({\cal O})}(\vecz_1;\vecz_2)\,  T(y_1) T(y_2) \right\rangle
  &= \left\langle \left(1+{\cal B}^{({\cal O})}_{T}(\vecz_1;\vecz_2) + {\cal B}^{({\cal O})}_{T^2}(\vecz_1;\vecz_2) \right)\,  T(y_1) T(y_2) \right\rangle \,.
\end{split}
\ee 
By definition of the regularization scheme all higher order terms involving $B_{T^k}^{({\cal O})}$ with $k>2$ have to vanish. It is also obvious that the first term on the r.h.s.\ gives $\langle T(y_1)T(y_2)\rangle$, which is one term that makes up $\langle {\cal O}{\cal O}TT\rangle / \langle {\cal O}{\cal O}\rangle$. However, it is non-trivial to see how the second and third terms combine into the form required to retrieve the full $\langle {\cal O}{\cal O}TT\rangle / \langle {\cal O}{\cal O}\rangle$. From the corresponding diamond integrals \eqref{eq:BTdef} and \eqref{eq:BTTdef} we find
\be
\label{eq:BTTcalc}
\begin{split}
\left\langle {\cal B}^{({\cal O})}_{T}(\vecz_1;\vecz_2) \,  T(y_1) T(y_2) \right\rangle
&= -\frac{6h}{\pi^2} \frac{1}{(y_1-y_2)^4} \left[ 1 + f(z_i,y_i) \ln \frac{(y_1-z_1)(y_2-z_2)}{(y_1-z_2)(y_2-z_1)} \right]
\,\\
\left\langle {\cal B}^{({\cal O})}_{T^2}(\vecz_1;\vecz_2) \,  T(y_1) T(y_2) \right\rangle
&= \frac{h^2}{4\pi^2} \frac{(z_1-z_2)^4}{(y_1-z_1)^2(y_2-z_2)^2(y_1-z_2)^2(y_2-z_1)^2} \\
 &\quad + \frac{h}{2\pi^2} \frac{(z_1-z_2)^2}{(y_1-y_2)^2(y_1-z_1)(y_2-z_2)(y_1-z_2)(y_2-z_1)} \\
 &\quad + \frac{6h}{\pi^2} \frac{1}{(y_1-y_2)^4} \left[ 1 + f(z_i,y_i) \ln \frac{(y_1-z_1)(y_2-z_2)}{(y_1-z_2)(y_2-z_1)} \right]
\,
\end{split}
\ee
where $f(z_i,y_i)$ is a rational function whose form will not matter. We can see that a non-trivial cancellation occurs between the two terms in \eqref{eq:BTTcalc}. The terms that survive are precisely those needed to complete $\langle {\cal O}{\cal O}TT\rangle / \langle {\cal O}{\cal O}\rangle$.


\bibliographystyle{JHEP}
\bibliography{references.bib}

\end{document}